\documentclass[twocolumn,useAMS,usenatbib]{mnras}

\usepackage{color}
\usepackage{amsmath}
\usepackage{graphicx}
\usepackage{amssymb}
\usepackage{psfrag}

\usepackage{etoolbox}
\makeatletter
\patchcmd\@combinedblfloats{\box\@outputbox}{\unvbox\@outputbox}{}{%
   \errmessage{\noexpand\@combinedblfloats could not be patched}%
}%
 \makeatother

\newcommand{\beq}{\begin{equation}}
\newcommand{\eeq}{\end{equation}}
\newcommand{\bay}{\begin{array}}
\newcommand{\eay}{\end{array}}
\newcommand{\beqa}{\begin{align}}
\newcommand{\eeqa}{\end{align}}
\newcommand{\beqy}{\begin{eqnarray}}
\newcommand{\eeqy}{\end{eqnarray}}
\newcommand{\nn}{\nonumber}
\newcommand{\rmd}{\mathrm{d}}
\newcommand{\brac}[1]{\left({#1}\right)}
\newcommand{\pd}[2]{\frac{\partial{#1}}{\partial{#2}}}
\newcommand{\td}[2]{\frac{\rmd{#1}}{\rmd{#2}}}
\newcommand{\curl}{\nabla\times}
\renewcommand{\div}{\nabla\cdot}

\newcommand{\bB}{{\boldsymbol B}}

\newcommand{\be}{{\boldsymbol e}}

\newcommand{\br}{{\boldsymbol r}}
\newcommand{\bv}{{\boldsymbol v}}

\newcommand{\skl}[1]{{\color{black}{#1}}}

\title[NS spindown and inclination-angle evolution]{Neutron-star spindown and magnetic inclination-angle evolution}
\author[S. K. Lander and D. I. Jones]{S. K. Lander${}^1$\thanks{skl@camk.edu.pl},
         D. I. Jones${}^2$\\ \\
         ${}^1$Nicolaus Copernicus Astronomical Centre, Polish Academy of Sciences, Bartycka 18, 00-716 Warsaw, Poland,\\
         ${}^2$Mathematical Sciences and STAG Research Centre, University of Southampton, Southampton SO17 1BJ, U.K.}

\begin{document}

\pagerange{\pageref{firstpage}--\pageref{lastpage}} \pubyear{0000}
\maketitle

\label{firstpage}

\begin{abstract}
A rotating fluid star, endowed with a magnetic field, can undergo a form of precessional motion: a sum of rigid-body free precession and a non-rigid response. On secular timescales this motion is dissipated
by bulk and shear viscous processes in the stellar interior and magnetospheric braking in the exterior, changing the inclination angle between the rotation and magnetic axes. Using our recent solutions for the non-rigid precessional dynamics, and viscous dissipation integrals derived in this paper, we make the only self-consistent calculation to date of these dissipation rates. We present the first results for the full coupled evolution of spindown and inclination angle for a model of a late-stage proto-neutron star with a strong toroidal magnetic field, allowing for both electromagnetic torques and internal dissipation when evolving the inclination angle. We explore this coupled evolution for a range of initial inclination angles, rotation rates and magnetic field strengths. For fixed initial inclination angle, our results indicate that the neutron-star population naturally evolves into two classes: near-aligned and near-orthogonal rotators -- with typical pulsars falling into the latter category.  Millisecond magnetars can evolve into the near-aligned rotators which mature magnetars appear to be, but only for small initial inclination angle and internal toroidal fields stronger than roughly $5\times 10^{15}$ G. Once any model has evolved to either an aligned or orthogonal state, there appears to be no further evolution away from this state at later times.
\end{abstract}


\section{Introduction}

Most neutron stars in the Universe were born in core collapse events associated with supernovae, while a small fraction were formed as the end result of the binary inspiral and merger of two stars. Both scenarios produce an extremely hot proto-neutron star, in conditions favourable for it to be rapidly rotating. Furthermore, although the star's magnetic field during and after the proto-neutron-star phase is poorly understood, there are various mechanisms with the potential to produce field strengths up to roughly $10^{15}$ G, a canonical value for magnetars (a very highly-magnetised class of neutron star). In fact, it is possible to produce magnetar-strength fields by magnetic-flux conservation alone during the core-collapse process, if this occurs in a very controlled way and does not disrupt the field \citep{woltjer,ferrario}; more realistically though, an additional mechanism like a dynamo \citep{TD93,bonanno} or the magneto-rotational instability \citep{obergau,guilet,rembiasz} is likely needed to amplify this remnant field to high strength. Strong differential rotation of the proto-neutron star would naturally lead to an intense toroidal-field component, roughly symmetric about the rotation axis. A rapidly-rotating, highly-magnetised neutron star 
is naturally prone to be a strong emitter of radiation -- and we are now in an era where it is plausible to anticipate detecting both electromagnetic and gravitational signals from such an object. This hope depends strongly on the degree of asymmetry of the star, and how misaligned the star's magnetic field and rotation are.

A newly-formed neutron star rapidly cools via neutrino emission, and spins down via electromagnetic (and possibly gravitational-wave) energy losses.  Prior to the formation of the solid crust, and prior to the onset of neutron superfluidity and proton superconductivity, the star's dynamics can be well approximated as those of a self-gravitating fluid body deformed by rotation and its magnetic field.  Then, any misalignment $\chi$ between the star's magnetic and rotation axes will result in the star undergoing a form of \emph{free precession}, in close analogy with such motion in a rigid body.  In a previous paper we studied this motion in detail for a star with a purely toroidal magnetic field, carrying out an analysis to second perturbative order in the stellar deformations -- at which the physics of rotation and magnetism couples -- to compute the departures from rigid-body precession that accompany this motion \citep{LJ17}.

This misalignment or \emph{inclination} angle $\chi$ will itself evolve over time.  As we will discuss later, in a star whose magnetic field-induced deformation is prolate (as is expected when the internal magnetic field is dominated by its toroidal component), viscous dissipation in the star will act to make $\chi$ increase.  By contrast, dissipation in stars with oblate deformations (poloidal internal fields) results in a decrease of $\chi$ over time.  The external magnetic torque acting on the star is also important, and also tends to make $\chi$ decrease.  It follows that the evolution in inclination angle early in a neutron star's life involves a complicated mixture of cooling, internal viscous dissipation,  and electromagnetic back-reaction.

There are strong motivations for trying to understand this interplay in detail.  Firstly, the distribution of inclination angles seen in the Galactic pulsar and magnetar population will be determined in part by the rapid early evolution described here,  an evolution  which is likely to come to a halt, or at least be greatly slowed down, once the star has cooled sufficiently to acquire a solid crust and/or superfluid components in the interior.  Secondly, the initial spin-down may be accompanied by gravitational radiation, with systems that become  orthogonal rotators being the strongest gravitational-wave emitters.  Knowledge of the inclination angle evolution is therefore useful in carrying out such gravitational wave searches, both following supernovae and binary merger events.

The problem of internal dissipation in precessing magnetic stars, and the consequent inclination-angle evolution, was first tackled in a series of papers by Leon Mestel and collaborators, who noted the existence of a non-rigid part of the motion, and made some suggestions for how to compute it \citep{mestel1, nittmann_wood_81, mestel2}, concentrating on main sequence stars.    \citet{jones_76_APSS} noted that such a mechanism could be relevant for neutron stars, possibly playing a role in determining the observed distribution of pulsar inclination angles.  Later, \citet{cutler_02} pointed out that the evolution towards orthogonality that a strong prolate deformation would produce would drive the motion towards one most efficient for long-lived continuous gravitational wave emission from a neutron star.  This idea was taken up in \citet{dallosso09}, who suggested that such evolution could be important in the early life of a rapidly spinning magnetar, the so-called \emph{millisecond magnetar} scenario popularly invoked to explain long gamma-ray bursts and superluminous supernovae \citep{metzger}.  \citet{lasky_glam} pursued this idea, making contact with the light curves of gamma-ray bursts that could accompany magnetar birth, and placing upper bounds on the accompanying emission of gravitational radiation.    Most recently, \citet{DP17}  noted that such inclination-angle evolution may naturally produce a roughly bimodal distribution of pulsar inclination angles.

There is -- however -- a major problem in applying Mestel's solutions to neutron-star physics, as done by these authors. A key simplification made by Mestel was to assume the non-rigid fluid response would be incompressible (i.e. divergence-free). As discussed in our previous paper \citep{LJ17}, this bypassed the need to tackle the highly complex system of equations which govern the general form of the non-rigid response, and instead can be used to argue for a `qualitative solution'. However, in the neutron-star case, as we will see later, the dominant dissipative mechanism at birth is bulk viscosity -- and the dissipation rate depends exactly on how compressible the motions are. If, therefore, these authors had consistently implemented Mestel's solution, \emph{bulk viscosity would be zero}; dissipation and inclination-angle changes would then occur on one of two far longer timescales: either that of Ohmic decay (considered by Mestel) or shear-viscous dissipation, and the orthogonalisation mechanism would be of little relevance in many cases (including newborn magnetars). No authors adopting Mestel's original solutions seem to have acknowledged this explicitly, but instead switched to making alternative arguments in a bid to produce very rough estimates of bulk dissipation. 

In contrast to the lesser-studied effect of internal dissipation, it has long been known that an external torque, like the electromagnetic one causing pulsar spin-down, acts to align the rotation and magnetic axes on the spin-down timescale \citep{davis_goldstein_70, michel_goldwire_70}, if one neglects contributions to the star's shape due to crustal strain \citep{goldreich70}.  In this paper, we will consider young hot stars, prior to crust formation, so this alignment mechanism will apply.

We advance this programme in two main ways.  Firstly, we use the detailed form of the non-rigid response computed in \citet{LJ17} to compute the internal viscous dissipation rates.  This was not possible in any of the previous studies as the non-rigid response was simply not known.  The response we found was \emph{not} divergence-free, and so we have been able to make a quantitative calculation of dissipation, including the bulk-viscous mechanism. Secondly, we simultaneously account for cooling, internal viscous dissipation, and electromagnetic torque effects, allowing for both spin-down and alignment for the latter.  This is the first time all of these ingredients have been combined self-consistently in this problem\footnote{As we were preparing to submit this paper, a new paper with similar focus to ours appeared online \citep{dallosso_spin}. It aims to study the coupled evolution of spin-down and inclination angle.  However, it does not make use of the relevant eigenfunctions describing the fluid motion of \citet{LJ17}.  It also misses a crucial term in the equations describing how spin-down causes a decrease in the inclination angle; as we will show later, this effect is important for a considerable portion of the neutron-star parameter space.}. In so doing, we build the most realistic model to date of the evolution of magnetic inclination angle in a newly-formed neutron star.

The structure of this paper is as follows. In section \ref{sect:precession_and_damping} we recall some classical results relating energy losses to precession damping, and discuss their relevance to our neutron-star model (postponing some detailed checks to Appendix \ref{sect:J_and_E_xi}). Next we discuss details of the mechanisms for energy loss: section \ref{fluidprec} describes internal viscous dissipation (due to both shear and bulk viscosity), and section \ref{EM_spindown} describes external losses related to the star's spindown. Since viscosities are highly temperature-dependent, we need a prescription for the cooling of the star; this is given in section \ref{cooling}. Before our numerical results, we first compare the timescales in the problem and use these back-of-the-envelope estimates to make some predictions about whether a given neutron star will end up as an aligned or an orthogonal rotator (section \ref{analytic}); this will help us to understand our results and provide an independent check on them. Section \ref{numerical} then presents details of our numerical method and the quantitative results obtained: we first study evolution under internal dissipation alone, to make contact with earlier work (section \ref{evol_nospindown}), then the full coupled evolution of inclination angle and spindown (section \ref{evol_spindown}), then explore whether a star is likely to undergo a second phase of evolution at later times, from an orthogonal to aligned state, or vice-versa (section \ref{revival}). We then consider the validity and possible limitations of our approach in section \ref{validity}, and the astrophysical implications of our results in section \ref{discussion}.  In appendix \ref{sect:diss_integrals} we derive explicit expressions for the shear and bulk dissipation integrals in terms of the fluid velocity for a \emph{compressible} star, in a form suited to our numerical computations.  Finally, in appendix \ref{sect:viscosity_coefficients} we present some details connected with the coefficient of bulk viscosity, to show how the expression we use connects with results in the literature.

\section{Non-rigid precession and damping} \label{sect:precession_and_damping}

In this section we first describe some basic properties of a precessing magnetic star (section \ref{sect:basic_features}), and then describe the formalism used to describe how precession evolves under a combination of viscous dissipation and electromagnetic radiation reaction (section \ref{sect:damp_form}).

\subsection{Basic features of free precession} \label{sect:basic_features}

We first recall a few basic features of the free precession of a biaxial body, without damping.  We will denote the principal moments of inertia as $I_1, I_2=I_1, I_3$, with the distortion $\Delta I \equiv I_3-I_1$ being sourced by (in our case) magnetic strains.  In the notation of \citet{jones_andersson_01}, the precession then consists of a rotation of $I_3$ about the invariant angular momentum axis $\boldsymbol J$, at a rate 
\beq
\label{eq:J_phi_dot} 
\dot\phi = \frac{J}{I_1} ,
\eeq
tracing out a cone of half-angle $\chi$ about $\boldsymbol J$.  Superimposed on this is a slow rotation about $I_3$ at a rate
\beq
\dot\psi = - \epsilon_B \dot\phi
\eeq
where we define the ellipticity
\begin{equation}
\label{eq:def_epsilon_B}
\epsilon_B = \frac{\Delta I}{I_3} .
\end{equation}
The angle $\chi$ (denoted as $\theta$ in \citet{jones_andersson_01}) is sometimes known as the \emph{wobble angle}, or, for our magnetised star, the \emph{inclination angle}.  



These results hold for a rigid star.  The basic picture remains the same in the more realistic case of a fluid star, deformed by magnetic strain (or elastic strain, not considered here), providing one interprets the ellipticity above as that caused by the magnetic strains \citep{munk_macdonald, jones_andersson_01, cutler_jones_01}, rather than by the centrifugal forces.  Note that we expect this deformation to be small; for a star of mass $M$, radius $R$, and magnetic field strength $B$, equation (20) of \citet{LJ17} gives:
\begin{equation}  
\label{eq:epsilon_B}
\epsilon_B  \sim  \frac{B^2 R^4}{GM^2} \sim 1.9 \times 10^{-6} \frac{B_{15}^2 R_6^4}{M_{1.4}^2} ,
\end{equation}
where $B_{15} = B / 10^{15}$ G, $M_{1.4} = M / 1.4 M_\odot$, and $R_6 = R / 10^6$ cm.    The free precession will have period $T_\omega$ given in terms of the spin period $T_\Omega$ by   
\begin{equation}
T_\omega =  \frac{T_\Omega}{\epsilon_B \cos\chi } .
\end{equation}
Using equation (\ref{eq:epsilon_B}) for $\epsilon_B$, we have
\begin{equation}
\label{eq:T_omega_parameterised}
T_\omega  =  5.2 \times 10^2 \, {\rm seconds} \frac{M_{1.4}^2}{f_{\rm kHz} B_{15}^2 R_6^4 \cos\chi} .
\end{equation}
Then the corresponding angular frequency is
\begin{align}
\label{omega}
\omega = \frac{2\pi}{T_\omega}  &= \Omega\epsilon_B\cos\chi\\
&= 1.2 \times 10^{-2} \, {\rm Hz} \frac{f_{\rm kHz} B_{15}^2 R_6^4 \cos\chi}{M_{1.4}^2} . 
\label{eq:omega_parameterised}
\end{align}

There will also be a deformation sourced by rotation of angular frequency\footnote{Mestel used $\alpha$ to denote the primary rotation rate, but this is both non-standard in modern work, and risks confusion with the inclination angle (itself denoted $\alpha$ in many observational studies).} $\Omega = 2\pi f$, of magnitude (equation (19) of \citet{LJ17}):
\begin{equation}
\label{eq:epsilon_alpha}
\epsilon_\Omega  \sim \frac{\Omega^2 R^3}{GM} \sim 0.21 \frac{R_6^3 f_{\rm kHz}^2}{M_{1.4}} , 
\end{equation}  
where $f_{\rm kHz} = f/$kHz.
Rigid-body precession would cause this centrifugal bulge, associated with the primary rotation, to be slowly rotated about the magnetic axis at rate $\omega$. Since our star is not solid, this clearly would not happen, and the precession gives rise to a non-rigid response: an additional velocity field of the fluid elements
sometimes called `$\xi$-motions', following the analysis of the problem by Mestel and others \citep{mestel1, mestel2, nittmann_wood_81}, where $\boldsymbol \xi$ was used to denote the Lagrangian displacement vector of a fluid element caused by the motion of the centrifugal bulge.  A calculation of these motions proved elusive, until the recent analysis of \citet{LJ17}, who exploited the smallness of $\epsilon_\Omega$ and $\epsilon_B$ to carry out a calculation to second order in perturbation theory.  Specifically, the calculation involved the solution for a background $\gamma=2$ polytropic model, whose density distribution is given by
\beq\label{back_rho}
\rho_0=\rho_c\frac{\sin(\pi r/R_*)}{\pi r/R_*};
\eeq
the solution to the order-$\epsilon_\Omega$ equations (details of which are not needed here); and
the solution to the order-$\epsilon_B$ equations for a purely toroidal magnetic field $\bB$:
\beq\label{back_B}
\bB=B_\phi\be_\phi=\Lambda\rho_0 r\sin\theta\be_\phi
\eeq
with an associated \emph{quantitive} solution for the magnetically-induced distortion
\beq\label{quant_epsB}
\epsilon_B=-0.019\frac{\Lambda^2}{G}.
\eeq
Note that in all our qualitative estimates we use the approximation for $\epsilon_B$ from equation \eqref{eq:epsilon_B}, but in our full numerical solutions we employ the more accurate expression above, equation \eqref{quant_epsB}. All of these solutions enter the full second-order equations of motion for terms proportional to $\epsilon_\Omega \epsilon_B$, which thereby describe the coupling between magnetic deformation and the rotation.  The lengthy expressions for $\dot{\boldsymbol \xi}$ are given in section 7 of \citet{LJ17}.  It is the calculation of these motions that will allow us to compute the viscous dissipation rates required in this study.

\subsection{Damping formalism} \label{sect:damp_form}

Note that if one only wishes to consider internal viscous damping (i.e.\ no radiation losses), then a simple diagnostic, used in the past \citep{DP17, lasky_glam} is to define a ``precessional energy''. This quantity represents the amount of kinetic energy which can be taken away from the star, at fixed angular momentum, just by changing the inclination angle. To define it, we start with the expression for the star's kinetic energy:
\beq
\label{eq:E_kinetic}
E_{\rm kin} = \frac{J^2}{2I_1} \brac{1 - \epsilon_B \cos^2\chi}.
\eeq 
For an oblate star ($\epsilon_B > 0$) this energy is minimised when $\chi = 0$, while for a prolate star it is minimised for $\chi = \pi/2$,  motivating the definition
\begin{equation}
E_{\rm prec} \equiv E(J, \chi) - E(J, \chi_{\rm min}) ,
\end{equation}
so that
\begin{align}
E_{\rm prec} &= \frac{J^2}{2I_1} \epsilon_B (\cos^2\chi_{\rm min} - \cos^2\chi) \nn\\
 & \approx  \frac{1}{2} I \Omega^2 \epsilon_B (\cos^2\chi_{\rm min} - \cos^2\chi),
\label{eq:E_prec_oblate}
\end{align}
where $\chi_{\rm min} = 0, \pi/2$ for $\epsilon_B > 0$, $\epsilon_B < 0$, respectively.  Here, and elsewhere in this section, we use the approximately-equal symbol to denote a result which is exact to our order of working (i.e. utilising the smallness of $\epsilon_\Omega$ and $\epsilon_B$).
Differentiation with respect to time (at fixed $J$, fixed $\Omega$) then immediately leads to a relation between the rate of inclination angle evolution $\dot\chi$ and the internal energy-dissipation rate due to viscous processes $\dot E_{\rm visc}$.  This relation may be used to define an evolution timescale:
\begin{equation}
\label{eq:tau_E_definition}
\tau_E \equiv   \frac{E_{\rm prec}}{\dot E_{\rm visc}} 
= \frac{I \Omega^2 \epsilon_{\rm B} (\cos^2\chi_{\rm min} - \cos^2\chi)      }{2\dot E_{\rm visc}} .
\end{equation}
This is the timescale on which the precessional motion is damped by internal viscous processes, and has been used in previous work, e.g. setting $\chi_{\rm min} = \pi/2$ gives equation (3) of \citet{DP17}.

In the more realistic case where radiation reaction effects (either electromagnetic or gravitational) are included, allowance must be made for angular momentum losses, $\dot J$, as described in \citet{cutler_jones_01}.  We begin by summarising the results for such
precession damping from rigid-body mechanics, as the main results go through to the non-rigid case, as argued in \citet{cutler_jones_01}.   
What follows is actually a slight extension of  \citet{cutler_jones_01}, where internal viscous dissipation was not considered.  Also, \citet{cutler_jones_01} considered a gravitational wave torque, not an electromagnetic one, but the argument is independent of this detail.   

If $E$ denotes the star's total energy, we have:
\begin{align}
\label{eq:E_dot_total}
\dot E &= \dot E_{\rm visc} + \dot E_{\rm EM}, \\
\dot J &= \dot J_{\rm EM}.
\end{align}
We can expect $E = E(J, \chi)$ to give
\beq
\dot E = \frac{\partial E}{\partial J} \dot J + \frac{\partial E}{\partial \chi} \dot\chi
\eeq
where the first and second partial derivatives on the right hand side are to be evaluated at fixed $\chi$ and fixed $J$, respectively. Rearranging the above gives us an expression for the evolution of the inclination angle:
\beq
\dot\chi = \frac{\dot E - \frac{\partial E}{\partial J} \dot J}{\frac{\partial E}{\partial \chi}}
\eeq

As noted above, in the absence of the spin-down torque, the magnetic dipole rotates at a constant rate $\dot\phi$ about the invariant angular momentum axis $\boldsymbol J$.    It then follows that
\beq 
\label{eq:E_dot_OG}
\dot E_{\rm EM} = \dot\phi \dot J_{\rm EM},
\eeq
as argued in \citet{ostriker_gunn_69}.  (An analogous expression holds for the gravitational-wave case considered in \citet{cutler_jones_01}.)  Then
\beq \label{chi_dot_general}
\dot\chi = \frac{\dot E_{\rm visc} + 
\dot J_{\rm EM}\left(\dot\phi - \frac{\partial E}{\partial J}\right) }{\frac{\partial E}{\partial \chi}}.
\eeq

Using exactly the same arguments as given in \citet{cutler_jones_01}) for elastic precessing stars, we can argue that for our magnetic precessing star the partial derivatives of $E$ with respect to $\chi$ and $J$ are, to leading order in $\epsilon_{\rm B}$, and for small $\chi$, given by including only the \emph{kinetic} contribution to $E$, i.e.\ neglecting the perturbations in internal energy, magnetic energy, and gravitational potential energy.
Strictly, this step in the argument, as formulated in  \citet{cutler_jones_01}, has been shown to hold only in the limit of small $\chi$, but we will follow other authors \citep{dallosso09, lasky_glam} in extending it to arbitrary $\chi$.   In support of this, we show in Appendix \ref{sect:J_and_E_xi}, that the perturbations in angular momentum, kinetic energy, and magnetic energy due to the non-rigid response (the `$\xi$-motions') are indeed of higher order in the rotational and magnetic parameters ($\epsilon_\Omega$ and $\epsilon_B$) than the kinetic energy term that we do retain.

Given this, we can use the known form of the kinetic energy for arbitrary $\chi$, equation \eqref{eq:E_kinetic}, to give: 
\beq
\frac{\partial E_{\rm kin}}{\partial \chi} = \frac{J^2}{I_1} \cos\chi \sin\chi \frac{\Delta I}{I_3}
\approx  \Omega^2 \cos\chi \sin\chi \Delta I.
\eeq
To evaluate equation \eqref{chi_dot_general}, we also need the result:
\beq
\frac{\partial E_{\rm kin}}{\partial J} = \frac{J}{I_1} \left(1-\cos^2\chi \frac{\Delta I}{I_3} \right),
\eeq
which is obtained by differentiating equation \eqref{eq:E_kinetic} and using the definition of $\epsilon_B$ from equation \eqref{eq:def_epsilon_B}.
Now making use of the above result and equation \eqref{eq:J_phi_dot}, we obtain:
\beq
\dot\phi - \frac{\partial E_{\rm kin}}{\partial J} = \frac{J}{I_1} \cos^2\chi \frac{\Delta I}{I_3}
 \approx \Omega \cos^2\chi \frac{\Delta I}{I}.
\eeq
Now we may recast equation \eqref{chi_dot_general} into a more explicit form for our problem:
\beq 
\label{eq:chi-dot}
\dot\chi = \frac{\dot E_{\rm visc} + 
\dot E_{\rm EM} \cos^2\chi \epsilon_{\rm B}}{I \Omega^2 \cos\chi \sin\chi \epsilon_{\rm B}},
\eeq
where we have used $\epsilon_{\rm B} = \Delta I / I$.  Note that the contribution to $\dot\chi$ from $\dot E_{\rm EM}$ is suppressed by a factor $\epsilon_{\rm B}$ relative to the $\dot E_{\rm visc}$ contribution.  Note that also that $\dot E_{\rm visc} < 0$ and $\dot E_{\rm EM} < 0$, so that internal damping gives $\dot\chi < 0$ for oblate deformations ($\epsilon_{\rm B} > 0$), and $\dot\chi > 0$ for prolate deformations ($\epsilon_{\rm B} < 0$), while the electromagnetic torque gives  $\dot\chi < 0$ for both oblate and prolate stars. 

The evolution in spin rate is easily obtained, by differentiation of equation (\ref{eq:J_phi_dot}):  
\beq
\ddot\phi = \frac{\dot J_{\rm EM}}{I_1},
\eeq
which to our order of working may be rewritten
\beq \label{eq:alpha-dot}
\dot\Omega = \frac{\dot J_{\rm EM}}{I}.
\eeq
The two coupled ODEs of equations (\ref{eq:chi-dot}) and (\ref{eq:alpha-dot}) describe, in general form, the evolution of $\Omega(t)$ and $\chi(t)$. One then needs a prescription for computing the actual form of the internal damping and the electromagnetic-radiation losses; these are described in sections \ref{fluidprec} and \ref{EM_spindown} respectively.

We can use the above to define evolution timescales for the inclination angle.  Following \citet{goldreich70} we will use $\sin\chi$ as the primary quantity to define 
\begin{equation}
\tau_\chi \equiv \frac{\sin\chi}{\frac{d}{dt} \sin\chi} .
\end{equation}
Then, using equation (\ref{eq:chi-dot}) we have
\begin{equation}
\frac{1}{\tau_{\chi}} = \frac{1}{\tau_{\chi}^{\rm visc}} + \frac{1}{\tau_{\chi}^{\rm EM}} ,
\end{equation}
where
\begin{equation} 
\label{eq:tau_sc_int_def}
\tau_{\chi}^{\rm visc} = \frac{I\Omega^2 \sin^2\chi \epsilon_B}{\dot E_{\rm visc}} ,
\end{equation}
\begin{equation} 
\label{eq:tau_sc_EM_def}
\tau_{\chi}^{\rm EM} = \frac{I\Omega^2}{\dot E_{\rm EM}} \frac{\sin^2\chi}{\cos^2\chi} .
\end{equation}
Note that the timescale of equation (\ref{eq:tau_sc_int_def}) is closely related to that of equation (\ref{eq:tau_E_definition}).  There is an additional factor of $2$ in  (\ref{eq:tau_sc_int_def}) as it essentially measures the evolution of an `amplitude' ($\sin\chi$) rather than an energy.  Equation (\ref{eq:tau_sc_int_def}) also has the advantage of taking exactly the same functional form in the oblate $\epsilon_B > 0$ and prolate $\epsilon_B < 0$ cases, with an overall sign that reflects whether the system tends to align ($\tau_{\chi}^{\rm visc} < 0$) or orthogonalise 
($\tau_{\chi}^{\rm visc} > 0$).

Note also that   the electromagnetic alignment timescale of equation (\ref{eq:tau_sc_EM_def}) is related to the more familiar electromagnetic spin-down timescale 
\beq
\label{eq:tau_EM}
\tau_{\rm spindown} \equiv \frac{\Omega}{2|\dot\Omega|} = \frac{I\Omega^2/2}{\dot E_{\rm EM}}
\eeq
by a $\chi$-dependent factor:
\begin{equation}
\label{eq:tau_chi_EM}
\tau_{\chi}^{\rm EM} = 2 \frac{\sin^2\chi}{\cos^2\chi} \tau_{\rm spindown} .
\end{equation}
We will make use of these definitions later to estimate timescales for the evolution in the inclination angle.

\section{Internal viscous dissipation}
\label{fluidprec}

\subsection{Dissipation integrals}
\label{diss_int}

In order to calculate the rate of dissipation of precessional energy by viscosity, one needs knowledge of the non-rigid part of the precessional motion. This velocity field was calculated in \citet{LJ17}, by solving the
equations of motion for a precessing magnetised star at second perturbative order, i.e. the equations for quantities at order $\epsilon_\Omega\epsilon_B$. The system of equations comprised the Euler equation
\begin{align}
\rho\left[ \frac{\partial{\boldsymbol v}}{\partial t} + ({\boldsymbol v} \cdot \nabla) {\boldsymbol v} \right]
&=- \nabla P - \rho\nabla \Phi + \frac{1}{4\pi} (\nabla \times {\boldsymbol B}) \times {\boldsymbol B},
\end{align}
coupled to the induction equation
\beq
\pd{\bB}{t}=\curl\brac{\bv\times\bB},
\eeq
where $\bv$ is the fluid velocity, $P$ pressure, $\rho$ mass density and $\Phi$ the gravitational potential.
The system was closed by the usual
equations: the Poisson equation, continuity equation, solenoidal
constraint on $\bB$ and an equation of state (for which we used a
simple polytropic relation $P=P(\rho)\propto \rho^2$). These do not enter
into the discussion here, so we refer the reader to \citet{LJ17} for
details. To our perturbative order the solutions were described by oscillatory velocity and magnetic-field perturbations, $\dot\bxi$ and $\delta \bB$, confined to the core region alone, i.e. non-zero for radii satisfying $0\leq r\leq 0.9R_*$. This was necessary since the perturbative ordering breaks down close to the surface.

In this paper, we now 
wish to understand the secular evolution of the precessing
star --  i.e. the dissipation of the steady-state non-rigid precession. Since, by
assumption, the additional secular terms act over far
longer timescales than $T_\omega$, our dynamical solutions from \citet{LJ17}
will still be valid. We will augment our Euler equation with the two
viscous `force' terms; the result is the
compressible Navier-Stokes equation together with a Lorentz-force term:
\begin{align}
\rho\left[  \frac{\partial{\boldsymbol v}}{\partial t} + ({\boldsymbol v} \cdot \nabla) {\boldsymbol v} \right] 
=&- \nabla P - \rho\nabla \Phi + \frac{1}{4\pi} (\nabla \times {\boldsymbol B}) \times {\boldsymbol B}\nn\\
    &+2\div(\eta{\boldsymbol{\sigma}})+\nabla(\zeta\div\bv),
 \label{general_Euler}
\end{align}
where the tensor $\boldsymbol{\sigma}$ has components
\beq
\sigma_{ab}=\frac{1}{2}(\nabla_a v_b+\nabla_b v_a)-\frac{1}{3}\nabla_c v_c g_{ab}.
\eeq

We can simply take the scalar product of the above Euler equation with the velocity
$\bv$ and integrate to determine the rate at which work is done by the dissipative forces. Doing so, we see that the total kinetic energy loss from the 
precession is the sum of contributions from the
two viscous terms on the right-hand side of equation \eqref{general_Euler}
\beq
\td{E_{\rm prec}}{t}
=  \int 2\bv\cdot[\div(\eta{\boldsymbol{\sigma}})]\ \rmd V
   +\int \bv\cdot\nabla(\zeta\div\bv)\ \rmd V.
 \label{vdotEuler}
\eeq
We can also expect magnetic-energy loss due to Ohmic dissipation of $\delta\bB$:
\beq
\td{E_{\rm mag}}{t}
=-\brac{\frac{c}{4\pi}}^2\int \frac{1}{\sigma} |\curl\delta\bB|^2\ \rmd V,
\eeq
where $\sigma$ is the electrical conductivity \citep{mestel2}.
Whilst this term has been invoked as the cause of precession damping in main-sequence stars \citep{mestel2,nittmann_wood_81}, it becomes negligibly slow for the extremely high values of $\sigma$ expected for neutron-star matter \citep{goldreich_reisenegger_92}. The internal energy losses for our model are therefore given, to leading order, by the two dissipation integrals from \eqref{vdotEuler}. We will use the term $\dot{E}_{\rm shear}$ to denote the integral involving $\eta$, and $\dot{E}_{\rm bulk}$ to denote the one involving $\zeta$.  In Appendix \ref{sect:diss_integrals} we re-write the integrands of these two equations to give expressions more suitable for use in our numerical computations.

\subsection{Viscosity coefficients} \label{sect:vc}

Next, we assemble some formulae to calculate the coefficients of shear and bulk viscosity, required in order to evaluate the dissipation rates from equation (\ref{vdotEuler}) above. These dissipation integrals will be used in their exact forms for the time evolutions described in section \ref{numerical}, but before that we wish to make some simple back-of-the-envelope timescale estimates in order to gain a better intuition for the rather complex problem.

Firstly, we can derive a general timescale for viscous dissipation before needing to specify the details of the particular viscosity. To do so, we begin with equation (\ref{vdotEuler}), replacing all spatial derivatives with factors of $R^{-1}$, and volume integrals with factors of $R^3$, to give:
\begin{equation} 
\label{eq:E_dot_shear_schematic}
\dot E_{\rm shear} \sim \eta R^3 \left(\frac{v}{R}\right)^2 ,
\end{equation}
\begin{equation} 
\dot E_{\rm bulk} \sim \zeta R^3 \left(\frac{v}{R}\right)^2 .
\end{equation}
To save writing things out twice, let us define
\begin{equation}
A_X: A_{\rm shear} = \eta,  \hspace{10mm} A_{\rm bulk} = \zeta ,
\end{equation}
i.e. $X = \{$shear, bulk$\}$.  Then
\begin{equation} 
\label{eq:E_X_both}
\dot E_X \sim A_X R v^2 g(\chi).
\end{equation}
This neglects the possibility that typical values of the shear and compressional parts of $\nabla_a v_b$ might be very different, an assumption which we will check later with our quantitative, numerical results (section \ref{evol_nospindown}).  We have included a dimensionless $\chi$-dependent factor $g(\chi)$ in this, so that we can still keep track of $\chi$-dependent factors in our back-of-the-envelope estimates.  In fact, the results of \citet{LJ17} show that the leading-order $\chi$-dependences in the nearly aligned and nearly orthogonal limits are $g(\chi) \sim \sin^2\chi$ and  $g(\chi) \sim \cos^2\chi$ for $\chi \ll 1$ and $\pi/2 - \chi \ll 1$ respectively.  As expected, the dissipation rate goes to zero in these two limits.  Retention of these $\chi$-factors will be particularly useful in section \ref{analytic}, where we investigate the competition between electromagnetic and dissipative effects.

To proceed, we make use of the scaling (see Section 3.2 of \citet{LJ17}):
\begin{equation}
v \sim \dot\xi \sim \epsilon_\Omega R \omega \sim \epsilon_\Omega R \Omega \epsilon_{\rm B} ,
\end{equation}
substitution of which into equation (\ref{eq:E_X_both}) gives:
\begin{equation}
\label{eq:E_X_estimate}
\dot E_X \sim A_X \Omega^2 R^3 \epsilon_\Omega^2  \epsilon_{\rm B}^2 g(\chi) .
\end{equation}
Inserting this into equation (\ref{eq:tau_sc_int_def}), and approximating the moment of inertia by $I \sim MR^2$, we obtain:
\begin{equation}
\tau_{\chi, X} \sim \frac{M}{A_X  R} \frac{1}{ \epsilon_\Omega^2  \epsilon_{\rm B}} 
\frac{\sin^2\chi}{g(\chi)} .
\end{equation}
We can eliminate $\epsilon_\Omega$ and   $\epsilon_{\rm B}$ in favour of $\Omega$ and $B$ using equations (\ref{eq:epsilon_B}) and (\ref{eq:epsilon_alpha}):
\begin{equation}
\tau_{\chi, X} \sim \frac{G^3 M^5}{(2\pi)^4 R^{11}} \frac{1}{A_X} \frac{1}{f^4 B^2} \frac{\sin^2\chi}{g(\chi)} .
\end{equation}
Finally, we parametrise  the above expression using typical neutron-star values for all quantities (aside from the viscosity coefficient): 
\begin{equation}
\label{eq:tau_E_both_parameterised}
\tau_{\chi, X} \sim 1.0 \times 10^{27} {\, \rm yr \,} \frac{M_{1.4}^5}{R_6^{11}} 
\left(\frac{1 \, {\rm g \, cm}^{-1} \, {\rm s}^{-1}}{A_X}\right)
f_{\rm kHz}^{-4}B_{15}^{-2} \frac{\sin^2\chi}{g(\chi)} ,
\end{equation}
recalling that $f_{\rm kHz}$ is the rotation rate in units of kHz, and where $B_{15}$ the average magnetic-field strength in units of $10^{15}$ G.
Note that the scaling $g(\chi \ll 1) \sim \sin^2\chi$ means that there is simple exponential-in-time behaviour in this limit.

Next we consider the specific forms of the shear and bulk viscosity coefficients, and the associated energy-dissipation timescales.

\subsubsection{Shear} \label{sect:shear}

Shear viscosity arises from scattering between particle species. In the young neutron star, before the condensation of neutrons or protons into superfluid phases, \citet{flowers_itoh_79} calculated the different particle scattering coefficients and found that the main contribution to shear viscosity is neutron-neutron scattering. Their results for the kinematic shear viscosity $\nu$ in the young star may be fitted by the formula \citep{cutler_lindblom_87,bildsten_ushomirsky_00}:
\begin{equation}
\nu \approx 1.8 \times 10^4 \brac{\frac{\rho}{1.5\times 10^{14}\textrm{g cm}^{-3}}}^{5/4} T_8^{-2} \, {\rm cm}^2 \, {\rm s}^{-1}.
\end{equation}
 Converting this into a dynamic viscosity $\eta=\rho\nu$, we have:
\begin{equation}
\label{eq:eta_estimate}
\eta \approx 1.9 \times 10^{16} \rho_{15}^{9/4} T_{10}^{-2} \, {\rm g} \, {\rm cm}^{-1} \, {\rm s}^{-1} .
\end{equation}
Inserting into equation (\ref{eq:tau_E_both_parameterised}) we obtain
\begin{equation}
\label{eq:tau_E_shear}
\tau_{\chi, {\rm shear}} \sim 5.2 \times 10^{10} {\, \rm yr \,} \frac{M_{1.4}^5}{R_6^{11}} 
\frac{T_{10}^2}{\rho_{15}^{9/4}}
f_{\rm kHz}^{-4} B_{15}^{-2} \frac{\sin^2\chi}{g(\chi)} .
\end{equation}
This suggests that shear viscosity will not play a significant role in the evolutions considered in this paper.   Note, however, that the scalings involve some very high powers, and that the temperature will drop below $10^{10}$ K very quickly. The estimate also assumed the shear and compressional pieces of the fluid stress tensor to be comparable, which we will see later is not true (section \ref{evol_nospindown}).  
For these reasons we will retain shear viscosity in our numerical simulations.

\subsubsection{Bulk} \label{sect:bulk}

Bulk viscosity arises from departures from chemical equilibrium as the neutron star matter is compressed and expanded by the perturbation.  The basic formalism for describing this is derived in \citet{L_and_L_fluids_87}, and extended in \citet{lindblom_owen_02}.  
The full expression for $\zeta$ can be computed by taking the real part of equation (3.11) of \citet{lindblom_owen_02}, to give:
\begin{equation}
\label{eq:zeta_general}
\zeta \approx - \frac{n \tau}{1 + (\omega \tau)^2} \left.\frac{\partial P}{\partial x}\right|_n \frac{\rmd x}{\rmd n} ,
\end{equation}
where $\tau$ is the relaxation time of the microphysical process involved, $n$ the baryon number density, so that $n = \rho / m_{\rm B}$ (with $m_{\rm B}$ being the baryon mass), and $x$ the proton
fraction, i.e. $x = n_{\rm p} / n$, where $n_{\rm p}$ is the proton number density.  In computing this quantity, we have largely followed the calculation set out in \citet{DP17}, \skl{who in turn make use of many results from \citet{reisenegger_goldreich_92}}.  We give the details in Appendix \ref{sect:viscosity_coefficients}, where we point out a few discrepancies that we encountered in the literature.  Here we simply state the results, substitution of which into equation (\ref{eq:zeta_general}) give $\zeta$. Firstly, we have:
\begin{equation}
\td{x}{n} \approx  6 \times 10^{-3} \frac{m_{\rm B}}{\rho_{\rm nuc}},
\end{equation}
\begin{equation}
\left.\frac{\partial P}{\partial x}\right|_n \approx  -5.2 \times 10^{33} \, {\rm erg} \, {\rm cm}^{-3} \, 
\left(\frac{\rho}{\rho_{\rm nuc}}\right)^{5/3} .
\end{equation}
\skl{Note that these results were obtained by assuming a simplistic equation of state where the stellar pressure is just the sum of the degeneracy pressures of non-relativistic neutrons and ultrarelativistic electrons; see Appendix \ref{sect:viscosity_coefficients}. It is somewhat unrealistic in its neglect of particle interactions, and gives a proton fraction which is only $0.03$ at the centre of the star (a factor of $\sim 4$ too small). As well as deviating from the results of more detailed calculations, the pressure calculated in this manner also does not scale with $\rho$ in the same way as the $\gamma=2$ polytropic model used for calculating our non-dissipative precessional solutions, representing a degree of internal inconsistency.  On the other hand, the model used here -- that of \citet{reisenegger_goldreich_92}  -- has the virtues of being analytic, familiar to many, and allowing for direct comparison with previous work. We thus proceed, whilst warning that a more realistic calculation would probably lead to a larger $\zeta$ and somewhat faster viscous dissipation, through the increase of the proton fraction.}

The relevant microphysical process in calculating the bulk viscosity is modified Urca, with energy release in neutrinos, which has a relaxation timescale of \citep{sawyer_89,reisenegger_goldreich_92}:
\begin{equation}
\label{eq:RG_relaxation_timescale}
\tau \sim \frac{0.2}{T_9^6} \left(\frac{\rho}{\rho_{\rm nuc}}\right)^{2/3} \, {\rm yr} .
\end{equation}

We can then compute the dimensionless quantity $\omega \tau$ that  appears in the formula for bulk viscosity, as per equation (\ref{eq:zeta_general}).  Using $\omega = 2\pi/T_\omega$, and combining equations (\ref{eq:T_omega_parameterised}) and (\ref{eq:RG_relaxation_timescale}) we obtain
\begin{equation}
\label{eq:omega_tau}
\omega \tau \approx 0.14  \, \frac{f_{\rm kHz} B_{15}^2 \cos\chi}{T_{10}^6} \frac{R_6^2}{M_{1.4}^{4/3}} .
\end{equation}

Given the steep scalings involved (particularly in temperature) we will not make any assumption as to the size of $\omega \tau$ compared with unity in our numerical evolutions.  We will, however, present below some simple back-of-the-envelope estimates separately for the $\omega \tau \gg 1$ and $\omega \tau \ll 1$ regimes, as in these two limits the results take a particularly simple form.

Specifically, if we combine the above results, in the regime  $\omega \tau \ll 1$ we have:
\begin{equation}
\label{eq:zeta_low_freq}
\zeta(\omega \tau \ll 1) \approx 4.2 \times 10^{33} \, {\rm g \, cm}^{-1} \, {\rm s}^{-1} \, 
 \frac{1}{T_{10}^6} \left(\frac{M_{1.4}}{R_6^3}\right)^{10/3} ,
\end{equation}
\begin{equation} 
\label{eq:tau_E_npe_small}
\tau_{\chi,{\rm bulk}} (\omega \tau \ll 1)  = 7.9 {\, \rm seconds \,} \frac{M_{1.4}^{5/3}}{R_6} 
\frac{T_{10}^6}{f_{\rm kHz}^4  B_{15}^2} \frac{\sin^2\chi}{g(\chi)} .
\end{equation}
while in the regime  $\omega \tau \gg 1$ we have:
\begin{equation}
\zeta (\omega \tau \gg 1) = 1.7 \times 10^{35}  \, {\rm g \, cm}^{-1} \, {\rm s}^{-1} \, 
T_{10}^6
\frac{M_{1.4}^6}{\cos^2\chi B_{15}^4 f_{\rm kHz}^2 R_6^{14}},
\end{equation}
\begin{equation}
\label{eq:tau_E_npe_large}
\tau_{\chi,{\rm bulk}} (\omega \tau \gg 1) = 0.19 {\, \rm seconds \,} \frac{R_6^3}{M_{1.4}} \frac{B_{15}^2}{f_{\rm kHz}^2 T_{10}^6} \frac{\sin^2\chi}{g(\chi)} .
\end{equation}

Note that we can easily use the low-$\omega\tau$ relations to obtain the results for arbitrary $\omega\tau$:
\begin{equation}
\label{eq:zeta_chi_npe_general}
\zeta =  \frac{\zeta(\omega \tau \ll 1)}{1+(\omega \tau)^2} ,
\end{equation}
\begin{equation}
\label{eq:tau_chi_npe_general}
\tau_{\chi}^{\rm bulk} =  \big[ \tau_{\chi}^{\rm bulk}(\omega \tau \ll 1) \big]   \,\,    \big[ 1+(\omega \tau)^2 \big]    .
\end{equation}

\section{Electromagnetic spindown}
\label{EM_spindown}

The newly-born neutron star will be acted upon by a strong spin-down electromagnetic torque.  We will write the luminosity as
\beq
\label{eq:E_dot_EM}
\dot E_{\rm EM}  = \Omega \dot J_{\rm EM} = -\frac{R^6}{6c^3} \Omega^4 B_{\rm p}^2 \lambda(\chi)
\eeq
where $B_{\rm p}$ is the strength of the external dipole field and $\lambda(\chi)$ captures the $\chi$-dependence of the luminosity.  For a vacuum dipole $\lambda(\chi) =  \sin^2\chi$, but full numerical simulations of charge-filled pulsar magnetospheres indicate that this angular dependence may be quite different, tending to a non-zero value in the limit $\chi \rightarrow 0$ \citep{gruzinov,spit06}. We will consider both of these cases in this paper, so that
\beq
\label{eq:lambda_cases}
\lambda(\chi)=
\begin{cases}
\sin^2\chi\ \ \textrm{ vacuum}\\
1+\sin^2\chi\ \ \textrm{ magnetosphere.}
\end{cases}
\eeq

In the coupled evolution of $\Omega$ and $\chi$ we can therefore expect qualitative differences between the two spindown prescriptions: in the vacuum case, if the star aligns (due to, e.g., external torques or an internal poloidal field) then the star \emph{completely stops spinning down}. Regions of almost-aligned and almost-orthogonal rotators would therefore tend to develop very different spin rates after the $\chi$ evolution has finished. By contrast, in the magnetospheric-spindown prescription, all stars continue to spin down regardless of the value of $\chi$, with the spin-down rate only varying by a factor of 2 at most.

Using the definition of the electromagnetic spin-down timescale of equation (\ref{eq:tau_EM}) we obtain
\beq
\tau_{\rm spindown} 
= 2.05 \times 10^3 {\, \rm seconds \,}
\frac{I_{45}}{R_6^6 B_{15}^2 f_{\rm kHz}^2 \lambda(\chi)}.
\eeq
Note that the corresponding electromagnetic alignment timescale differs from this by a $\chi$-dependent factor, as given in equation (\ref{eq:tau_chi_EM}):
\beq\label{chi_EM}
\tau_\chi^{\rm EM} 
= 4.1 \times 10^3 {\, \rm seconds \,}
\frac{I_{45}}{R_6^6 B_{15}^2 f_{\rm kHz}^2}\frac{\sin^2\chi}{\cos^2\chi\lambda(\chi)}.
\eeq

\section{Cooling}
\label{cooling}

The viscosity coefficients are temperature-dependent, and so will vary
considerably over the early life of a neutron star, a phase in which it cools rapidly from an initial temperature (at the end of the proto-neutron-star phase) of around $10^{11}$ K. Simulations of neutron-star cooling demonstrate that different regions of the star cool at different rates, resulting in a temperature profile which can be quite non-uniform and involve sharp gradients in transition regions (see e.g. \citet{gnedin}). The canonical cooling mechanism is modified Urca, and we will assume the cooling proceeds due to this mechanism alone -- but note that if the proton fraction becomes high enough, the much faster direct Urca mechanism can operate.

Although a realistic NS temperature distribution is position-dependent, its profile is rather flat within the core alone, which is the relevant one for us (recall that our analysis is only applicable in the region $0\leq r\leq 0.9R_*$). 
Furthermore, our results do not directly depend on local differences in $T$ or its gradient; the energy losses from the precession are given by volume integrals over the whole star. For these reasons, it is reasonable at our level of working to approximate the star as cooling isothermally, so that it is only time- and not position-dependent.

\begin{figure}
\begin{center}
\begin{minipage}[c]{\linewidth}
\includegraphics[width=\textwidth]{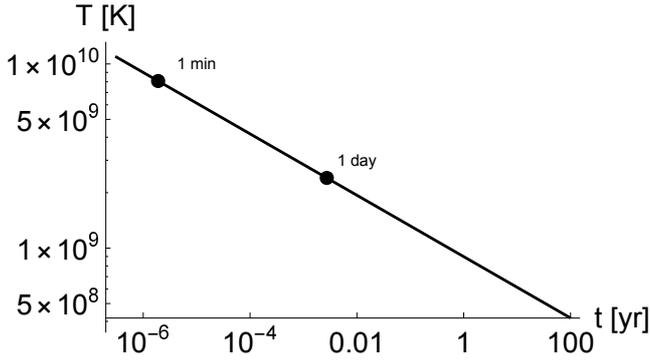}
\caption{ Temperature evolution given by the approximation to modified
  Urca which we adopt. The time scale is in years, but we indicate with points the values of temperature after one minute, and after one day. We choose $T(0)=10^{11}$ K, but the later evolution is fairly insensitive to the exact choice, since the star drops an order
  of magnitude from this value after less than one minute. The cooling
  is easily the shortest timescale to begin with, and so as a first
  approximation one can consider there to be no evolution of $\Omega$
  or $\chi$ for the phase in which $T\gtrsim 10^{10}$ K. \label{fig:T_of_t} }
\end{minipage}
\end{center}
\end{figure}

For the cooling evolution, we
adopt the analytic approximation to modified-Urca cooling given by
\citet{pgw_06} in which neither the protons nor the
neutrons are Cooper-paired (i.e. that the star is assumed to be too hot for
the protons to form a superconductor, or the neutrons a superfluid):
\beq\label{murca_approx}
T(t)=\brac{\frac{6 N^s}{C}t+\frac{1}{T_0^6}}^{-1/6},
\eeq
where $T_0$ is the temperature at time $t=0$, and $N^s$ and $C$ are
numerical constants, which for unpaired baryons are
$C=10^{30}$ erg K${}^{-2}$ and $N^s=10^{-32}$ erg s${}^{-1}$
K${}^{-8}$. In this paper we fix $T_0=10^{11}$ K, roughly corresponding to the
end of the proto-neutron-star phase.  Figure \ref{fig:T_of_t} gives a plot of temperature versus time.
We use this cooling prescription as input for the
temperature-dependent viscosity coefficients $\eta$ and $\zeta$ defined in section \ref{fluidprec}.   

The characteristic timescale for this cooling is given in  equation (14) of \citet{pgw_06}:
\beq
\label{eq:tau_cool}
\tau_\nu^{\rm slow} 
\approx {\, 16 \, \rm seconds \,} \frac{C_{30}}{N_{-32}^{\rm s} T_{10}^6} 
\eeq
where $C_{30} = C / 10^{30}$ erg K${}^{-2}$ and $N_{-32} = N / 10^{-32}$ erg s${}^{-1}$ K${}^{-8}$.

\section{Analytic estimates and timescales}
\label{analytic}

Before trying to solve the equations, let us see how much understanding for the evolution we
can garner from timescale comparisons.  We have a multi-dimensional parameter space to explore, involving spin frequency $\Omega$, inclination angle $\chi$, temperature $T$, and magnetic field strength $B$.  Within the context of our model at least, the magnetic field strength does not evolve in time.

\begin{figure*}
\begin{center}
\begin{minipage}[c]{\linewidth}
\includegraphics[width=\textwidth]{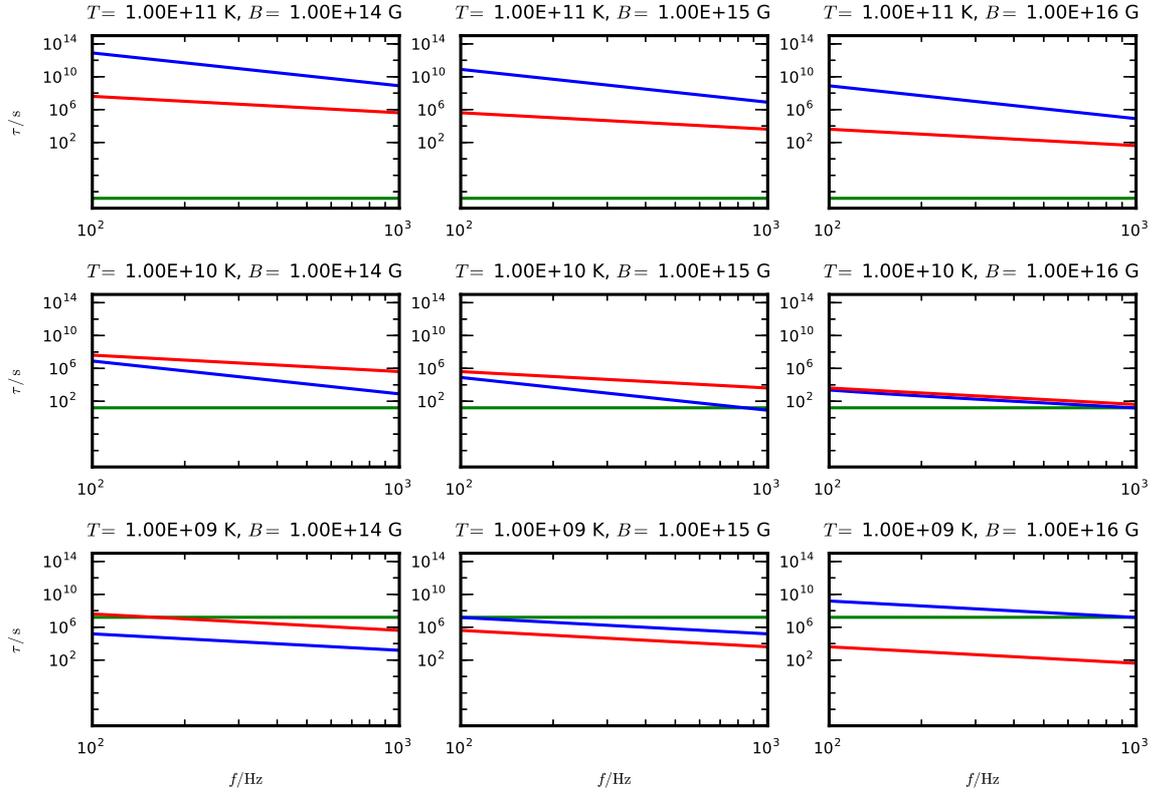}
\caption{  Plots of cooling timescale (green), spindown timescale (red) and orthogonalisation timescale due to bulk viscosity  (blue), versus spin frequency for a range of magnetic field strengths and temperatures.  
 \label{fig:timescales}}
\end{minipage}
\end{center}
\end{figure*}

We have already collected key timescales relevant to those quantities that do vary.  Specifically, we have back-of-the-envelope estimates for inclination evolution due to shear viscosity (equation \eqref{eq:tau_E_shear}), inclination evolution due to  bulk viscosity (equation \eqref{eq:tau_chi_npe_general}),  cooling (equation \eqref{eq:tau_cool}), and electromagnetic torques (equation \eqref{eq:tau_chi_EM}).  The first of these will generally be too long to be relevant.  To help make sense of the parameter space, we plot the other three timescales in Figure \ref{fig:timescales}.
The angular factor $\lambda(\chi)$ has been set equal to unity in estimating the spindown timescale.  
Figure \ref{fig:timescales} shows that, for $T \gtrsim 10^{10}$ K, the timescale for cooling is significantly shorter than the others, so that early in the star's life the temperature will fall at roughly constant $\Omega$ and $\chi$.  However, as is apparent from Figure \ref{fig:timescales}, the ordering of the remaining timescales is dependent upon the exact values of $\Omega$, $B$ and $T$.

To gain insight into the rather complex parameter space, it is useful to  look at the competition between the orthogonalising effect of bulk viscosity and alignment due to the EM torque. To do so, let us set the corresponding timescales equal:   
\begin{equation}
\tau_{\chi}^{\rm EM} = \tau_{\chi}^{\rm bulk} .
\end{equation}
We can then solve for the critical frequency $f_{\rm crit}$ above which orthogonalisation wins out over alignment (for a star with a prolate magnetic deformation).  Equating equations (\ref{eq:tau_chi_EM}) and (\ref{eq:tau_chi_npe_general})  we find:
\begin{equation} \label{fcrit}
f^{\rm crit}_{\rm kHz} \approx \frac{2.8 T_{10}^6}{[4.1 \times 10^3 \tilde\lambda(\chi) T_{10}^6 - 0.155 B_{15}^4 \cos^2\chi]^{1/2} },
\end{equation}
where for convenience we have defined
\begin{equation}
\tilde\lambda(\chi) = \frac{g(\chi)}{\cos^2\chi \lambda(\chi)}.
\end{equation}

\begin{figure}
\begin{center}
\begin{minipage}[c]{\linewidth}
\includegraphics[width=\textwidth]{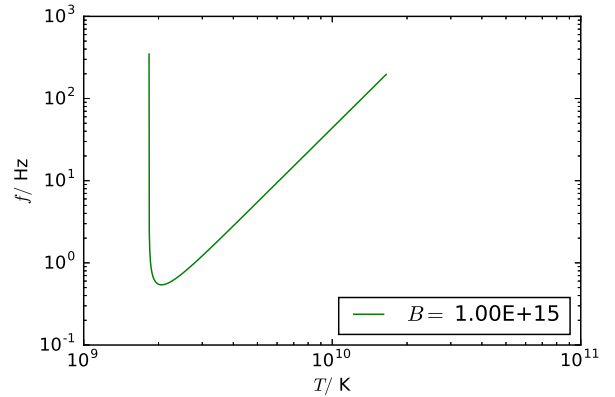}
\caption{  Critical curve dividing orthogonalisation (due to bulk viscosity) from alignment (due to the EM torque) for a $B= 10^{15}$ G star, with the $\chi$-dependent trigonometric factors set equal to unity.    Stars above the curve tend to orthogonalise. \label{fig:f_crit_v_T_single}}
\end{minipage}
\end{center}
\end{figure}

In the context of neutron-star oscillations driven unstable by the emission of gravitational radiation, it is typical to plot curves (or `windows') in the frequency-temperature plane, above which an instability is active. In close analogy with this, we will now fix $B$ in equation \eqref{fcrit} and plot $f_{\rm crit}$ as a function of $T$, to generate an  `orthogonalisation' curve; see Figure \ref{fig:f_crit_v_T_single} for the $B=10^{15}$ G case assuming vacuum-dipole spindown (i.e. $\lambda(\chi)=\sin^2\chi$).  Stars \emph{above} the curve tend to orthogonalise, while stars below align.

\begin{figure}
\begin{center}
\begin{minipage}[c]{\linewidth}
\includegraphics[width=\textwidth]{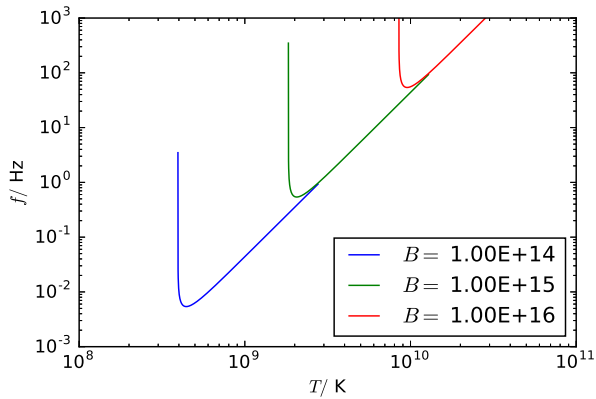}
\caption{  Same as Figure \ref{fig:f_crit_v_T_single}, but now for different magnetic field strengths.  \label{fig:f_crit_v_T_multi}}
\end{minipage}
\end{center}
\end{figure}

This curve has the property that
\begin{equation}
f_{\rm crit} \rightarrow \infty {\hspace{10mm} \rm for \hspace{10mm}} T \rightarrow T_{\rm singular}^+ 
\end{equation}
where
\begin{equation}
\label{eq:first_scaling}
T_{\rm singular} \approx 0.201 B_{15}^{2/3} \left[\frac{\cos^2\chi}{\tilde\lambda(\chi)}\right]^{1/6}.
\end{equation}
Below this temperature there are no solutions for $f_{\rm crit}$.  The curve has a minimum with coordinates
\begin{align}
T^{\rm min}_{10} \approx & 0.206 B_{15}^{2/3}  \left[\frac{\cos^2\chi}{\tilde\lambda(\chi)}\right]^{1/6}, \\
f_{\rm kHz}^{\rm min} \approx & 5.4 \times 10^{-4} \frac{B_{15}^2 \cos\chi}{\tilde\lambda(\chi)}.
\end{align}
In the large $T$, $\omega\tau \ll 1$ regime, the curve has the asymptotic form
\begin{equation}
\label{eq:last_scaling}
f^{\rm crit}_{\rm kHz} \approx 0.044 T_{10}^3 \frac{1}{\tilde\lambda(\chi)^{1/2}} .
\end{equation}
In Figure \ref{fig:f_crit_v_T_multi} we plot orthogonalisation curves for three different values of $B$, again with the $\chi$-dependent trigonometric  factors set equal to unity.

Study of the eigenfunctions in \citet{LJ17} shows that $\chi$ enters the displacement vector $\boldsymbol{\xi}$ via trigonometric factors $\sin\chi$ and $\sin2\chi$.  It follows that in the small-$\chi$ limit, we can expect the scaling $\xi\sim \sin\chi \sim \chi$.  Given that the viscous damping rate is quadratic in $\xi$, we then have  $g(\chi) \sim \sin^2\chi \sim  \chi^2$ in this limit; see  Section \ref{sect:vc} and equations (\ref{eq:E_dot_shear_schematic})--(\ref{eq:E_X_estimate}) above.   For the vacuum dipole torque, $\lambda(\chi) = \sin^2\chi$, as per equation (\ref{eq:lambda_cases}).  It follows that, in the small-$\chi$ limit, $\tilde\lambda(\chi) \approx 1$, and so the plots of Figures \ref{fig:f_crit_v_T_single} and \ref{fig:f_crit_v_T_multi}, where the trigonometric factors of equation (\ref{fcrit}) were set to unity,  apply directly.  However, for the magnetospheric torque, we have $\lambda(\chi) = 1 + \sin^2\chi$ (equation (\ref{eq:lambda_cases})), resulting in $\tilde\lambda \rightarrow \infty$ in the small-$\chi$ limit.  The scalings of equations (\ref{eq:first_scaling})--(\ref{eq:last_scaling}) show that in this case, the orthogonalisation curve moves upwards and to the right in the $(T, f)$ plane, i.e.\ the size of the orthogonalisation region effectively decreases.

Of course, the derivation of the orthogonalisation curves is based on rough timescale estimates, but we can nevertheless hope that their basic shape is robust, and that their location in the $(T, f)$ plane for the vacuum EM dipole and magnetospheric torques is approximately correct, with the understanding that the curves move to higher $T$ and $f$ in the small-$\chi$ limit for the magnetospheric case.

\begin{figure}
\begin{center}
\begin{minipage}[c]{\linewidth}
\includegraphics[width=\textwidth]{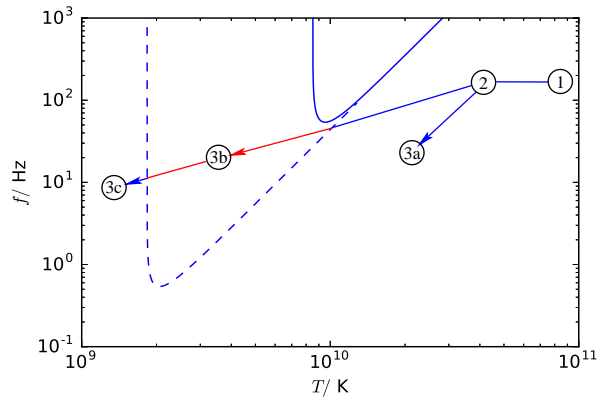}
\end{minipage}
\caption{\label{track_in_window} 
  Possible scenarios for the evolution of a star's inclination angle, shown as trajectories through the orthogonalisation window in the frequency-temperature plane. Take a star with a certain birth spin, starting life at (1). Its initial evolution is dominated by cooling, so its trajectory is horizontally left towards (2). Its evolution from that point then starts to involve spindown too, so its trajectory is diagonally left and downwards in the $f-T$ plane. For a stronger magnetic field the trajectory's gradient is steeper and the orthogonalisation window (solid curve) smaller, so there is a greater chance of it missing this window and therefore ending life as an aligned rotator (3a). For a weaker magnetic field the star enters its now-larger orthogonalisation window (dashed curve) and begins to orthogonalise. If it completes the process whilst in the window, its trajectory ends there (3b), and the star's final state is an orthogonal rotator. If it does not finish orthogonalising in time, there is a risk of it cooling/spinning down enough to exit the window on the other side, at which point the $\chi$-evolution reverses direction, and the star ends up as an aligned rotator (3c).}
\end{center}
\end{figure}

With these orthogonalisation curves, we can now anticipate possible evolution scenarios for $\chi$. In
Figure \ref{track_in_window} we sketch two different trajectories a star could take through the $(f,T)$ plane over time, depending on its magnetic-field strength. In both cases the star starts at (1) and first moves horizonally left, since cooling is the shortest timescale initially. After a little while the star reaches a point (2) where spindown also becomes important, and so its trajectory is now diagonally downward (spindown) and to the left (cooling). For the more highly magnetised star, the downward trajectory is steeper, and its orthogonalisation window is also smaller; it fails to enter the window at all, and its $\chi$-evolution finishes at point (3a) once the star has become an aligned rotator. Instead, with a weaker magnetic field the star enters its orthogonalisation window (which is now larger) and so $\chi$ starts to increase towards $\pi/2$. It will either become an orthogonal rotator somewhere within the window (3b), or exit the window on the other side, in which case $\chi$ decreases again, and the star's ineluctable fate is to become an aligned rotator (3c). We therefore expect stars born with higher magnetic fields and lower rotation rates to evolve into aligned rotators, and others to become orthogonal rotators -- unless their evolution through the window is slow enough that they exit it again. Seeing how small the orthogonalisation window becomes at $10^{16}$ G, it is clear that even the most rapidly-rotating models at this field strength may fail to enter the orthogonalisation window, and will instead align.

\section{Numerical evolutions}
\label{numerical}

\subsection{Summary of equations and numerical solution}

Armed with the intuition gained from the previous section, we now build up to solving the full coupled evolution numerically.
The equations to solve are a pair of coupled ODEs in the variables $\Omega$ and $\sin\chi$. Combining equations \eqref{eq:alpha-dot} and \eqref{eq:E_dot_EM} gives us the explicit form for one of our two ODEs, for the evolution of $\Omega$:
\beq \label{alpha_explicit}
\td{\Omega}{t}=-\frac{R B_{\rm p}^2}{6c^3 I}\Omega^3\lambda(\chi),
\eeq
where we recall that $\lambda=\sin^2\chi$ in vacuum and $\lambda=1+\sin^2\chi$ for a charge-filled magnetosphere.
The other ODE, for the evolution of $\sin\chi$, comes from combining equations \eqref{eq:chi-dot} and \eqref{vdotEuler}:
\beq \label{chi_explicit}
\td{(\sin\chi)}{t}=\frac{\dot{E}_{\textrm{bulk}}+\dot{E}_{\textrm{shear}}
                                    +\epsilon_B(1-\sin^2\chi)\dot{E}_{\textrm{EM}}}
                                  {I\epsilon_B\Omega^2\sin\chi},
\eeq
where we recall that in the denominator we will now be using the quantitative solution for $\epsilon_B$ given by equation \eqref{quant_epsB} and not the qualitative estimate from equation \eqref{eq:epsilon_B}. The numerator of the above equation contains the three mechanisms for energy dissipation. The first is due to bulk viscosity acting on the internal motions:
\beq
\label{eq:E_dot_bulk_general}
\dot{E}_{\rm bulk} = -\int \zeta (\div\bv)^2\ \rmd V,
\eeq
and the second is internal dissipation due to shear viscosity:
\begin{align}
\label{eq:E_dot_shear_general}
\dot E_{\rm shear}
 = -\int & \bigg[
            \eta |\nabla \times {\boldsymbol v}|^2
            + \frac{4}{3} \eta (\nabla \cdot {\boldsymbol v})^2
            - \nabla\eta\cdot\nabla(v^2) \nn\\
   &+2(\nabla\eta \cdot {\boldsymbol v}) (\nabla\cdot {\boldsymbol v}) 
+2 (\nabla\eta\times{\boldsymbol v})\cdot(\nabla\times{\boldsymbol v})
\bigg]  \ \rmd V.
\end{align}
The expression for bulk dissipation of equation (\ref{eq:E_dot_bulk_general}) is well known (see e.g.  \citet{ipser_lindblom_91}), while the expression for shear dissipation was derived specifically for our numerical calculations.  Both expressions for viscous dissipation are derived in appendix \ref{sect:diss_integrals}.   The third mechanism for energy loss is due to the effect of the external electromagnetic torque, and is given in equation \eqref{eq:E_dot_EM}.
The viscosity coefficients are temperature-dependent, and so we need a prescription for the star's cooling. The time evolution of the temperature is given by equation \eqref{murca_approx}; $T$ depends neither on $\Omega$ nor $\chi$, and so evolves passively rather than being coupled with the other ODEs.

The coupled ODEs are evolved using the software package {\scshape Mathematica}. We expect our solutions to asymptotically approach one of the two limiting values of $\chi$: zero or $\pi/2$ (except in the case of magnetospheric spindown, where the $\chi=0$ limit is reached in finite time). We work with the variable $\sin\chi$, which typically evolves until it reaches a value very close to $0$ or $1$, at which point the evolution proceeds progressively more slowly -- effectively stalling for several characteristic timescales. Gradually, however, numerical error builds up and results in pathological behaviour of the solution -- e.g. $\sin\chi$ evolves to become greater than $1$ -- at which point the simulation blows up. This pathological behaviour can be postponed by using {\scshape Mathematica}'s algorithms for handling stiff systems, but cannot be eliminated indefinitely; this is natural for a numerical evolution. For this reason we stop all evolutions when $\chi$ reaches a value of $\pi/360$ ($0.5^\circ$) or $179\pi/360$ ($89.5^\circ$), and report the time taken for it to reach that value as the alignment or orthogonalisation time, respectively. We explore the possible longer-term behaviour of $\chi$, and whether an almost-aligned or almost-orthogonal rotator evolution can revive after several alignment/orthogonalisation timescales and change direction (e.g. from almost-orthogonal back towards aligned), in section \ref{revival}.

We work in variables made dimensionless through division by the
requisite combination of powers of Newton's gravitational constant $G$, the central density $\rho_c$ and the stellar radius $R_*$. We also
normalise the temperature to be in units of $10^{10}$ K. This aids
in producing accurate evolutions, since the fundamental variables are then (very roughly) of
order unity, whereas in physical units one may have parameters varying
by many orders of magnitude. The zero-dissipation oscillatory solutions from \citet{LJ17} may
be redimensionalised to a physical stellar model\footnote{Since we work in a perturbation scheme with $\epsilon_B,\epsilon_\Omega\ll 1$, the stellar structure is just a spherical hydrostatic equilibrium to our order of working.} by using the physical values of $\rho_c$ and $R_*$ for the desired model. This `scale invariance' comes from the fact that no physical constant enters the equations other than $G$. Now however, for the secular evolution problem, the computations involve
additional physical quantities, e.g. the nuclear
density $\rho_{\rm nuc}$ used in the viscosities. Clearly, in dimensionless units this must take the value
$\rho_{\rm nuc}/\rho_c\approx 2.8\times 10^{14} \mathrm{g\ cm}^{-3}/\rho_c$, but this then requires us to specify $\rho_c$ for each evolution instead of being able to rescale results afterwards.
In all the results presented in this paper we fix $\rho_c$ to give us a fiducial neutron-star model
with a mass $1.4$ times that of the Sun, and a radius of 12km. We have, however, performed additional simulations for a 2-solar-mass, 12-km model, finding results barely distinguishable from those of the $1.4$-solar-mass model. This gives us confidence that this aspect of our model does not need further exploration.

Since the
dissipative terms involving $\delta\bB$ are negligible for our problem (see section \ref{diss_int}), we only
need the solutions for $\dot\bxi$, which we established in
\citet{LJ17}. These are composed of radial functions associated with the toroidal and poloidal pieces of the perturbed magnetic field, spherical
harmonics $Y_l^m$ in the angular coordinates, the precession frequency
$\omega$, the background magnetic field $\bB_0$ whose strength is determined by the prefactor $\Lambda$ (see equation \eqref{back_B}) and the spherical background density distribution $\rho_0$ (see equation \eqref{back_rho}).
The star's internal motions have position-dependence through the coordinates $r,\theta,\phi$, and additionally depend on the position-independent quantities $\Lambda,\Omega,\chi$. Over the dynamical timescales considered in \citet{LJ17} these latter three quantities could be regarded as constants, but over the secular dissipative timescales we now consider, $\Omega$ and $\chi$ also have time-dependence. Looking at, e.g., equations (203) and (228) of \citet{LJ17} and comparing, we see that the internal motions have the following scalings:
\begin{align}
\dot\bxi \sim \frac{\omega W_l^m(r) Y_l^m(\theta,\phi)}{B}
 &\propto \frac{\omega\Omega^2\Lambda\sin^2\chi Y_l^m(\theta,\phi)}{\Lambda\rho_0}\nn\\
 &\propto  \frac{\Omega^3\Lambda^2\sin^2\chi Y_l^m(\theta,\phi)}{\rho_0},
\end{align}
where we have also used results from equations  \eqref{eq:epsilon_B}, \eqref{omega} and \eqref{back_B} of this paper. In this equation, the radial functions $W_l^m$ come from a vector-spherical-harmonic expansion of $\delta\bB$. From the above scaling we see that the non-rigid piece of the internal motions depends on several quantities -- some position-dependent and some time-dependent:
\beq
\dot\bxi=\dot\bxi(r,\theta,\phi,\Omega(t),\chi(t),\Lambda)=\dot\bxi({\boldsymbol r},t).
\eeq
The viscosity coefficients also have both position- and time-dependence:
\begin{align}
\zeta &=\zeta(\rho_0,\omega(\Lambda,\Omega,\chi),\tau(\rho_0,T))
    =\zeta(\rho_0(r),\Lambda,\Omega(t),\chi(t),T(t)),\\
\eta &= \eta(\rho_0(r),T(t)).
\end{align}
As a result, the integrands for both the bulk- and shear-viscous dissipation cases depend on position $(r,\theta,\phi)$ and time $(\Omega,\chi,T)$ -- which suggests that we might need to re-evaluate the integrals at each timestep with the new values of $\Omega(t),\chi(t)$ and $T(t)$. The integrals are relatively time-consuming to evaluate just once, because they involve vector operations and then integration of rather complicated expressions for $\dot\bxi$; having to evaluate them at every timestep in the evolution of a single model with fixed $\Lambda$ and $\Omega(0)$ (i.e. magnetic-field strength and birth spin) would already be prohibitively slow. In a survey of the $\Lambda,\Omega(0)$ parameter space the problem would be worse still.

Fortunately, for the shear-viscous dissipation integrand the position- and time-dependent pieces separate neatly. Take, for example, the first term:
\begin{align}
\eta|\curl\dot\bxi|^2\propto & \rho_0^{9/4}T^{-2}(\Omega^3\Lambda^2\sin^2\chi Y_l^m/\rho_0)^2\nn\\
 \propto & T^{-2}(t)\Omega^6(t)\Lambda^4\sin^4\chi(t) \times \rho^{1/4}(r) [Y_l^m(\theta,\phi)]^2.
\end{align}
This allows us to pull the time-dependent pieces out of the integrand, and integrate the position-dependent piece only once at the start, a quantity which we will denote as $\mathfrak{I}$. This quantity, which is just a constant, can then be used at every timestep, and in fact in every evolution regardless of the value of $\Omega,T$:
\begin{align}
\int \eta|\curl\dot\bxi|^2\ \rmd V
\propto & T^{-2}(t)\Omega^6(t)\Lambda^4\sin^4\chi(t)\int \rho^{1/4} (Y_l^m)^2\ \rmd V\nn\\
 =& T^{-2}(t)\Omega^6(t)\Lambda^4\sin^4\chi(t)\mathfrak{I}.
\end{align}
We would also like to separate the integrand of the bulk dissipation integral in a similar manner, into one position-dependent but time-independent piece $\mathcal{I}_{\rm pos}$, and one constant-position but time-dependent
piece $\mathcal{I}_{\rm time}$:
\beq
\mathcal{I}=\mathcal{I}_{\rm pos}(r,\theta,\phi)\mathcal{I}_{\rm time}(T,\chi,\Omega,\Lambda),
\eeq
where $\Lambda$ is not itself a function of time, but is convenient to
assign to the time-dependent piece of $\mathcal{I}$. This time, though, we have a problem: the bulk viscosity coefficient $\zeta$ includes the term $1/(1+\omega^2\tau^2)$, where the relaxation timescale $\tau$ depends both on time (through a temperature dependence) and position (through a density dependence). The position- and time-dependent pieces of $\zeta$ cannot, therefore, be disentangled. Instead, we consider in the next section a suitable approximation to $\zeta$ for which it is possible to effect this split, whilst retaining an acceptable level of accuracy.

\subsection{Approximating the bulk-viscosity coefficient}

\begin{figure}
\begin{center}
\begin{minipage}[c]{\linewidth}
\includegraphics[width=\textwidth]{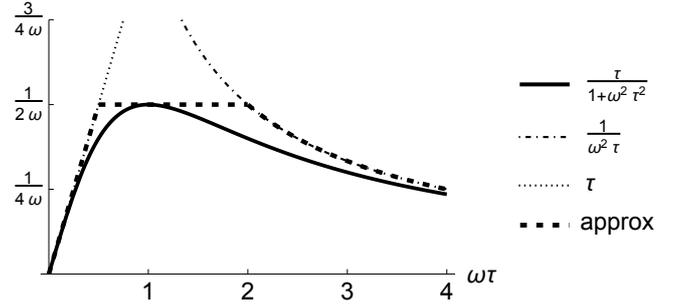}
\end{minipage}
\caption{\label{bulk_approx}
 Approximating the true time- and position-dependent piece (the bold solid line) of the bulk viscosity coefficient $\zeta$ with a piecewise function (labelled `approx'). This approximation function is made up of the standard $\omega\tau\ll 1$ and $\omega\tau\gg 1$ limiting cases, equal to $\tau$ and $1/(\omega^2\tau)$ respectively, joined by an $\omega\tau\sim 1$ approximation: a constant-valued piece given by the maximum of the true function, $1/(2\omega)$. Within the approximation $\zeta$ may be written as a time-dependent function multiplied by a position-dependent function -- an attribute which allows us to speed up our computations by a vast factor, at little cost to the accuracy. The maximum error is 25\%, at the two points joining different pieces of the approximation, but only very small ranges of $\omega\tau$ have errors greater than 10\%, and the changes in the evolution of $\chi$ when crossing between different $\omega\tau$ regimes are virtually unnoticeable.}
\end{center}
\end{figure}

As described above, our problem is how to deal with the function
\beq
f(\omega,\tau)=\frac{\tau}{1+(\omega\tau)^2}
\eeq
within the coefficient $\zeta$. The situation would simplify greatly if the star were always in the regime $\omega \tau \ll 1$ or $\omega \tau \gg 1$, as the space- and time-dependence of $\zeta$ would then separate.   However, equation (\ref{eq:omega_tau})  shows that $\omega\tau$ scales very  steeply with magnetic field strength, spin frequency and (particularly) temperature; we can expect hot young stars to start in the regime $\omega\tau\ll 1$, and evolve to the $\omega\tau\gg 1$ regime as they cool.  It follows that we need to allow for the full $\omega$ and $\tau$ dependence of $f(\omega,\tau)$ in our evolutions.  To do so exactly would be numerically very expensive.  We can instead proceed as follows.  First note the standard approximations in the $\omega\tau\ll 1$ and $\omega\tau\gg 1$ cases:
\beq
f(\omega,\tau)\approx
\begin{cases}
  \tau \ \ & (\omega\tau\ll 1),\\
  \frac{1}{\omega^2\tau} \ \ & (\omega\tau\gg 1).
\end{cases}
\eeq
Whilst these two approximations do match at $\omega\tau=1$, and together allow a split of space- and time-dependent parts, it is clear from
figure \ref{bulk_approx} that they deviate considerably from the
original function at that point. Instead, for intermediate values $\omega\tau\sim 1$
we approximate $f(\omega,\tau)$ by a constant:
\beq
f(\omega,\tau)\approx \max[f(\omega,\tau)]=\frac{1}{2\omega}
 \ \ (\omega\tau\sim 1).
\eeq
Across the full range of $\omega\tau$, we therefore approximate
$f(\omega,\tau)$ by the continuous function given by matching together
the separate approximations for each of the three $\omega\tau$
regimes: 
\beq
f(\omega,\tau)\approx
\begin{cases}
  \tau \ \                               & 0\leq \omega\tau\leq \frac{1}{2},\\
  \frac{1}{2\omega}\ \           & \frac{1}{2}\leq\omega\tau\leq 2,\\
  \frac{1}{\omega^2\tau} \ \ & \omega\tau\geq 2,
\end{cases}
\eeq
where the transition values can be seen graphically in figure \ref{bulk_approx}, or found through an elementary calculation.

One issue remains: $\omega\tau$ is position-dependent, due to the $\rho^{2/3}$ factor in the relaxation timescale $\tau$ (recall that $\omega$ is constant in space), but we are not able to allow for different regions of the star to be in different $\omega\tau$ regimes within the prescription given here. Instead, we treat the whole star as always being in the same regime, and to choose which piece of the piecewise approximation for $\zeta$ to apply at any time we use as a diagnostic the volume-averaged value of $\omega\tau$, i.e. we calculate $\omega\tau$ at the average value of the function $\rho^{2/3}$ (which occurs at a radius $r_{\rm av}=17R_*/30$ for our $\gamma=2$ polytrope).

We do not expect the use of this diagnostic to have a serious effect on our evolutions, however. To understand this, suppose that in a cooling star the value of $\omega\tau$ is unity at the diagnostic radius $r_{\rm av}$, but is greater than $2$ in the inner region of the star and less than $1/2$ in the outer region (since $\omega\tau$ increases with density); the star therefore has regions in which each of the three pieces of our $\zeta$ approximation should be applied. The value of $\omega\tau$ increases everywhere over time as the star cools, however, due to its $T^{-6}$ dependence. Therefore, the $\omega\tau\sim 1$ regime is applied `too late' to the inner region (which has already passed through this regime and out to the $\omega\tau\gg 1$ regime), and `too early' to the outer region (which is still in the $\omega\tau\ll 1$ regime, but will enter the $\omega\tau\sim 1$ regime next). Since the dissipation -- and hence the evolution of $\chi$ -- depends on the volume integral of $\zeta$, these minor inconsistencies should then average out to be rather small.

\subsection{Inclination-angle evolution in a cooling star without spindown}
\label{evol_nospindown}

\begin{figure}
\begin{center}
\begin{minipage}[c]{\linewidth}
\includegraphics[width=\textwidth]{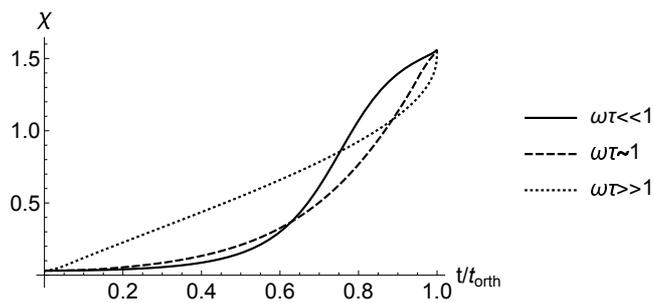}
\end{minipage}
\caption{\label{compare_wt}
 Comparing the evolution of the inclination angle $\chi$ due to bulk
 viscosity in the three regimes $\omega\tau\ll 1$, $\omega\tau\sim 1$
 and $\omega\tau\gg 1$ -- i.e. the three different pieces of the
 function used to approximate the original bulk-viscosity
 expression. Over time $\omega\tau$ itself evolves and can cross
 between regimes, but we have picked initial data for each evolution
 in which $\omega\tau$ spends almost all its time in one regime only. To facilitate direct
 comparison, we have normalised the simulation time for each model by
 the corresponding time taken to orthogonalise.}
\end{center}
\end{figure}

In this section we assume $\Omega$ is constant, so that only equation \eqref{chi_explicit} is evolved. This will prepare us for the full evolutions to follow, and enable us to make contact with earlier studies.

To begin with, we look at the evolution of $\chi$ due to bulk viscous
dissipation alone, comparing how this proceeds in the different
$\omega\tau$ regimes; see figure \ref{compare_wt}. The point here is to compare these regimes, in order to check that the $\chi$-evolution is not dramatically different in the three cases.  Were it to be very different, our approximation to the bulk-viscosity coefficient would risk introducing serious deviations from results using the exact coefficient. Results in figure \ref{compare_wt} are
normalised by the time taken for each evolution to orthogonalise, or more specifically to reach the value $\chi=179\pi/360$ ($89.5^\circ$),
so that the shape of the curves can be directly compared; they do not orthogonalise
in the same real time. For figure \ref{compare_wt} we chose three
simulations in which $\omega\tau$ remained in the same regime for virtually the
entire evolution, but in simulations where a transition is made
between regimes the change is almost unnoticeable, and gives us confidence in our approximation.

\begin{figure}
\begin{center}
\begin{minipage}[c]{0.9\linewidth}
\includegraphics[width=\textwidth]{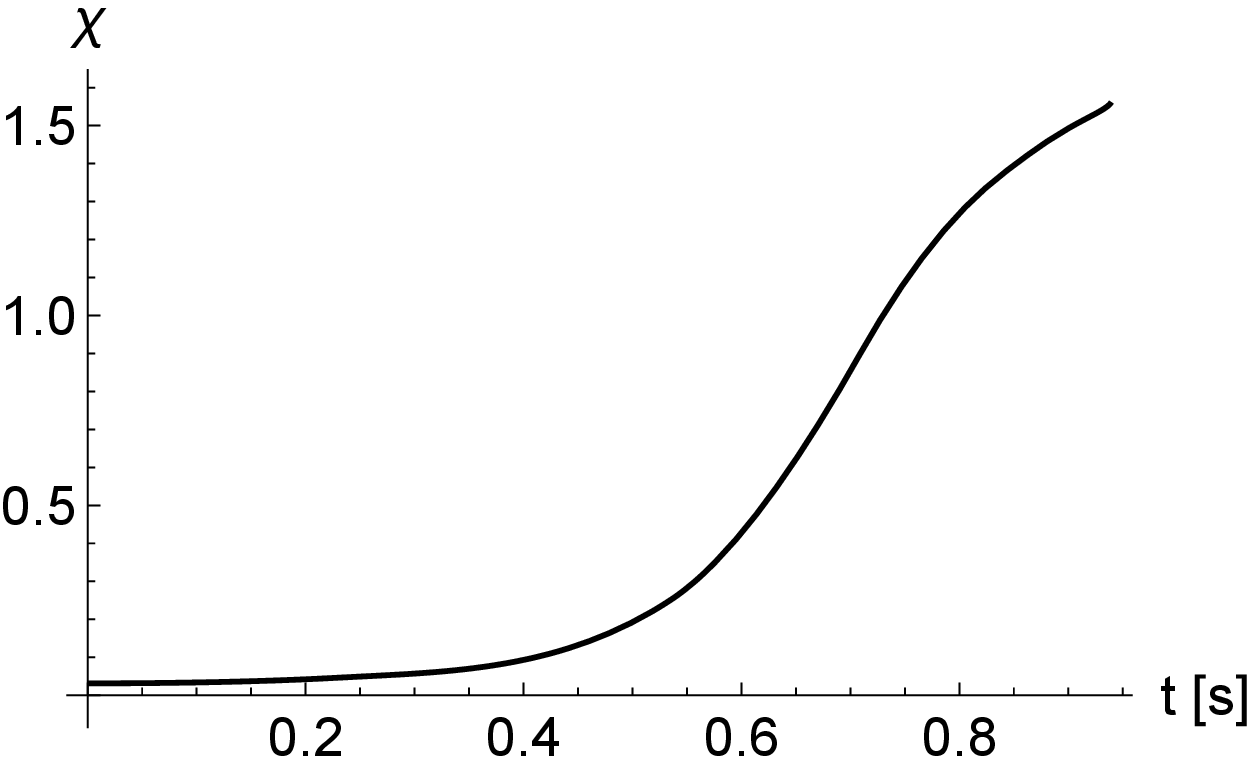}
\includegraphics[width=\textwidth]{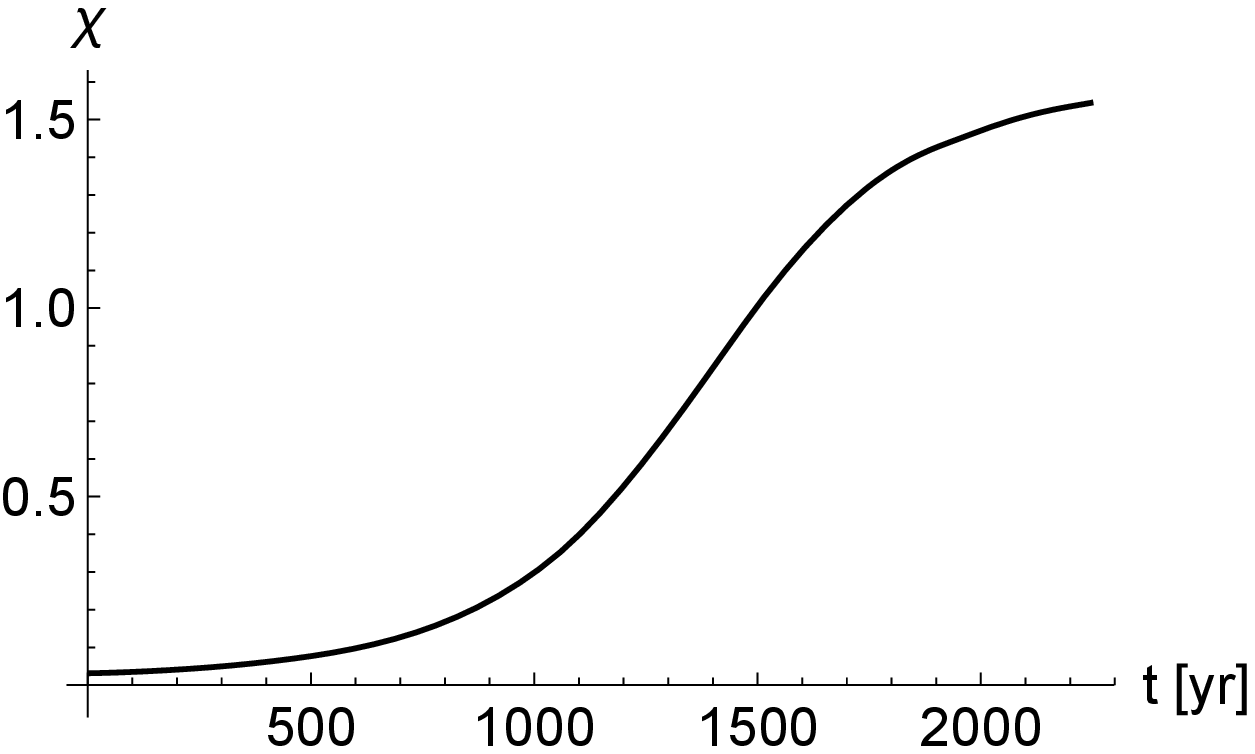}
\end{minipage}
\caption{\label{bulk_shear}
 Comparing the evolution of the inclination angle $\chi$ due to bulk
 viscosity alone (upper panel) and shear viscosity alone (lower
 panel), for a stellar model with $B=10^{15}$ G and $f=1$ kHz. Note that the upper time scale is in seconds, whereas the lower one is in years.}
\end{center}
\end{figure}

Next we compare the effects of bulk and shear viscosity on the evolution of $\chi$. Figure \ref{bulk_shear} shows the results of two evolutions: one where we set $\eta=0$, so that only bulk viscosity is operative, and a second where $\zeta=0$, so that the only dissipation is due to shear viscosity. It is immediately obvious that bulk viscosity is far more efficient at dissipating energy for the chosen model, as expected given the back-of-the-envelope estimates of Sections \ref{sect:shear} and \ref{sect:bulk}.  Given the steep scalings of both effects with spin frequency, magnetic field strength and temperature, we will nevertheless retain shear viscosity in our numerical evolutions, in case it proves important in some portion of our parameter space.

Let us now compare our numerical results with the earlier analytic timescale estimates, beginning with that for bulk-viscous dissipation.
Figure \ref{bulk_shear} suggests that the evolution under bulk viscosity proceeds on a timescale of roughly a second, so, following Figure \ref{fig:T_of_t},  let us take $T=10^{10}$ K in our estimate. For our model, with $R_6=1.2, f=1$ kHz and $B=10^{15}$ G, equation \eqref{eq:omega_tau} suggests that $\omega\tau$ is less than unity, though not appreciably. We only have timescale estimates for the limiting cases of very small and very large $\omega\tau$, so let us adopt the former, equation \eqref{eq:tau_E_npe_small}. This gives us a prediction of $6.6$ seconds, which is acceptably similar to our numerical value of $0.94$ seconds (they differ by a factor of $7$). For the shear viscosity it is not appropriate to take $T=10^{10}$ K, since the star only spends a small fraction of its life at such a high temperature; instead, over the longer timescales on which shear viscosity acts, $T=10^8$ K is a more suitable value (see Figure \ref{fig:T_of_t}). Then, setting $\rho_{15}=1.3$ (i.e. using the central density), equation \eqref{eq:tau_E_shear} predicts orthogonalisation over a timescale of $3.9\times 10^5$ yr, a hundred times slower than the numerical result of $\sim 2 \times 10^3$ yr from the simulation shown in Figure \ref{bulk_shear}. This time the difference between the estimated and actual timescales is large enough to be of concern.

The source of the discrepancy comes from an assumption made early on in deriving the timescales: that the shear and compressional pieces of the fluid motion $\dot\bxi$ are comparable. Any difference between the two is not something which could be anticipated from analytic estimates, but with our quantitative results for $\dot\bxi$ we may now check this. From numerical evaluations, looking at the dominant $m=1$ contributions to $\dot\bxi$, we find that
\beq
\frac{\int|\curl\dot\bxi|\ \rmd V}{\int|\div\dot\bxi|\ \rmd V} \approx 15.4 .
\eeq
Since the bulk viscosity dissipation integral of equation (\ref{eq:E_dot_bulk_general}) depends only on the square of $|\div\dot\bxi|$, while the shear viscosity integral of equation  (\ref{eq:E_dot_shear_general}) depends upon the square of  $|\curl\dot\bxi|$ (together with various other terms), this translates to shortening the shear timescale by a factor of $15.4^2\approx 240$ compared with our original estimate. Accounting for this factor, the shear timescale estimate drops to $1.6\times 10^3$ yr; as for the bulk viscosity estimate, very close to the numerical result.

\begin{figure}
\begin{center}
\begin{minipage}[c]{\linewidth}
\includegraphics[width=\textwidth]{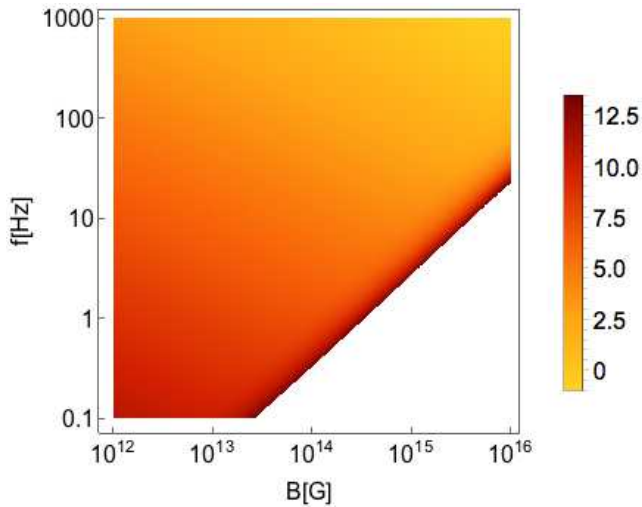}
\end{minipage}
\caption{\label{orthogtime_nospindown}
 Evolution timescales for $\chi$ due to internal dissipation (shear and bulk viscosity) alone, with no spindown. In this density plot, the colourscale shows the time (in log seconds) required for a neutron star's rotation and magnetic axes to orthogonalise, from a starting inclination angle of $\chi_0\equiv\chi(t=0)=\pi/100$. We consider models with (fixed) rotation rates from $0.1$ to $1000$ Hz, and average internal magnetic field strengths from $10^{12}$ up to $10^{16}$ G. Since we have not yet included spindown, all models eventually orthogonalise. Times range from under a second up to a million years (where we cut off the scale). In most of the coloured region $\chi$ evolves predominantly under the action of bulk viscosity, but in the dark-red strip along the bottom right edge, bulk-viscous dissipation becomes very slow. Here dissipation due to shear viscosity eventually becomes a significant effect, once the star has cooled sufficiently.}
\end{center}
\end{figure}

Comparison with the analytic estimates has given us a useful check of our numerical code. We next explore a parameter space of evolutions, with (fixed) rotation rates in the range $0.1-1000$ Hz, average internal magnetic fields in the range $10^{12}-10^{16}$ G, and an initial inclination angle fixed at $\chi_0\equiv\chi(t=0)=\pi/100$. We produce tables of simulation results, splitting the $f = \Omega/(2\pi)$ and $B$ ranges into 64 points equally spaced in $\log f$ and $\log B$, so that in total we perform $64^2=4096$ evolutions. We use these results in Figure \ref{orthogtime_nospindown} to produce a density plot where the colourscale shows the time taken for different models to reach orthogonalisation, going up to times of one million years.  Note that this is much longer than the hundred-year timescale over which we think our model is applicable; we go to these long timescales simply to better understand the interplay of the various factors.  In the majority of the parameter space where orthogonality is reached within the million-year cutoff, bulk viscosity is the dominant effect driving the precession dissipation and hence the $\chi$-evolution. Towards the edge of the shaded region, however, a band of slowly-orthogonalising models appears, especially at higher magnetic fields. For these models shear viscosity begins to become the dominant evolution mechanism.

\begin{figure*}
\begin{center}
\begin{minipage}[c]{\linewidth}
\psfrag{chi-pi2}{$\mathcal{O}$}
\psfrag{chi-0}{$\mathcal{U}$}
\includegraphics[width=\textwidth]{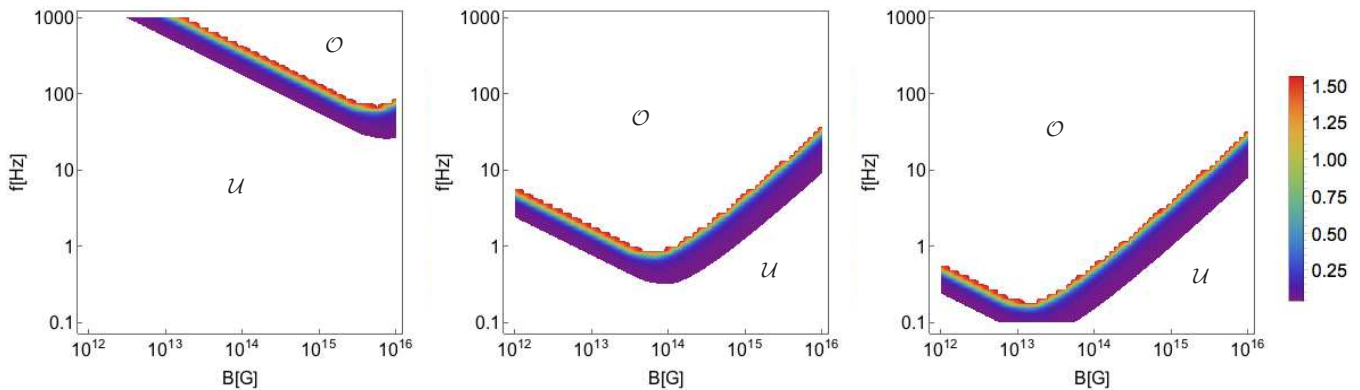}
\end{minipage}
\caption{\label{rainbow_nospindown}
 Evolutions of $\chi$ with internal dissipation but no spindown, for initial rotation rates and average magnetic-field strengths as shown on the axes. The colourscale
 shows values of $\chi$ after one minute, one year and one hundred years (from
 left to right). Since the majority of each plot is covered by one region
 where the inclination angle is essentially unchanged from its
 original value of $\chi_0=\pi/100$ and another where the star has
 reached orthogonality, we truncate the
 colour scale to exclude these models and to highlight the intermediate regions. Specifically, the colourscale shows $\pi/100+\pi/360\leq\chi\leq\pi/2-\pi/360$ rad ($2.3^\circ\leq\chi\leq 89.5^\circ$); we mark the regions of orthogonal models
 $\mathcal{O}$ and those which are essentially unevolved, $\chi\approx\chi_0$, by $\mathcal{U}$. Clearly, there is only ever a narrow region
 with intermediate values of $\chi$.}
\end{center}
\end{figure*}

Figure \ref{orthogtime_nospindown} gave us a broad idea of how long it would take for any given model to orthogonalise. Figure \ref{rainbow_nospindown} is complementary to this, and shows the value of $\chi$ for each model after a fixed amount of time; we show the $\chi$ distribution after one minute, one year and a hundred years for a set of models which all start at $\chi=\pi/100$. The parameter space is dominated by models which have reached orthogonalisation and models which are essentially unevolved (i.e. for which $\chi\approx\chi_0$). In order to highlight those models with intermediate inclination angles, we truncate the colourscale -- showing the value of $\chi$ -- to avoid any model within $\pi/360$ rad (i.e. $0.5^\circ$) of $\chi_0$ or $\pi/2$. At any given time we see that there is only a rather narrow band of models midway through orthogonalising, in agreement with the findings of \citet{DP17}.

The main features of these plots can be understood in some detail, using the back-of-the-envelope estimates of Section \ref{analytic}.  First consider Figure \ref{rainbow_nospindown}.  Each panel shows a snapshot at constant time $t$, and therefore also at constant temperature $T(t)$.  The coloured bands basically separate systems that have orthogonalised on the respective timescales from those that have not, and so should be defined by $\tau_\chi = $ constant, where the constant is one minute, one year, or one hundred years, as one reads the panels from left to right.  From equation (\ref{eq:omega_tau}) we see that, by virtue of the scaling $\omega \tau \sim B^2$, the star will be in the regime $\omega \tau  \ll 1$ for small $B$.  Then we can use equation (\ref{eq:tau_E_npe_small})  to set $\tau_\chi (\omega \tau \ll 1) =$ constant.  The scalings of equation (\ref{eq:tau_E_npe_small})  show that this defines a curve of the form $f \sim B^{-1/2}$, in agreement with the low-$B$ portions of the plots.  Conversely, the same reasoning shows that the star will be in the regime $\omega \tau  \gg 1$ for large $B$.  Then we can use equation (\ref{eq:tau_E_npe_large}) to set $\tau_\chi (\omega \tau \gg 1) =$ constant.  The scalings of equation (\ref{eq:tau_E_npe_large}) show that this defines a curve of the form $f \sim B$, in agreement with the high-$B$ portions of the plots. The back-of-the-envelope estimates therefore do a good job of explaining the qualitative features of Figure \ref{rainbow_nospindown}.  Note, however, that if one inserts actual numbers, we see that in fact equations (\ref{eq:tau_E_npe_small}) and (\ref{eq:tau_E_npe_large}) systematically underestimate the strength of bulk viscosity in producing orthogonalisation, by about a factor of $10$ as measured by the curves $f = f(B)$ defined by $\tau_\chi =$ constant.

This line of reasoning also helps us interpret Figure \ref{orthogtime_nospindown}.  Contours of constant colourscale correspond to constant ages, so as expected have the same U-shaped form as the curves of Figure \ref{rainbow_nospindown}; this feature is just visible to the eye, being most apparent towards the top right of the plot.

\subsection{Coupling to spindown}
\label{evol_spindown}

We now have a coupled system of ODEs, equations \eqref{alpha_explicit} and \eqref{chi_explicit}, describing the joint evolution of $\Omega$ and $\chi$.

Note that the rate of spindown depends on the external magnetic field
strength. However, a peculiarity of a purely toroidal magnetic field
is that it vanishes outside the star, so formally speaking there would
be \emph{no} spindown at all. As discussed in \citet{LJ17} though, a
purely toroidal field is unrealistic, partly on stability grounds and
partly because any mechanism for field generation or evolution is
likely to result in a mixed poloidal-toroidal field. Instead, our
model is intended to be an approximation to a realistic stellar
magnetic field in which the toroidal component dominates. So, we would
expect a `reasonably strong' exterior field to accompany the interior
field we approximate as being purely toroidal -- but at this point in
the calculation our model ceases to be completely self-consistent, and
we must choose the exterior field strength in an arbitrary manner. We
believe the simplest starting point is to assume the interior and exterior
fields are comparable, and therefore to set:
\beq
B_{\rm p}=\bar{B}_{\rm int}
\equiv\frac{3}{4\pi R_*^3}\int|\bB|\ \rmd V
=\frac{3}{4\pi R_*^3}\int\Lambda\rho r\sin\theta\ \rmd V.
\eeq
The ratio $B_{\rm p}/\bar{B}_{\rm int}$ is really the only significant tuneable parameter within our model. In all the results presented in this paper we set the ratio to unity, but it should be noted that models of neutron-star magnetic equilibria suggest an internal field appreciably stronger than the external one; we discuss this more in section \ref{validity}.

\subsubsection{Evolution of $\chi$ under spindown alone} \label{sect:spindown_alone}

\begin{figure*}
\begin{center}
\begin{minipage}[c]{\linewidth}
\psfrag{A}{$\mathcal{A}$}
\psfrag{U}{$\mathcal{U}$}
\includegraphics[width=\textwidth]{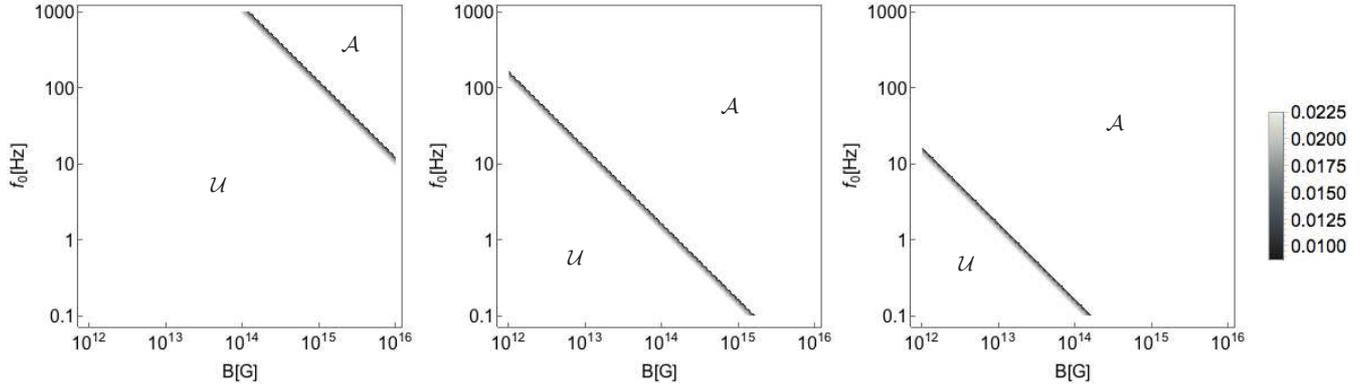}
\end{minipage}
\caption{\label{rainbow_purespindown}
 Evolutions of $\chi$ under (magnetospheric) spindown but without internal dissipation, for initial rotation rates and average magnetic-field strengths as shown on the axes and with $\chi_0=\pi/100$. The greyscale
 shows values of $\chi$ after one minute, one year and one hundred years (from
 left to right). As for the previous plot, we highlight the currently-evolving region of models by truncating the range of $\chi$ shown: specifically, the greyscale shows $\pi/360\leq\chi\leq\pi/100-\pi/360$ rad ($0.5^\circ\leq\chi\leq 1.3^\circ$); we mark the regions with aligned models
 $\mathcal{A}$ and those which are essentially unevolved, $\chi\approx\chi_0$, by $\mathcal{U}$. As for the case of internal dissipation,  there is only ever a narrow region
 with intermediate values of $\chi$.}
\end{center}
\end{figure*}

To help us to understand the main results of this paper, for the coupled evolution of $\Omega$ and $\chi$ under both internal dissipation and spindown, we pause briefly to study the evolution of $\chi$ under the effect of spindown alone, with internal dissipation turned off. We employ the magnetospheric-spindown prescription, i.e. with $\lambda(\chi)=1+\sin^2\chi$ in equation \eqref{eq:E_dot_EM}. In figure \ref{rainbow_purespindown} we produce a plot analogous to Figure \ref{rainbow_nospindown}: from an initial inclination angle of $\chi=\pi/100$ we show the later $\chi$ distribution at snapshots taken at one minute, one year and one hundred years. A simple diagonal band separates models which have reached alignment (defined for our numerical results as occurring at $\chi=\pi/360$, i.e. $0.5^\circ$) from those which are essentially unevolved from their starting value. Because we start close to alignment the band of currently-evolving inclination angles, shown by the greyscale, is very thin.

These spindown-only evolution results agree well with our expectations from earlier. The predicted alignment timescale $\tau_\chi$ from equation \eqref{chi_EM} is $B^{-2}f^{-2}$, so that a line of constant $\tau_\chi$ should have a slope of $-1$ in the (logarithmic)  $f-B$ plane; this is indeed seen in Figure \ref{rainbow_purespindown}. Let us also check the magnitude of the timescale from this figure, not just its scaling. We see from the numerical results plotted that a star with $\chi_0=\pi/100$, $f_0=10$ Hz and $B=10^{16}$ G aligns after one minute. Putting these same values into equation \eqref{chi_EM}, and using standard small-angle approximations ($\sin\chi\approx\chi$,$\cos\chi\approx 1$) we get a back-of-the-envelope estimate that alignment should occur after
\begin{align}
\tau_\chi^{\rm EM}
&\approx 4 \times 10^5 {\, \rm seconds \,}
   \frac{\sin^2\chi_0}{\cos^2\chi_0(1+\sin^2\chi_0)}\nn\\
&\approx 4 \times 10^5 {\, \rm seconds \,}
   \brac{\frac{\pi}{100}}^2\approx 400\ {\rm s},
\end{align}
in satisfactory agreement with the numerical results.

\subsubsection{Evolution of $\chi$ under spindown and viscosity}

\begin{figure}
\begin{center}
\begin{minipage}[c]{\linewidth}
\includegraphics[width=\textwidth]{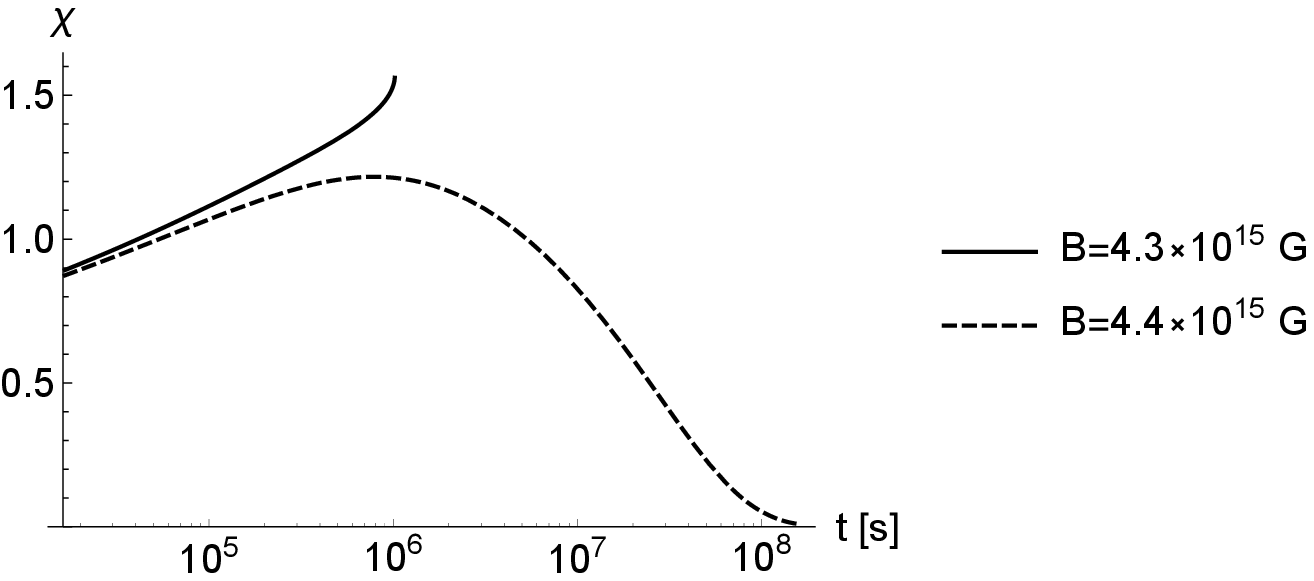}
\includegraphics[width=\textwidth]{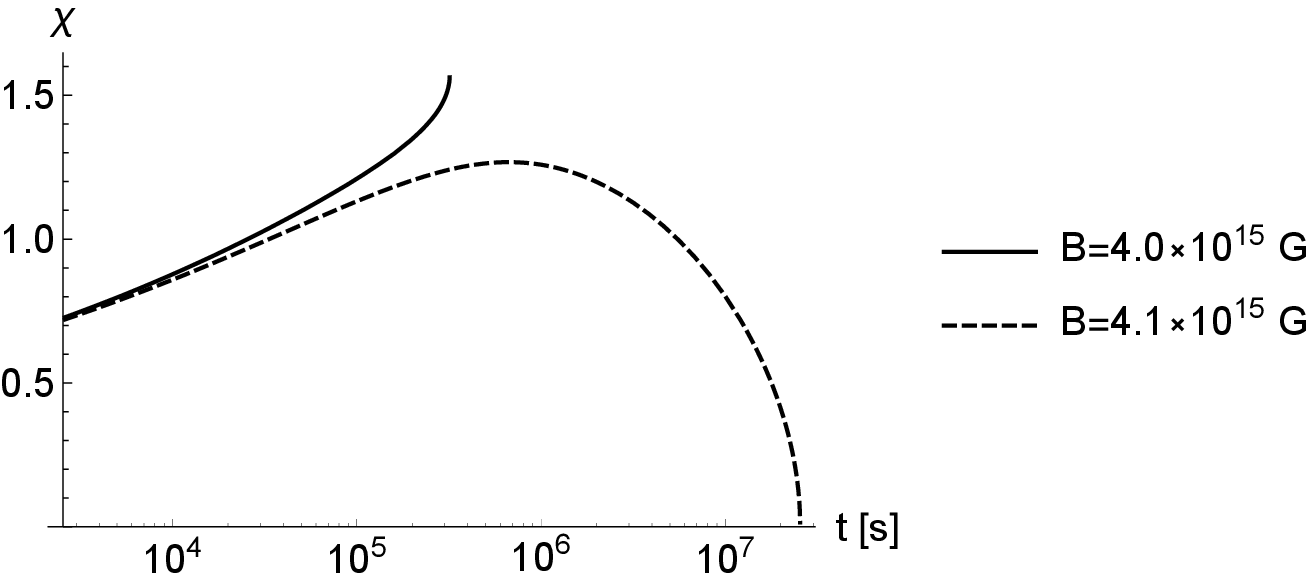}
\end{minipage}
\caption{\label{intext_balance}
 The evolution of $\chi$ for fixed initial inclination angle of $\pi/6$ and initial
 rotation rate of $10$ Hz. We show results for two slightly different magnetic field
 strengths that demonstrate the possible evolution scenarios,
 and the fine balance between them. Top: vacuum-dipole spindown; bottom:
 magnetospheric spindown.}
\end{center}
\end{figure}

We now move to the full problem involving the coupled evolution of $\Omega$ and $\chi$ under internal (viscous dissipation) and external (electromagnetic torque) processes.
In both the vacuum and magnetospheric spindown prescriptions, there will be a competition between the action of the electromagnetic torque in driving the star's axes into alignment, and internal dissipation driving them orthogonal. We had some hints from our analytic estimates in section \ref{analytic} of the joint effect of these two on the parameter space of models, but let us now investigate this quantitatively. 

First, in Figure \ref{intext_balance} we give examples of evolutions on the threshold between the aligned and orthogonal regimes, where the internal and external dissipative mechanisms almost balance. Fixing the initial rotation rate $f_0=10$ Hz and the initial $\chi$ as $\pi/6$, we find that for weaker magnetic fields $\chi\to\pi/2$. Increasing the magnetic-field strength, however, we enter the regime of models which align. Such a model, just into the aligned regime, shows the same initial evolution of $\chi$ towards $\pi/2$ as a neighbouring orthogonalising model, but then the  $\chi$ curve bends downwards and the angle continues decreasing until it reaches zero. The qualitative behaviour is the same for both spindown prescriptions (vacuum dipole and magnetospheric), but because the magnetospheric prescription represents a stronger spindown for a given field strength, alignment can occur at slightly weaker field strengths than in the corresponding vacuum-dipole case.  We show these plots as they illustrate the rather rich behaviour that the competition between alignment and orthogonalisation can produce, with non-monotonic evolution in $\chi(t)$.  

\begin{figure}
\begin{center}
\begin{minipage}[c]{\linewidth}
\includegraphics[width=\textwidth]{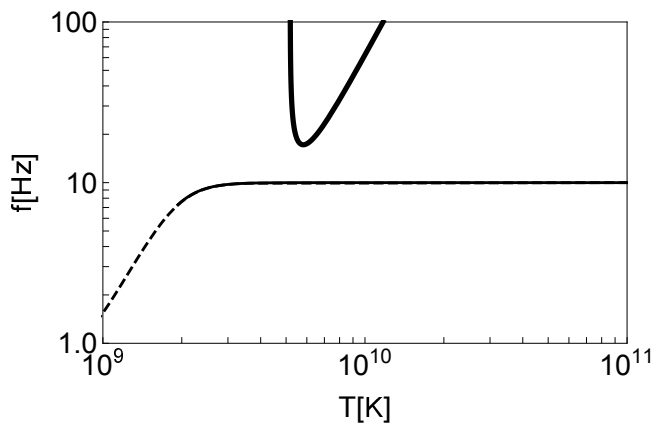}
\end{minipage}
\caption{\label{fig:traj_example}
 Numerical trajectories in the $(T, f)$ plane for the two stars whose $\chi$-evolution is shown in the bottom panel of Figure \ref{intext_balance}, together with the qualitatively predicted orthogonalisation curve for a star with $\chi = \pi/6$, $B = 4 \times 10^{15}$ G under a magnetospheric spin-down torque.  \skl{With the understanding that this latter curve would move slightly lower and to the left if it could be calculated quantitatively, we can envisage that the solid (weaker-field) trajectory would finish in the window upon reaching orthogonalisation, whilst the dashed trajectory would fail to complete its evolution within the window, then exit and later reach alignment.}}
\end{center}
\end{figure}

To gain further insight, a plot of the trajectories in the spin frequency--temperature plane is shown in Figure \ref{fig:traj_example}, for the  magnetospheric torque, corresponding to the two stars of the bottom panel of Figure \ref{intext_balance} (i.e. the evolution of longer duration, shown with the dashed line, corresponds to a model which eventually aligns). Together with these trajectories -- quantitative results obtained from numerical simulations -- we plot in bold the corresponding orthogonalisation curve, corresponding to a $B = 4 \times 10^{15}$ G star with a magnetospheric torque, and with all $\chi$-dependent trigonometric factors set to unity.
Strictly speaking Figure \ref{fig:traj_example} is not a quantitative version of Figure \ref{track_in_window}, since the orthogonalisation curve itself evolves over time (via its dependence on $\chi(t)$). Nonetheless, a comparison between these numerical trajectory solutions and the analytic critical curve is instructive. 
We can now attempt to connect these trajectories and the orthogonalisation window to explain the behaviour of $\chi$ in the lower plot of Figure \ref{intext_balance}. The stars are `born' on the right hand side of Figure \ref{fig:traj_example}, in the region where alignment wins out over orthogonalisation.  Given that Figure \ref{intext_balance} shows that -- at the level of time resolution employed in the plot -- $\chi$ initially increases for both stars, we can infer that the two stars almost immediately cool into the orthogonalisation window, before any significant alignment occurs.  Unfortunately, we see the trajectories do not quite penetrate the orthogonalisation window -- they pass just below it.  This is presumably due to the use of rough timescale and constant-$\chi$ estimates in obtaining the orthogonalisation curve.
With this understanding, we can see that upon entry to the window, the $\chi$ of both stars then grows steadily. The lower $B$-field star, with its slightly larger orthogonalisation window, reaches orthogonality whilst inside its window, and we terminate the evolution. However, for the higher-field star,  orthogonality is not quite reached while the star's trajectory is within its window.  Instead, the star cools and spins down out of the window again, back into the alignment region, and $\chi$ then steadily decreases.

\begin{figure*}
\begin{center}
\begin{minipage}[c]{\linewidth}
\includegraphics[width=\textwidth]{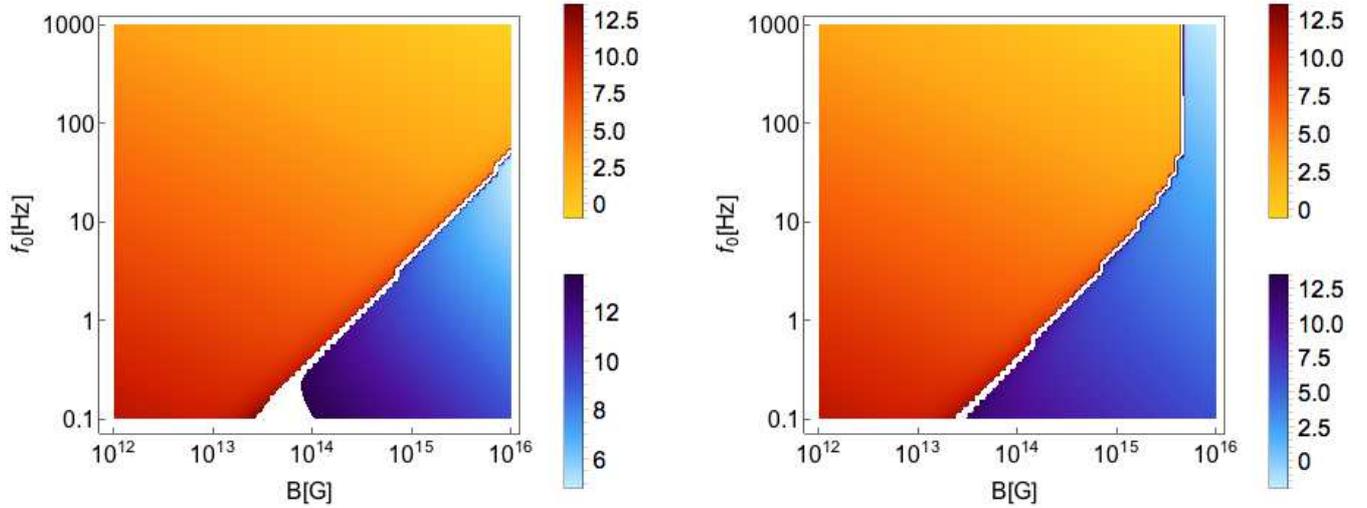}
\end{minipage}
\caption{\label{times_vac_vs_spit}
 Coupled evolutions with internal dissipation and spindown, for
 initial $\chi=\pi/100$. Density plot showing the time (in log seconds) taken for models
 either to align (blue colourscale) or orthogonalise (red-yellow
 colourscale). Left panel: vacuum spindown, right: magnetospheric spindown. Note that in the latter case every model with
  $B\gtrsim 5\times 10^{15}$ G aligns.}
\end{center}
\end{figure*}

Next we make plots analogous to Figure \ref{orthogtime_nospindown}, showing the time taken for a model to finish evolving to a limiting case -- either an aligned or an orthogonal rotator (more precisely, recall that our simulations finish once $\chi<\pi/360$ or $\chi>179\pi/360$).   As before, to help us understand our model, we follow our evolutions up to an age of a million years, bearing in mind that the effects of forming the crust and  superfluidity/superconductivity will have made themselves felt long before this time.  We now need to differentiate between models aligning and orthogonalising, so we use the colourscale from Figure \ref{orthogtime_nospindown} for the latter, and a second blue colourscale for models which align. 

The first of these figures in the case coupled to spindown is Figure \ref{times_vac_vs_spit}, showing the difference in effects of a vacuum spindown prescription or a magnetospheric one. For $\chi_0=\pi/100$, we see that for both spindown prescriptions a large portion of the parameter space is occupied by models which orthogonalise, especially for more rapid rotation and weaker magnetic fields. Models with slow initial rotation and weaker magnetic fields always orthogonalise, but over long timescales. Rapid rotation and strong magnetic fields are optimal conditions for both internal dissipation and external torques to act rapidly, so in this region of parameter space there is a competition between these two effects, and an abrupt transition between aligned and orthogonal models. For very strong magnetic fields and slower rotation rates, the external torque tends to be more effective. For magnetospheric spindown -- but not for the vacuum case -- there is a cut-off field strength of $B\approx 5\times 10^{15}$ G beyond which every model aligns.

\begin{figure}
\begin{center}
\begin{minipage}[c]{\linewidth}
\includegraphics[width=\textwidth]{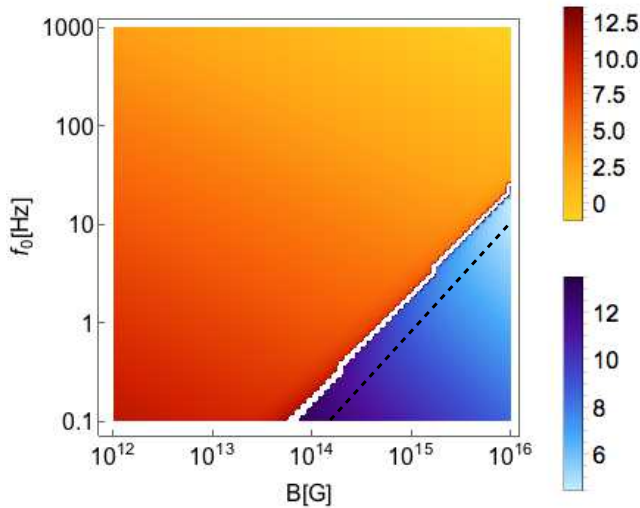}
\end{minipage}
\caption{\label{times_chi6-3}
 The effect of different initial $\chi$ on the distribution of models. The right-hand side of Figure \ref{times_vac_vs_spit} showed times taken for models to align/orthogonalise under the joint effect of internal dissipation and magnetospheric spindown, for $\chi_0=\pi/100$. Here we show the same plot but for $\chi_0=\pi/6$; as expected, since we begin closer to $\chi=\pi/2$ the region of orthogonal models grows. The corresponding plot for $\chi_0=\pi/3$ is very similar, so for this case we simply indicate the aligned-orthogonal boundary by a dashed line.}
\end{center}
\end{figure}

Figure \ref{times_chi6-3} complements the right-hand panel of the previous figure, again showing alignment and orthogonalisation times for models with magnetospheric spindown, but this time for a larger initial inclination angle, $\chi_0=\pi/6$. The effect is what one would expect: having started closer to orthogonality, more of the parameter space evolves to become orthogonal rotators. There is no longer a cut-off field strength beyond which every model aligns. The corresponding plot for $\chi_0=\pi/3$ is qualitatively very similar to the $\chi_0=\pi/6$ case, with the region of orthogonal rotators broadening further. Since there is little additional information to be gleaned from the $\chi_0=\pi/3$ plot, we omit it, and simply indicate with a dashed line on the $\chi_0=\pi/6$ plot the demarcation between  orthogonal and aligned rotators when $\chi_0=\pi/3$.

\begin{figure*}
\begin{center}
\begin{minipage}[c]{\linewidth}
\psfrag{chi100}{$\chi_0=\pi/100$}
\psfrag{chi6}{$\chi_0=\pi/6$}
\psfrag{chi3}{$\chi_0=\pi/3$}
\psfrag{onemin}{one minute}
\psfrag{oneyr}{one year}
\psfrag{hunyr}{one hundred years}
\psfrag{A}{$\mathcal{A}$}
\psfrag{O}{$\mathcal{O}$}
\psfrag{U}{$\mathcal{U}$}
\includegraphics[width=\textwidth]{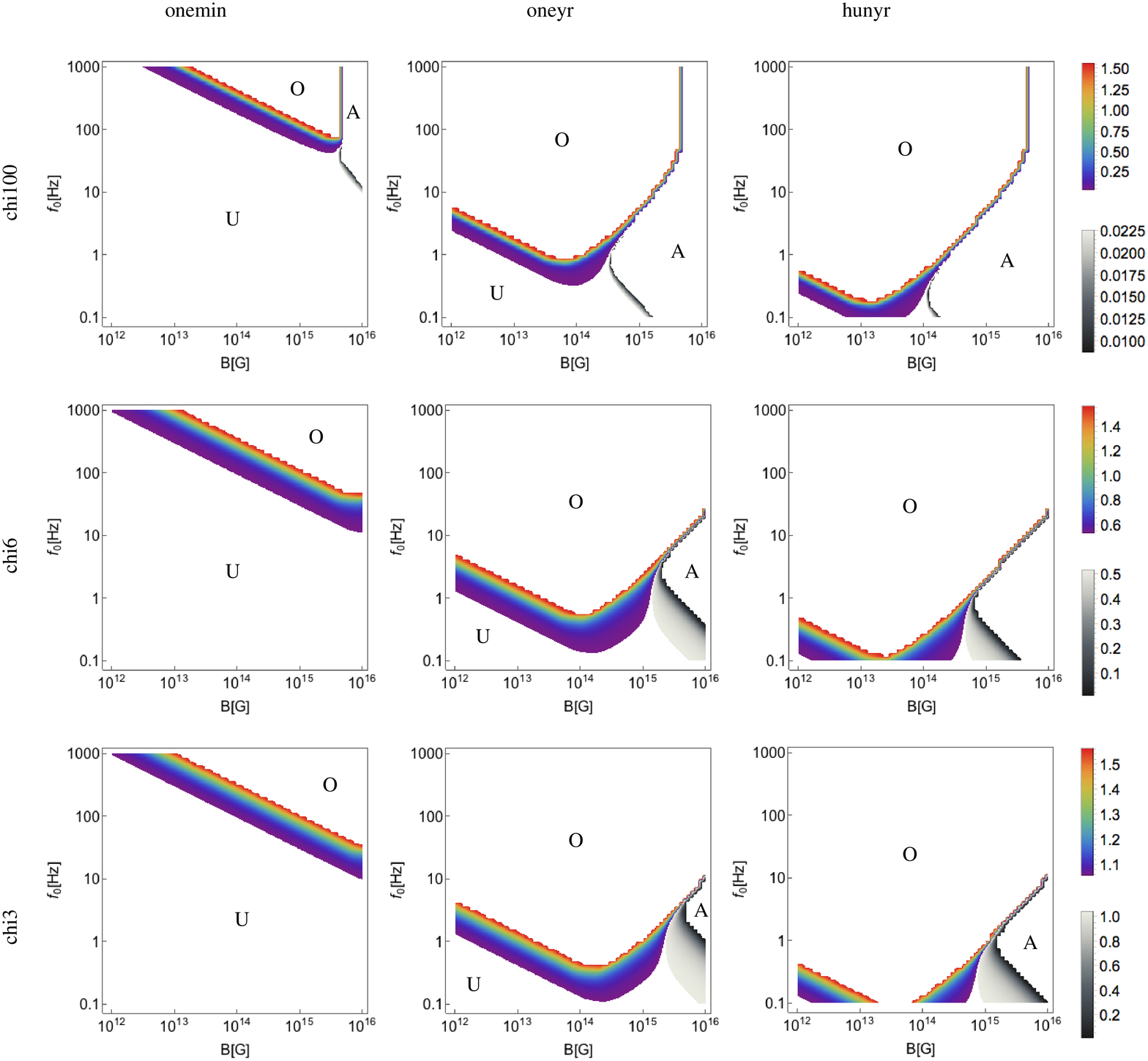}
\end{minipage}
\caption{\label{rainbow_spindown}
 Snapshots in time showing the distribution of $\chi$ values for stars with internal viscous dissipation and magnetospheric spindown. Time increases from left to right, and initial $\chi$ increases from top to bottom, as shown. We have cut off a (small) interval around the initial
 $\chi$ in each case, colour-coding the models which orthogonalise from that point
 with the rainbow colour scheme, and the models which align from that
 point with the greyscale. For clarity we have not plotted certain
 regions, but have instead marked them with a letter: these are where
 the star has either reached alignment ($\mathcal{A}$), orthogonality ($\mathcal{O}$), or remains
 unchanged ($\mathcal{U}$) from its original inclination angle. The same pair of colourscales is adopted for all plots for the same initial $\chi$. We see that bulk-viscous dissipation acts to carve out a region of orthogonal rotators, beginning from the top right of the plot and progressing diagonally downwards towards the bottom left region over time. For the highest field strengths, however, it has to compete with the effect of the magnetospheric torque, which instead carves out a region of aligned rotators for slow initial rotation (and in some cases, also for rapid initial rotation). After a hundred years, almost the entire parameter space has reached one limiting case for $\chi$.}
\end{center}
\end{figure*}

Figure \ref{rainbow_spindown} shows values of $\chi$ at snapshots in time for our usual range of $f_0$ and $B$ values. As for the earlier figures \ref{rainbow_nospindown} and \ref{rainbow_purespindown}, we exclude all regions within $0.5^\circ$ of orthogonality, alignment, or the initial $\chi$ value -- marking the three excluded regions $\mathcal{O}$, $\mathcal{A}$ and $\mathcal{U}$. We again use the rainbow colourscale to show models where the star is close to orthogonality, and the greyscale for those stars close to alignment. 

The first main point to note from this figure is that by the age of one hundred years, almost all models have evolved to become either aligned or orthogonal rotators, with only a small region of oblique rotators (those with intermediate values of $\chi$). Note that once a given model has reached one limit -- aligned or orthogonal -- the evolution of $\chi$ ceases, so that no $\mathcal{O}$ or $\mathcal{A}$ region will ever decrease in size as time progresses, and any $\mathcal{A}$-$\mathcal{O}$ boundary line is formed permanently; this is discussed in more detail in section \ref{revival}. The spindown, however, continues indefinitely after the $\chi$-evolution has ceased, since the limit $\Omega=0$ is not reached in finite time.

While displaying a lot of structure, many of the features of Figure \ref{rainbow_spindown} can be readily understood in terms of earlier results.  The thick U-shaped curve is simply the curve that featured prominently in Figure \ref{rainbow_nospindown}.  It is a curve of constant $\tau_{\chi}^{\rm bulk}$, dividing those stars that have orthogonalised due to bulk viscosity from those that have not; see section \ref{evol_nospindown} for details.  The lines of gradient $-1$ visible at the bottom right of most plots in Figure \ref{rainbow_spindown} are the same ones which appeared in Figure \ref{rainbow_purespindown}: lines of constant $\tau_\chi^{\rm EM}$. They divide those stars that have aligned due to electromagnetic torques from those that have not; see section \ref{sect:spindown_alone} for details.

The main new feature of Figure \ref{rainbow_spindown} is the curves separating the aligned and orthogonalised stars.  These can be best understood in terms of the evolutionary paths through the orthogonalisation window of Figure \ref{track_in_window}. There we anticipated that a star born with a very strong magnetic field could miss the window altogether and therefore evolve to become an aligned rotator; this was also more likely to happen for slower initial rotation. Models with weaker magnetic fields were predicted to enter this window and become orthogonal rotators. This expectation is broadly consistent with the results displayed in Figure \ref{rainbow_spindown}.

\begin{figure}
\begin{center}
  \begin{minipage}[c]{\linewidth}
\includegraphics[width=\textwidth]{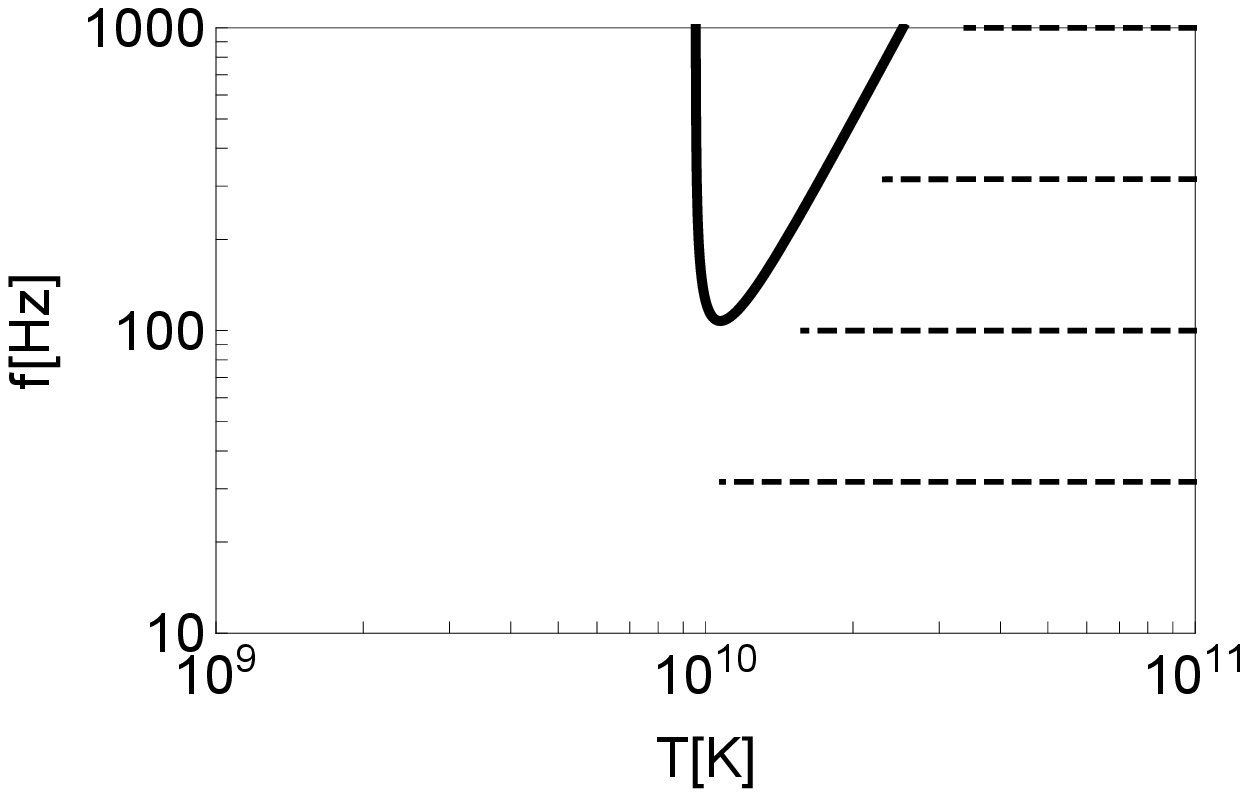}
\end{minipage}
\caption{\label{windowmiss}
 Four trajectories (dashed lines) in the $f-T$ plane, all for a $10^{16}$ G star with $\chi_0=\pi/100$, but with four different initial rotation rates: $1000$, $316$, $100$ and $31.6$ Hz, from top to bottom. Spindown is negligible over these short timescales, so the trajectory direction is dominated by cooling, and terminates once the evolution of $\chi$ finishes. The trajectories are shown together with the estimated orthogonalisation window (solid curve) for a $10^{16}$ G star under magnetospheric spindown. From the window alone it is clear that any trajectory starting at $f_0\lesssim 100$ Hz will miss the window and result in an aligned rotator. The less obvious result is that the $f_0=1000$ Hz star becomes an aligned rotator, because its $\chi$-evolution finishes before reaching the window. Lowering $f_0$, the $\chi$-evolution timescale slightly increases, but the window is also further away. This explains the almost vertical dividing line between aligned and orthogonal rotators in the top-right region of the $\chi_0=\pi/100$ panels in Figure \ref{rainbow_spindown}.}
\end{center}
\end{figure}

As before, we can gain some additional insight by looking at trajectories of different evolutions within the $f-T$ plane, as shown in figure \ref{windowmiss}, and comparing these with the predicted orthogonalisation curve (recalling the caveat that the latter is a constant-$\chi$ analytic estimate, and so we are not quite comparing like with like). 
Firstly, we see that spindown does not significantly alter these  trajectories, which are almost horizontal (since they finish once $\chi$ has finished evolving -- which takes a matter of seconds for these models). Secondly, and more subtly, we gain an understanding for the almost vertical line separating $\mathcal{A}$ and $\mathcal{O}$ regions in the top row of panels in figure \ref{rainbow_spindown}. The most rapidly-rotating model ($f_0=1000$ Hz) from Figure \ref{windowmiss}, although on course to enter the orthogonalisation window, finishes its evolution -- it becomes an aligned rotator -- before reaching the window. Decreasing $f_0$, the trajectories increase in length, but also have further to go to reach the orthogonalisation window, simply from the diagonal shape of the right-hand part of the window's curve. This naturally leads to the section of the $\mathcal{A}-\mathcal{O}$ dividing line which is virtually independent of birth rotation rate.

Figure \ref{rainbow_spindown} shows only cases with magnetospheric spindown; the corresponding case for vacuum spindown is qualitatively rather similar and does not warrant the extra space needed to show it. The major change is that the vacuum spindown is weaker, so the regions with orthogonal rotators are larger.

\begin{figure*}
\begin{center}
\begin{minipage}[c]{\linewidth}
\includegraphics[width=\textwidth]{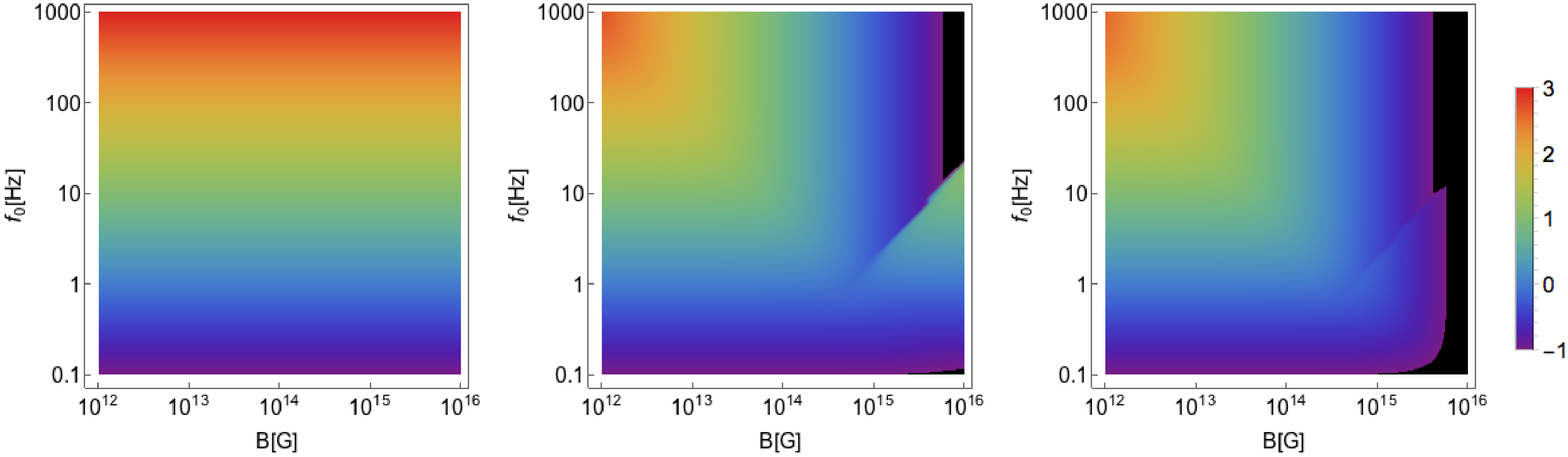}
\end{minipage}
\caption{\label{spindown_hunyr}
 Effect of the coupled evolution of $f$ and $\chi$ on the star's
 spindown, with $\chi_0=\pi/6$. The colourscale shows the present-day spin rate in log
 Hz.  Left: a reference plot to guide the eye, of $f$ at
 time zero. Middle plot: distribution of rotation rates after a hundred
 years, for spindown in vacuum. Right-hand plot: same plot but for
 the case of a charge-filled magnetosphere.}
\end{center}
\end{figure*}

Throughout this section we have been solving the coupled evolution of $\chi$ and rotation rate $\Omega$, though all results we have presented so far have been for the inclination angle. Next we wish to look at the evolution of the spin rate, at late times, after $\chi$ has ceased evolving significantly.     To do so, we have had first to evolve the coupled system of ODEs until $\chi$ reaches one of the two limiting cases, alignment or orthogonality, at which point there is no further evolution of the angle; the star remains as either an aligned or orthogonal rotator indefinitely. The spindown continues, however, so we use the stellar parameters at the endpoint of each coupled $\chi$-$\Omega$ evolution to start a new evolution involving spindown alone, with the appropriate \emph{fixed} end-state value of $\chi$ ($0$ or $\pi/2$) inserted in the spindown formula. The full evolution of $\Omega$ is then given by joining together these two separate evolutions, and representative results are given in Figure \ref{spindown_hunyr}, for $\chi_0=\pi/6$. Here we plot the distribution of rotation rates with both the vacuum-dipole and magnetospheric prescriptions after a hundred years. The vacuum-dipole case features a dramatic transition between regions of rapidly-rotating and slowly-rotating models. This is a consequence of the fact that $\dot\Omega\to 0$ as $\chi\to 0$. In particular, some of the stars with magnetic field strengths near to $10^{16}$ G align very rapidly at a point when their spin rate is still relatively high, and then remain with this value of $\Omega$ indefinitely afterwards (within the context of this model alone, of course). Other neighbouring models, which instead become orthogonal rotators, continue to spin down after the $\chi$ evolution is over. The transitions between aligned- and orthogonal-rotator regions may still be discerned from the corresponding plot showing magnetospheric spindown, but now they are much less dramatic -- as expected from the fact that the difference in spindown rate between the $\chi=0$ and $\chi=\pi/2$ limits is only a factor of two.

\subsection{Late-time evolution of an aligned or orthogonal rotator}
\label{revival}

A natural question to ask is whether the evolution of $\chi$ ceases forever once the star reaches alignment or orthogonality. In the two strict mathematical limits, it is clear from equation \eqref{eq:chi-dot} that the answer is yes: the equation for the inclination angle reduces to $\dot\chi=0$, and the value of $\chi$ thus remains constant ever after. The physical problem is more complicated, though, because $\chi$ will only tend asymptotically to this value. The one exception to these statements is the case of a star evolving towards alignment under the action of our magnetospheric spindown prescription: in this case the limit $\chi=0$ is attained in finite time, and \eqref{eq:chi-dot} does not reduce to $\dot\chi=0$. We regard this as an artefact of the phenomenological nature of the magnetospheric spindown prescription -- which is not an exact, self-consistent result like the vacuum-dipole case. One fix that could be envisaged would be a similar phenomenological modification to the $\chi$-evolution equation, such that it does indeed reduce to $\dot\chi=0$ for $\chi=0$. In the limit $\chi\to\pi/2$, however, $\dot\chi\to 0$ for both spindown prescriptions.

Given this behaviour of $\chi$, one could then imagine a situation where it evolves until it is very close to $\pi/2$ -- because internal dissipation dominates over spindown to begin with -- but then, as the star cools and spins down further, it reaches a point where spindown becomes more effective than internal dissipation.  At this point the evolution, having stalled near $\chi=\pi/2$, could revive, and at a much later time the star could evolve to become an aligned rotator. In fact, this scenario had already been envisaged in one of the early papers on inclination-angle evolution \citep{jones_76_APSS}.  It is difficult to study this revival scenario by direct numerical simulation, since it entails evolving the governing equations over a very long period of time for which $\dot\chi\approx 0$, which in turn results in an accumulation of numerical error. In practice, we see unphysical behaviour in our evolutions where $\sin\chi$ grows bigger than unity and then diverges.

Before performing any additional checks, however, we already have reason to doubt whether the revival scenario would ever actually occur. Our numerical results have shown us that beginning with a large $\chi$ greatly reduces the fraction of stars which evolve to become aligned rotators, and that a small initial $\chi$ reduces the eventual fraction of orthogonal rotators; a star with (e.g.) $\chi\approx\pi/2$ would therefore need the spindown timescale to become \emph{far} shorter than the internal-dissipation timescale in order to evolve back to an aligned rotator, and vice-versa for a star with $\chi\approx 0$. Figure \ref{fig:timescales}, however, shows us that these two timescales are not actually significantly different at late times.

A numerical experiment reinforces our suspicions that the revival scenaio is unlikely. We take a model which orthogonalises, but whose parameters place it very close to the $\mathcal{A}-\mathcal{O}$ dividing line between aligned and orthogonal rotators, so that if it had a very slightly stronger magnetic field it would align. We first run the evolution until it reaches $\chi=179\pi/360$ (our standard cut-off value), then terminate it as usual.
 We wish to know if, at some later time, spindown can overcome the effect of viscosity and thus drive the star towards alignment. To this end, we assume that $\chi$ stops evolving for some time and remains fixed at $179\pi/360$; during this period the star continues to cool as usual, and the rotation also evolves according to the usual prescription, but $\chi$ is fixed. This means that both bulk viscosity and the spindown torque weaken in this interim period; if bulk viscosity weakens at a higher rate, then when we switch on the evolution of $\chi$ again, the star might evolve towards an aligned rotator. In practice though, the two weaken at a similar rate, and however long we wait before restarting the coupled evolution of $\Omega$ and $\chi$, the final phase of the $\chi$ evolution is to continue increasing -- from its initial restart state of $\chi=179\pi/360$ towards $\pi/2$. We have verified that this also happens for less extreme restart values of $\chi$ than $179\pi/360$: a model just on the orthogonal-rotator side of the $\mathcal{A}-\mathcal{O}$ boundary still orthogonalises eventually if its $\chi$-evolution is stopped and restarted in the above manner at a restart value of $\chi=9\pi/20$, and a model just on the aligned-rotator still aligns at late times even if its restart value is set to a relatively large value of $\chi=\pi/20$. Combining these arguments with the results shown in the right-hand panels of Figure \ref{rainbow_spindown}, we think it likely that in the absence of some additional physical mechanism,  most of the parameter space of models we have surveyed in this paper will not experience any evolution in their inclination angle after a hundred years.

\section{Validity of our approach}
\label{validity}

We have solved for the motion of a neutron star with misaligned rotation and magnetic axes (in \citet{LJ17}), and now have studied the damping of this motion when coupled to spindown. Through this programme of work, we believe we have the most complete picture to date of the evolution of $\chi$. The problem is, however, a complex one that required a number of approximations for certain technical steps of the calculation, and also physical assumptions that went into our neutron-star model. The technical-type approximations include the truncation of our non-dissipative solutions from \citet{LJ17} at the $l=4,m=2$ multipole, and approximating the bulk-viscosity coefficient in this paper by a more convenient piecewise function -- but we do not believe any of these will have introduced serious qualitative errors in our results. There are aspects of our physical model which are not completely general, however, and so we wish first to review briefly the salient features of our model, and then discuss how universal our results are likely to be.

Briefly, then, our stellar model consist of a precessing star, with a fixed purely-toroidal field assumed not to evolve in time; it is distorted away from sphericity by both rotation and its magnetic field. Its equation of state, used to solve for the steady-state motions, is a polytrope with $P=P(\rho)\propto\rho^2$. There is no additional buoyancy force acting on fluid elements, as might appear in a neutron star at high temperature or with gradients in its chemical composition. In the outermost shell of the star, $0.9R_*<r\leq R_*$, we effectively assume the motions to vanish, since the ordering of our perturbation scheme breaks down there \citep{LJ17}. For the viscosity coefficients we need to make some assumptions about the microphysics: we assume the star to be composed of neutrons, protons and electrons only, with neither neutrons nor protons being in a superfluid state. We assume cooling is due to the modified Urca mechanism. We utilise two different spindown prescriptions: the analytic result for a rotating dipole in vacuum, and a fit to numerical results for a charge-filled NS magnetosphere. Although our internal magnetic-field model is purely toroidal, and therefore would not extend outside the star (it would require an exterior current to do so), we assume it matches to some dipolar poloidal exterior field and set the value of the exterior dipole field to be the volume-averaged interior-field strength.

Let us discuss some of these assumptions in more detail.

\emph{Purely toroidal field:} fields of this type are known to be generically unstable \citep{tayler}. We regard this choice as a first approximation to a stable mixed poloidal-toroidal field whose toroidal component is the dominant one, but the issue of finding physically-realistic and stable neutron-star magnetic equilibrium models remains unresolved. We know, however, that the magnetic field must be dominantly toroidal in order to distort the star into a prolate shape (a poloidal field, by contrast, deforms a star in an oblate manner), and this prolate shape is the fundamental reason that any of the models we consider are able to evolve to become orthogonal rotators.
 
\emph{Ratio of internal to external field:} the ratio between these two fields is the only serious free parameter in our model. We have set it to unity for the sake of definiteness, but it is likely that neutron stars harbour interior toroidal fields somewhat stronger than the exterior poloidal field whose strength we infer from spindown. The only inconsistent choice for us would be to assume an external field significantly \emph{stronger} than the interior toroidal field; this would suggest that the poloidal component of the field would dominate in the interior, and therefore that the star's magnetic distortion would be oblate. No such model would ever evolve to become an orthogonal rotator.

\emph{No superfluidity or crust:} because of these assumptions, our models are only directly applicable to the early era in a neutron star's life. Crust-freezing and condensation of the core into superfluid phases are processes which happen gradually, but very roughly one can regard their effects as becoming significant after one hundred years \citep{kruger_ho_andersson_15}.  Once the crust forms, there will be elastic contributions to the effective moment of inertia tensor, with the shape-change induced by the centrifugal bulge adding to the magnetic distortions, to an extent determined by the shear modulus of the crust.  The problem of a precessing biaxial star with an elastic crust was considered in detail in \citet{goldreich70}, who showed that the effect of the electromagnetic torque is then to damp (or pump) the precessional motion on the electromagnetic timescale, with no secular variation in the angle between the magnetic axis and spin axis.  The onset of neutron superfluidity in the interior complicates the picture further, opening up the possibility of the neutron rotational vortices `pinning' to the crust, effectively adding a gyroscopic term to the precessional motion \citep{shaham77}.  It therefore seems likely that, after the onset of crust formation and superfluidity, secular variations in inclination angle can only proceed on longer timescales connected with plastic flow or vortex creep, substantially arresting the evolutions considered in our analysis.  It is possible that the associated  small- or large-scale failure events might be observable as discrete changes in the inclination angle. We note, however, that most of our models have completed their $\chi$-evolution (and become either aligned or orthogonal rotators) by the hundred-year mark.

\emph{Buoyancy forces:} these have the effect of making the motion of fluid elements closer to incompressible, $\div\dot\bxi\approx 0$, which is a major concern: if this term were to be precisely zero, there would be no dissipation due to bulk viscosity, and so evolution would proceed on the far slower shear-viscous or Ohmic-diffusion timescale. In fact, Mestel's original solutions \emph{assumed} incompressible flow in order to simplify the problem; if one consistently adopted his results, one would conclude that almost no neutron stars would ever orthogonalise. Since our evolutions begin when the star is still very hot, we will first check how quickly thermal buoyancy forces become negligible. \citet{prakash97} give an expression for the ratio of thermal pressure $P_{\rm th}$ to the standard zero-temperature degeneracy pressure $P_0$ of the form:
\beq
\frac{P_{\rm th}}{P_0}\approx \frac{5}{3\pi^2} s_B^2,
\eeq
where $s_B$ is the entropy per baryon in units of the Boltzmann constant, and where we have dropped some correction terms not relevant to our discussion. Now, by the end of the proto-neutron-star phase $s_B$ is already below unity; adopting the value $s_B=0.5$ taken from the end of the simulations of \citet{BL86} gives
\beq
\frac{P_{\rm th}}{P_0}\approx 0.04,
\eeq
meaning that even for our initial state of $T=10^{11}$ K thermal buoyancy forces are not large, and that after less than a minute (in which the temperature drops below $10^{10}$ K) they become completely negligible. We have confirmed numerically that our results are little affected if we turn off bulk viscosity for temperatures above $10^{10}$ K.

Buoyancy forces may also originate from chemical gradients in the star -- in particular, its density-dependent ratio of protons to neutrons. If a fluid element is abruptly displaced from its equilibrium position, it retains its original composition for a timescale $\tau$, corresponding to how long chemical reactions take to re-equilibrate the element to its new surroundings \citep{reisenegger_goldreich_92}. Over that timescale, buoyancy forces will be present, acting to push the fluid element back to its previous position. This reaction timescale is the same $\tau$ as that of equation \eqref{eq:RG_relaxation_timescale}; and if it is shorter than the precession timescale $T_\omega$, some degree of buoyancy will affect the motion and hence the evolutions we describe in this paper. Equivalently, composition-gradient buoyancy becomes important in the regime $\omega\tau\gg 1$.

  Now turning to equation \eqref{eq:omega_tau}, we see that any $\chi$-evolution which is still ongoing once the star drops below $T\sim 10^{10}$ K is likely to be affected after this point, given the steep temperature scaling of $\omega\tau$. Buoyancy forces will also play a role in stars born slowly rotating and/or with weak magnetic fields, even above $10^{10}$ K. On the other hand, stars with $B\gtrsim 10^{15}$ G and $f_0\gtrsim 100$ Hz are likely to complete their $\chi$-evolutions whilst still in the $\omega\tau\slash\!\!\!\!\!\gg 1$ regime.

\section{Discussion}
\label{discussion}

The inclination angle of a neutron star is a key parameter for understanding neutron-star emission -- both electromagnetic and gravitational. This is the first study to account for the competition between the most fundamental effects which can drive this angle towards either zero or $\pi/2$, and the first to treat the internal-dissipation problem self-consistently. A strong internal toroidal field is the \emph{only} feasible mechanism, to our knowledge, which allows for a neutron star to evolve towards a state where its rotation and magnetic symmetry axes are orthogonal. \citet{gourg_holl} showed that crustal magnetic field evolution could produce features mimicking the wandering of the magnetic axis, but this would be a transient feature and not one with a clear directionality towards an orthogonal-rotator state. It is also obvious that there is a bias towards detecting pulsars with larger $\chi$ values, since this geometry increases the chance that the star's radio beam will sweep past a terrestrial telescope. What we study here, however, is a genuine evolution of the axes towards orthogonality. It can only operate for an internal magnetic field whose toroidal component dominates; if the poloidal component is the larger one, there is no mechanism for the inclination angle to increase over time.

Generally speaking, we find that for the majority of the parameter space corresponding to attributes of known neutron stars, the star is expected to evolve to become an orthogonal rotator. The range of birth parameters which lead to aligned rotators is much smaller: our prediction is that alignment only occurs if the magnetic field is greater than roughly $10^{14}$ G and the birth rotation rate less than around $100$ Hz. There is also a very narrow region of millisecond magnetars, with rotation and magnetic axes almost aligned at birth -- which evolve to become aligned rotators too.

Since observations suggest that magnetars are generally close to being aligned rotators \citep{welt_john}, our results can be used to infer details about the possible birth spins of magnetars and their internal toroidal magnetic fields. In particular, millisecond magnetars are commonly invoked as the central engine of some gamma-ray bursts and superluminous supernovae \citep{metzger} -- so it is natural to ask whether these neutron stars later evolve into the older, slowly rotating (but still highly magnetised) magnetars we observe: those which appear to be near-aligned rotators. Our results indicate that these observed magnetars could indeed have started life as a population with millisecond periods, but only for extremely strong toroidal magnetic fields, above roughly $5\times 10^{15}$ G. Such millisecond magnetars, which appear to represent the optimal scenario for a neutron star to orthogonalise and emit copious gravitational radiation \citep{cutler_02,dallosso09}, therefore evolve extremely quickly into stationary, non-radiating configurations. Magnetar spindown is -- however -- more complicated than the simple prescriptions we have used here, involving additional braking due to the stellar wind; we intend to return to this problem in future, to check whether our tentative conclusions are confirmed by more detailed modelling.

At any given snapshot in time, it is notable that there are only ever thin bands of oblique rotators (i.e. those stars with $\chi$ close to neither zero nor $\pi/2$); most models are either aligned or orthogonal already, or have not evolving significantly from their initial state. Furthermore, by the age of a hundred years (at which our model becomes less trustworthy) most models corresponding to typical neutron-star values have finished their $\chi$-evolution, and only stars along the narrow delineation between orthogonal and aligned rotators are predicted to be oblique rotators. A similar qualitative result was obtained by \citet{DP17}, although they did not account for the aligning effect of the electromagnetic torque ($\chi$ can only \emph{increase} in their work), and so their dividing line is really between orthogonal and unevolved models. This suggests that either there is some additional physical input to our model which would increase the region of intermediate inclination angles, or that there is simply an observational bias towards detecting these oblique-rotator neutron stars (clearly, an entirely aligned rotator could not be seen as a pulsar). Our broad result that inclination angles should be clustered around the two limiting cases of $0$ and $\pi/2$ is borne out by various observation hints of bimodality \citep{taur_man,rookyard}. Quantitatively, though, it is more problematic to reconcile our results with observations: our results predict that a typical pulsar will always be an orthogonal rotator, with low inclination angles only expected for field strengths greater than typical pulsar values (around $10^{14}$ G) and birth rotation rates which are rather low (below roughly $10$ Hz). Still, at least the orthogonal-rotator configuration is compatible with being a pulsar -- had we found typical pulsar parameters to correspond to stars which evolve into aligned rotators, our model would have been in serious conflict with observations.

The most obvious physical ingredient missing from our model at early times after the star's birth, is the presence of buoyancy forces related to composition gradients.  For magnetic fields below typical magnetar values and initial spins below $100$ Hz, these buoyancy forces are likely to cause the inclination-angle evolution to stall and thus to broaden the region of parameter space with intermediate inclination angles. At later times, when the crust has formed, the $\chi$-evolution will not be smooth as for our fluid-star model here. Instead, we anticipate intermittent evolution in discrete increments related to seismic activity (i.e. small-scale elastic failure events) of the crust. This process needs to be modelled in order to understand any observations of changes in $\chi$ in mature neutron stars. In particular, we speculate that such changes are more likely to be observed in frequently-glitching pulsars, and could be related to the apparent $0.6^\circ$/century increase in $\chi$ for the Crab pulsar \citep{crab_angle}.

The inclination angle and the birth spin of a neutron star are both difficult quantities to determine, and both rely on significant extrapolations of theoretical models or observational data. The internal field strength of a neutron star is also not known, though the most reasonable estimates would suggest that it is comparable or a little larger than the exterior value. The evolutions described in this paper give the opportunity to combine information from these partially-known parameters to glean interesting hints about aspects of neutron stars which are difficult to probe.  For example, a reasonable guess for a given neutron star's internal field strength and observational information that it is approximately an orthogonal rotator would allow us to put a bound on its birth spin rate. Our work is complementary to other recent studies invoking a decaying inclination angle to explain properties of the observed pulsar population distribution \citep{gullon,john_kara}; given that we find many neutron stars will instead experience an \emph{increasing} inclination angle in their early lives, it will be interesting to see the effect this has on population syntheses.

In this discussion section we have outlined some ways in which our work can be used to connect theoretical and observed properties of rotating, magnetised neutron stars, though our conclusions are very preliminary. The main aim of this paper was instead to lay down the theory of precession dissipation in young neutron stars, and we will explore the observational implications of our results in more detail elsewhere.


\section*{Acknowledgements}

SKL acknowledges support from the European Union's Horizon 2020
research and innovation programme under the Marie Sk\l{}odowska-Curie
grant agreement No. 665778, via fellowship UMO-2016/21/P/ST9/03689 
of the National Science Centre, Poland.  DIJ acknowledges funding from STFC through grant number ST/M000931/1.

We would like to dedicate this paper to the memory of Leon Mestel, a true pioneer in the study of astrophysical magnetism.


\bibliographystyle{mnras}


\bibliography{references}

\begin{thebibliography}{}
\makeatletter
\relax
\def\mn@urlcharsother{\let\do\@makeother \do\$\do\&\do\#\do\^\do\_\do\%\do\~}
\def\mn@doi{\begingroup\mn@urlcharsother \@ifnextchar [ {\mn@doi@}
  {\mn@doi@[]}}
\def\mn@doi@[#1]#2{\def\@tempa{#1}\ifx\@tempa\@empty \href
  {http://dx.doi.org/#2} {doi:#2}\else \href {http://dx.doi.org/#2} {#1}\fi
  \endgroup}
\def\mn@eprint#1#2{\mn@eprint@#1:#2::\@nil}
\def\mn@eprint@arXiv#1{\href {http://arxiv.org/abs/#1} {{\tt arXiv:#1}}}
\def\mn@eprint@dblp#1{\href {http://dblp.uni-trier.de/rec/bibtex/#1.xml}
  {dblp:#1}}
\def\mn@eprint@#1:#2:#3:#4\@nil{\def\@tempa {#1}\def\@tempb {#2}\def\@tempc
  {#3}\ifx \@tempc \@empty \let \@tempc \@tempb \let \@tempb \@tempa \fi \ifx
  \@tempb \@empty \def\@tempb {arXiv}\fi \@ifundefined
  {mn@eprint@\@tempb}{\@tempb:\@tempc}{\expandafter \expandafter \csname
  mn@eprint@\@tempb\endcsname \expandafter{\@tempc}}}

\bibitem[\protect\citeauthoryear{{Bildsten} \& {Ushomirsky}}{{Bildsten} \&
  {Ushomirsky}}{2000}]{bildsten_ushomirsky_00}
{Bildsten} L.,  {Ushomirsky} G.,  2000, \mn@doi [\apjl] {10.1086/312454}, \href
  {http://adsabs.harvard.edu/abs/2000ApJ...529L..33B} {529, L33}

\bibitem[\protect\citeauthoryear{{Bonanno}, {Rezzolla}  \& {Urpin}}{{Bonanno}
  et~al.}{2003}]{bonanno}
{Bonanno} A.,  {Rezzolla} L.,   {Urpin} V.,  2003, \mn@doi [\aap]
  {10.1051/0004-6361:20031459}, \href
  {http://adsabs.harvard.edu/abs/2003A%26A...410L..33B} {410, L33}

\bibitem[\protect\citeauthoryear{{Burrows} \& {Lattimer}}{{Burrows} \&
  {Lattimer}}{1986}]{BL86}
{Burrows} A.,  {Lattimer} J.~M.,  1986, \mn@doi [\apj] {10.1086/164405}, \href
  {http://adsabs.harvard.edu/abs/1986ApJ...307..178B} {307, 178}

\bibitem[\protect\citeauthoryear{{Cutler}}{{Cutler}}{2002}]{cutler_02}
{Cutler} C.,  2002, \mn@doi [\prd] {10.1103/PhysRevD.66.084025}, \href
  {http://ukads.nottingham.ac.uk/abs/2002PhRvD..66h4025C} {66, 084025}

\bibitem[\protect\citeauthoryear{{Cutler} \& {Jones}}{{Cutler} \&
  {Jones}}{2001}]{cutler_jones_01}
{Cutler} C.,  {Jones} D.~I.,  2001, \mn@doi [\prd]
  {10.1103/PhysRevD.63.024002}, \href
  {http://adsabs.harvard.edu/abs/2001PhRvD..63b4002C} {63, 024002}

\bibitem[\protect\citeauthoryear{{Cutler} \& {Lindblom}}{{Cutler} \&
  {Lindblom}}{1987}]{cutler_lindblom_87}
{Cutler} C.,  {Lindblom} L.,  1987, \mn@doi [\apj] {10.1086/165052}, \href
  {http://adsabs.harvard.edu/abs/1987ApJ...314..234C} {314, 234}

\bibitem[\protect\citeauthoryear{{Dall'Osso} \& {Perna}}{{Dall'Osso} \&
  {Perna}}{2017}]{DP17}
{Dall'Osso} S.,  {Perna} R.,  2017, \mn@doi [\mnras] {10.1093/mnras/stx2097},
  \href {http://adsabs.harvard.edu/abs/2017MNRAS.472.2142D} {472, 2142}

\bibitem[\protect\citeauthoryear{{Dall'Osso}, {Shore}  \& {Stella}}{{Dall'Osso}
  et~al.}{2009}]{dallosso09}
{Dall'Osso} S.,  {Shore} S.~N.,   {Stella} L.,  2009, \mn@doi [\mnras]
  {10.1111/j.1365-2966.2008.14054.x}, \href
  {http://adsabs.harvard.edu/abs/2009MNRAS.398.1869D} {398, 1869}

\bibitem[\protect\citeauthoryear{{Dall'Osso}, {Stella}  \&
  {Palomba}}{{Dall'Osso} et~al.}{2018}]{dallosso_spin}
{Dall'Osso} S.,  {Stella} L.,   {Palomba} C.,  2018, preprint, \href
  {http://adsabs.harvard.edu/abs/2018arXiv180611164D} {} (\mn@eprint {arXiv}
  {1806.11164})

\bibitem[\protect\citeauthoryear{{Davis} \& {Goldstein}}{{Davis} \&
  {Goldstein}}{1970}]{davis_goldstein_70}
{Davis} L.,  {Goldstein} M.,  1970, \mn@doi [\apjl] {10.1086/180482}, \href
  {http://adsabs.harvard.edu/abs/1970ApJ...159L..81D} {159}

\bibitem[\protect\citeauthoryear{{Ferrario}, {Melatos}  \& {Zrake}}{{Ferrario}
  et~al.}{2015}]{ferrario}
{Ferrario} L.,  {Melatos} A.,   {Zrake} J.,  2015, \mn@doi [\ssr]
  {10.1007/s11214-015-0138-y}, \href
  {http://adsabs.harvard.edu/abs/2015SSRv..191...77F} {191, 77}

\bibitem[\protect\citeauthoryear{{Flowers} \& {Itoh}}{{Flowers} \&
  {Itoh}}{1979}]{flowers_itoh_79}
{Flowers} E.,  {Itoh} N.,  1979, \mn@doi [\apj] {10.1086/157145}, \href
  {http://adsabs.harvard.edu/abs/1979ApJ...230..847F} {230, 847}

\bibitem[\protect\citeauthoryear{{Gnedin}, {Yakovlev}  \& {Potekhin}}{{Gnedin}
  et~al.}{2001}]{gnedin}
{Gnedin} O.~Y.,  {Yakovlev} D.~G.,   {Potekhin} A.~Y.,  2001, \mn@doi [\mnras]
  {10.1046/j.1365-8711.2001.04359.x}, \href
  {http://adsabs.harvard.edu/abs/2001MNRAS.324..725G} {324, 725}

\bibitem[\protect\citeauthoryear{{Goldreich}}{{Goldreich}}{1970}]{goldreich70}
{Goldreich} P.,  1970, \mn@doi [\apjl] {10.1086/180513}, \href
  {http://adsabs.harvard.edu/abs/1970ApJ...160L..11G} {160, L11}

\bibitem[\protect\citeauthoryear{{Goldreich} \& {Reisenegger}}{{Goldreich} \&
  {Reisenegger}}{1992}]{goldreich_reisenegger_92}
{Goldreich} P.,  {Reisenegger} A.,  1992, \mn@doi [\apj] {10.1086/171646},
  \href {http://adsabs.harvard.edu/abs/1992ApJ...395..250G} {395, 250}

\bibitem[\protect\citeauthoryear{{Gourgouliatos} \&
  {Hollerbach}}{{Gourgouliatos} \& {Hollerbach}}{2018}]{gourg_holl}
{Gourgouliatos} K.~N.,  {Hollerbach} R.,  2018, \mn@doi [\apj]
  {10.3847/1538-4357/aa9d93}, \href
  {http://adsabs.harvard.edu/abs/2018ApJ...852...21G} {852, 21}

\bibitem[\protect\citeauthoryear{{Gruzinov}}{{Gruzinov}}{2005}]{gruzinov}
{Gruzinov} A.,  2005, \mn@doi [Physical Review Letters]
  {10.1103/PhysRevLett.94.021101}, \href
  {http://adsabs.harvard.edu/abs/2005PhRvL..94b1101G} {94, 021101}

\bibitem[\protect\citeauthoryear{{Guilet} \& {M{\"u}ller}}{{Guilet} \&
  {M{\"u}ller}}{2015}]{guilet}
{Guilet} J.,  {M{\"u}ller} E.,  2015, \mn@doi [\mnras] {10.1093/mnras/stv727},
  \href {http://adsabs.harvard.edu/abs/2015MNRAS.450.2153G} {450, 2153}

\bibitem[\protect\citeauthoryear{{Gull{\'o}n}, {Miralles}, {Vigan{\`o}}  \&
  {Pons}}{{Gull{\'o}n} et~al.}{2014}]{gullon}
{Gull{\'o}n} M.,  {Miralles} J.~A.,  {Vigan{\`o}} D.,   {Pons} J.~A.,  2014,
  \mn@doi [\mnras] {10.1093/mnras/stu1253}, \href
  {http://adsabs.harvard.edu/abs/2014MNRAS.443.1891G} {443, 1891}

\bibitem[\protect\citeauthoryear{{Ipser} \& {Lindblom}}{{Ipser} \&
  {Lindblom}}{1991}]{ipser_lindblom_91}
{Ipser} J.~R.,  {Lindblom} L.,  1991, \mn@doi [\apj] {10.1086/170039}, \href
  {http://adsabs.harvard.edu/abs/1991ApJ...373..213I} {373, 213}

\bibitem[\protect\citeauthoryear{{Johnston} \& {Karastergiou}}{{Johnston} \&
  {Karastergiou}}{2017}]{john_kara}
{Johnston} S.,  {Karastergiou} A.,  2017, \mn@doi [\mnras]
  {10.1093/mnras/stx377}, \href
  {http://adsabs.harvard.edu/abs/2017MNRAS.467.3493J} {467, 3493}

\bibitem[\protect\citeauthoryear{{Jones}}{{Jones}}{1976}]{jones_76_APSS}
{Jones} P.~B.,  1976, \mn@doi [\apss] {10.1007/BF00642671}, \href
  {http://ukads.nottingham.ac.uk/abs/1976Ap%26SS..45..369J} {45, 369}

\bibitem[\protect\citeauthoryear{{Jones} \& {Andersson}}{{Jones} \&
  {Andersson}}{2001}]{jones_andersson_01}
{Jones} D.~I.,  {Andersson} N.,  2001, \mn@doi [\mnras]
  {10.1046/j.1365-8711.2001.04251.x}, \href
  {http://adsabs.harvard.edu/abs/2001MNRAS.324..811J} {324, 811}

\bibitem[\protect\citeauthoryear{{Kr{\"u}ger}, {Ho}  \&
  {Andersson}}{{Kr{\"u}ger} et~al.}{2015}]{kruger_ho_andersson_15}
{Kr{\"u}ger} C.~J.,  {Ho} W.~C.~G.,   {Andersson} N.,  2015, \mn@doi [\prd]
  {10.1103/PhysRevD.92.063009}, \href
  {https://ui.adsabs.harvard.edu/#abs/2015PhRvD..92f3009K} {92, 063009}

\bibitem[\protect\citeauthoryear{{Landau} \& {Lifshitz}}{{Landau} \&
  {Lifshitz}}{1987}]{L_and_L_fluids_87}
{Landau} L.~D.,  {Lifshitz} E.~M.,  1987, {Fluid mechanics}.
Butterworth-Heinemann

\bibitem[\protect\citeauthoryear{{Lander} \& {Jones}}{{Lander} \&
  {Jones}}{2017}]{LJ17}
{Lander} S.~K.,  {Jones} D.~I.,  2017, \mn@doi [\mnras] {10.1093/mnras/stx349},
  \href {http://adsabs.harvard.edu/abs/2017MNRAS.467.4343L} {467, 4343}

\bibitem[\protect\citeauthoryear{{Lasky} \& {Glampedakis}}{{Lasky} \&
  {Glampedakis}}{2016}]{lasky_glam}
{Lasky} P.~D.,  {Glampedakis} K.,  2016, \mn@doi [\mnras]
  {10.1093/mnras/stw435}, \href
  {http://adsabs.harvard.edu/abs/2016MNRAS.458.1660L} {458, 1660}

\bibitem[\protect\citeauthoryear{{Lindblom} \& {Owen}}{{Lindblom} \&
  {Owen}}{2002}]{lindblom_owen_02}
{Lindblom} L.,  {Owen} B.~J.,  2002, \mn@doi [\prd]
  {10.1103/PhysRevD.65.063006}, \href
  {http://adsabs.harvard.edu/abs/2002PhRvD..65f3006L} {65, 063006}

\bibitem[\protect\citeauthoryear{{Lyne}, {Graham-Smith}, {Weltevrede},
  {Jordan}, {Stappers}, {Bassa}  \& {Kramer}}{{Lyne} et~al.}{2013}]{crab_angle}
{Lyne} A.,  {Graham-Smith} F.,  {Weltevrede} P.,  {Jordan} C.,  {Stappers} B.,
  {Bassa} C.,   {Kramer} M.,  2013, \mn@doi [Science]
  {10.1126/science.1243254}, \href
  {http://adsabs.harvard.edu/abs/2013Sci...342..598L} {342, 598}

\bibitem[\protect\citeauthoryear{{Mestel} \& {Takhar}}{{Mestel} \&
  {Takhar}}{1972}]{mestel1}
{Mestel} L.,  {Takhar} H.~S.,  1972, \mn@doi [\mnras]
  {10.1093/mnras/156.4.419}, \href
  {http://adsabs.harvard.edu/abs/1972MNRAS.156..419M} {156, 419}

\bibitem[\protect\citeauthoryear{{Mestel}, {Nittmann}, {Wood}  \&
  {Wright}}{{Mestel} et~al.}{1981}]{mestel2}
{Mestel} L.,  {Nittmann} J.,  {Wood} W.~P.,   {Wright} G.~A.~E.,  1981, \mnras,
  \href {http://adsabs.harvard.edu/abs/1981MNRAS.195..979M} {195, 979}

\bibitem[\protect\citeauthoryear{{Metzger}, {Giannios}, {Thompson},
  {Bucciantini}  \& {Quataert}}{{Metzger} et~al.}{2011}]{metzger}
{Metzger} B.~D.,  {Giannios} D.,  {Thompson} T.~A.,  {Bucciantini} N.,
  {Quataert} E.,  2011, \mn@doi [\mnras] {10.1111/j.1365-2966.2011.18280.x},
  \href {http://adsabs.harvard.edu/abs/2011MNRAS.413.2031M} {413, 2031}

\bibitem[\protect\citeauthoryear{{Michel} \& {Goldwire}}{{Michel} \&
  {Goldwire}}{1970}]{michel_goldwire_70}
{Michel} F.~C.,  {Goldwire} Jr. H.~C.,  1970, \aplett, \href
  {http://adsabs.harvard.edu/abs/1970ApL.....5...21M} {5, 21}

\bibitem[\protect\citeauthoryear{{Munk} \& {MacDonald}}{{Munk} \&
  {MacDonald}}{1975}]{munk_macdonald}
{Munk} W.~H.,  {MacDonald} G.~J.~F.,  1975, {The rotation of the earth: a
  geophysical discussion.}.
Cambridge University Press

\bibitem[\protect\citeauthoryear{{Nittmann} \& {Wood}}{{Nittmann} \&
  {Wood}}{1981}]{nittmann_wood_81}
{Nittmann} J.,  {Wood} W.~P.,  1981, \mn@doi [\mnras]
  {10.1093/mnras/196.3.491}, \href
  {http://adsabs.harvard.edu/abs/1981MNRAS.196..491N} {196, 491}

\bibitem[\protect\citeauthoryear{{Obergaulinger}, {Cerd{\'a}-Dur{\'a}n},
  {M{\"u}ller}  \& {Aloy}}{{Obergaulinger} et~al.}{2009}]{obergau}
{Obergaulinger} M.,  {Cerd{\'a}-Dur{\'a}n} P.,  {M{\"u}ller} E.,   {Aloy}
  M.~A.,  2009, \mn@doi [\aap] {10.1051/0004-6361/200811323}, \href
  {http://adsabs.harvard.edu/abs/2009A%26A...498..241O} {498, 241}

\bibitem[\protect\citeauthoryear{{Ostriker} \& {Gunn}}{{Ostriker} \&
  {Gunn}}{1969}]{ostriker_gunn_69}
{Ostriker} J.~P.,  {Gunn} J.~E.,  1969, \mn@doi [\apj] {10.1086/150160}, \href
  {http://adsabs.harvard.edu/abs/1969ApJ...157.1395O} {157, 1395}

\bibitem[\protect\citeauthoryear{{Page}, {Geppert}  \& {Weber}}{{Page}
  et~al.}{2006}]{pgw_06}
{Page} D.,  {Geppert} U.,   {Weber} F.,  2006, \mn@doi [Nuclear Physics A]
  {10.1016/j.nuclphysa.2005.09.019}, \href
  {http://adsabs.harvard.edu/abs/2006NuPhA.777..497P} {777, 497}

\bibitem[\protect\citeauthoryear{{Prakash}, {Bombaci}, {Prakash}, {Ellis},
  {Lattimer}  \& {Knorren}}{{Prakash} et~al.}{1997}]{prakash97}
{Prakash} M.,  {Bombaci} I.,  {Prakash} M.,  {Ellis} P.~J.,  {Lattimer} J.~M.,
   {Knorren} R.,  1997, \mn@doi [\physrep] {10.1016/S0370-1573(96)00023-3},
  \href {http://adsabs.harvard.edu/abs/1997PhR...280....1P} {280, 1}

\bibitem[\protect\citeauthoryear{{Reisenegger} \& {Goldreich}}{{Reisenegger} \&
  {Goldreich}}{1992}]{reisenegger_goldreich_92}
{Reisenegger} A.,  {Goldreich} P.,  1992, \mn@doi [\apj] {10.1086/171645},
  \href {http://adsabs.harvard.edu/abs/1992ApJ...395..240R} {395, 240}

\bibitem[\protect\citeauthoryear{{Rembiasz}, {Guilet}, {Obergaulinger},
  {Cerd{\'a}-Dur{\'a}n}, {Aloy}  \& {M{\"u}ller}}{{Rembiasz}
  et~al.}{2016}]{rembiasz}
{Rembiasz} T.,  {Guilet} J.,  {Obergaulinger} M.,  {Cerd{\'a}-Dur{\'a}n} P.,
  {Aloy} M.~A.,   {M{\"u}ller} E.,  2016, \mn@doi [\mnras]
  {10.1093/mnras/stw1201}, \href
  {http://adsabs.harvard.edu/abs/2016MNRAS.460.3316R} {460, 3316}

\bibitem[\protect\citeauthoryear{{Rookyard}, {Weltevrede}  \&
  {Johnston}}{{Rookyard} et~al.}{2015}]{rookyard}
{Rookyard} S.~C.,  {Weltevrede} P.,   {Johnston} S.,  2015, \mn@doi [\mnras]
  {10.1093/mnras/stu2083}, \href
  {http://adsabs.harvard.edu/abs/2015MNRAS.446.3356R} {446, 3356}

\bibitem[\protect\citeauthoryear{{Sawyer}}{{Sawyer}}{1989}]{sawyer_89}
{Sawyer} R.~F.,  1989, \mn@doi [\prd] {10.1103/PhysRevD.39.3804}, \href
  {http://adsabs.harvard.edu/abs/1989PhRvD..39.3804S} {39, 3804}

\bibitem[\protect\citeauthoryear{{Shaham}}{{Shaham}}{1977}]{shaham77}
{Shaham} J.,  1977, \mn@doi [\apj] {10.1086/155249}, \href
  {http://adsabs.harvard.edu/abs/1977ApJ...214..251S} {214, 251}

\bibitem[\protect\citeauthoryear{{Spitkovsky}}{{Spitkovsky}}{2006}]{spit06}
{Spitkovsky} A.,  2006, \mn@doi [\apjl] {10.1086/507518}, \href
  {http://adsabs.harvard.edu/abs/2006ApJ...648L..51S} {648, L51}

\bibitem[\protect\citeauthoryear{{Tauris} \& {Manchester}}{{Tauris} \&
  {Manchester}}{1998}]{taur_man}
{Tauris} T.~M.,  {Manchester} R.~N.,  1998, \mn@doi [\mnras]
  {10.1046/j.1365-8711.1998.01369.x}, \href
  {http://adsabs.harvard.edu/abs/1998MNRAS.298..625T} {298, 625}

\bibitem[\protect\citeauthoryear{{Tayler}}{{Tayler}}{1973}]{tayler}
{Tayler} R.~J.,  1973, \mn@doi [\mnras] {10.1093/mnras/161.4.365}, \href
  {http://adsabs.harvard.edu/abs/1973MNRAS.161..365T} {161, 365}

\bibitem[\protect\citeauthoryear{{Thompson} \& {Duncan}}{{Thompson} \&
  {Duncan}}{1993}]{TD93}
{Thompson} C.,  {Duncan} R.~C.,  1993, \mn@doi [\apj] {10.1086/172580}, \href
  {http://adsabs.harvard.edu/abs/1993ApJ...408..194T} {408, 194}

\bibitem[\protect\citeauthoryear{{Weltevrede} \& {Johnston}}{{Weltevrede} \&
  {Johnston}}{2008}]{welt_john}
{Weltevrede} P.,  {Johnston} S.,  2008, \mn@doi [\mnras]
  {10.1111/j.1365-2966.2008.13382.x}, \href
  {http://adsabs.harvard.edu/abs/2008MNRAS.387.1755W} {387, 1755}

\bibitem[\protect\citeauthoryear{{Woltjer}}{{Woltjer}}{1964}]{woltjer}
{Woltjer} L.,  1964, \mn@doi [\apj] {10.1086/148028}, \href
  {http://adsabs.harvard.edu/abs/1964ApJ...140.1309W} {140, 1309}

\makeatother
\end{thebibliography}


\appendix

\section{Non-rigid corrections to precession-damping results} \label{sect:J_and_E_xi}

In this appendix we show that the contributions of the non-rigid response, or `$\xi$-motions', to the star's angular momentum, kinetic energy, and magnetic energy, are of higher order than the kinetic and angular momentum terms that we retain in section  \ref{sect:damp_form}.
See \citet{LJ17} for more detail on the results used here, particularly Section 3.2 for the scalings of $\xi$ and its derivatives with the small dimensionless parameters $\epsilon_\Omega$ and $\epsilon_B$.

First let us assess the contribution of the $\xi$-motions to the angular momentum:
\beq
{\boldsymbol J}={\boldsymbol J}_\Omega+{\boldsymbol J}_\omega+{\boldsymbol J}_\xi
 = \int\rho\br\times\Big[({\boldsymbol\Omega}_{\rm pri}+\omega\be_z)\times\br+\dot\bxi \Big]\ \rmd V,
\eeq
where ${\boldsymbol\Omega}_{\rm pri}$ denotes the piece of the star's rigid rotation proportional to the primary rotation; $\omega$ denotes the slow secondary rotation about the magnetic axis.
The first two
terms in ${\boldsymbol J}$, with subscripts $\Omega$ and $\omega$, add up to
give the full rigid solution, and the final term ${\boldsymbol J}_\xi$ is the
correction due to the fluid response $\bxi$. But we know that
\beq
\dot\bxi\sim\omega\xi\sim\epsilon_\Omega \omega,
\eeq
meaning that the non-rigid piece is a factor of $\epsilon_\Omega$
smaller than the piece due to the secondary rotation $\omega$, and so
may safely be neglected.

As a second check, let us compare the kinetic energy associated with
the non-rigid response with the rigid-body precessional kinetic
energy. The latter has scaling:
\beq
E_{\rm prec}\sim \frac{J^2}{I}\sim\frac{I^2\Omega^2}{I}\sim\Omega^2\sim\epsilon_\Omega.
\eeq
By contrast, the kinetic energy associated with the $\xi$-motions is:
\beq
E^{\rm kin}_\xi=\frac{1}{2}\int\rho\dot\xi^2\ \rmd V\sim\omega^2\xi^2\sim\epsilon_\Omega^3\epsilon_B^2,
\eeq
which is clearly of far higher order than $E_{\rm prec}$.
The magnetic energy associated with the non-rigid fluid response is
also of higher order than $E_{\rm prec}$:
\beq
E^{\rm mag}_\xi=\frac{1}{8\pi}\int(\delta B)^2\ \rmd V\sim\xi^2 B^2\sim\epsilon_\Omega^2\epsilon_B^2.
\eeq
We have thus established that for the purposes of calculating the
evolution of $\chi$, the relevant energy to dissipate is the standard
expression for precessional kinetic energy of a rigid body, whilst the
energies associated with the complex fluid response to precession is
of higher order and does not need to be accounted for. This justifies the neglect of these small corrections in section  \ref{sect:damp_form}.

\section{Shear and bulk viscosities}

\subsection{Dissipation integrals} \label{sect:diss_integrals}

For our work, we require explicit expressions for energy losses from compressible fluid motions due to both shear and bulk viscosity.  Somewhat surprisingly, we have been unable to find derivations or unequivocal statements about the nature of these dissipation integrals, written in terms of the fluid velocity as we require. For example, \citet{ipser_lindblom_91} define an `energy' [their quotation marks] of fluid perturbations -- whose exact physical meaning is not discussed -- and present its dissipation rate in terms of the fluid stress tensor. \citet{L_and_L_fluids_87} give a more explicit derivation for dissipation due to shear viscosity, but only for the case of an incompressible fluid. Here we wish to present full derivations for both kinds of viscous dissipation in the general case in terms of vector operations on the velocity field, making it explicit where terms have been dropped (e.g. surface integrals), and comparing with previous studies. We will regard the energy loss due to viscous dissipation as being given by the work done by each of the viscous `force' terms in the compressible Navier-Stokes equation \eqref{general_Euler}. 
Firstly then, the energy dissipation through bulk viscosity is given by
\begin{align}
\dot{E}_{\rm bulk} &= \int\bv\cdot\nabla(\zeta\div\bv)\ \rmd V\nn\\
 &= \int\left\{\div[\zeta\bv(\div\bv)]-\zeta(\div\bv)^2\right\}\ \rmd V\nn\\
 &= \int \zeta (\div\bv)\bv\cdot\hat{\boldsymbol n}\ \rmd S-\int \zeta (\div\bv)^2\ \rmd V\nn\\
 &= -\int \zeta (\div\bv)^2\ \rmd V,
\end{align}
where we have used the vector identity
$\div(\phi {\boldsymbol a})={\boldsymbol a}\cdot\nabla\phi-\phi\div{\boldsymbol a}$ and the
divergence theorem to convert one term into a surface integral, which
is assumed to vanish. We can justify this by noting that if $\rho=0$ at the surface, $\zeta$ should be zero too. The
above expression is indeed the standard one for bulk-viscosity dissipation.

Now we do the same for the shear term. Again, directly from our definition of viscous dissipation as the work done by the `force' term in \eqref{general_Euler}, we see that
\begin{align}
\dot{E}_{\rm shear} &= 2\int v_a\nabla_b(\eta \sigma_{ab})\ \rmd V\nn \\
& = 2\int \Big\{\nabla_b[v_a \eta \sigma_{ab}] - (\nabla_b v_a) \eta \sigma_{ab}\Big\}\ \rmd V.
\end{align}
The first term can be converted to a surface integral:
\beq
\dot{E}_{\rm shear} = 2\int  v_a \eta \sigma_{ab} \ dS_b  -   2 \int (\nabla_b v_a) \eta \sigma_{ab} \ \rmd V,
\eeq
and will be zero if, at the surface, we have $\eta=0$ or $v_a  \sigma_{ab} dS_b = 0$. The latter condition would be satisfied if the surface traction due to the shear stresses, $ \sigma_{ab} dS_b$, were zero.  This leaves:
\beq
\dot{E}_{\rm shear} =  -   2 \int  \eta (\nabla_b v_a) \sigma_{ab} \ \rmd V.
\eeq
The symmetry of $\sigma_{ab}$ allows us to write this as:
\beq
\dot{E}_{\rm shear} =  -   2 \int  \eta \frac{1}{2} (\nabla_b v_a + \nabla_a v_b)  \sigma_{ab} \ \rmd V,
\eeq
and by the trace-free nature of $\sigma_{ab}$
\beq
\dot{E}_{\rm shear} =  -   2 \int   \eta \left[\frac{1}{2}(\nabla_b v_a + \nabla_a v_b) -\frac{1}{3}(\nabla\cdot {\boldsymbol v}) g_{ab}\right]  \sigma_{ab} \ \rmd V,
\eeq
so we have
\beq
\label{eq:E_dot_shear_IL}
\dot{E}_{\rm shear} =  -   2 \int  \eta \sigma_{ab}\sigma_{ab} \ \rmd V,
\eeq
in agreement with \citet{ipser_lindblom_91}.  No assumption has been made about the constancy of $\eta$. 

We now proceed to write this result explicitly in terms of the velocity.  Let us start by partially `undoing' the last two steps above:
\begin{align}
\sigma_{ab} \sigma_{ab} &=  \left[\frac{1}{2}(\nabla_a v_b + \nabla_b v_a) -\frac{1}{3}(\nabla\cdot {\boldsymbol v}) g_{ab}\right]  (\nabla_a v_b) \nn\\
&= \frac{1}{2} (\nabla_a v_b) (\nabla_b v_a) + \frac{1}{2} (\nabla_b v_a) (\nabla_a v_b) - \frac{1}{3} (\nabla \cdot {\boldsymbol v})^2  
\end{align}
and note that
\begin{align}
|\curl\bv|^2=&(\curl\bv)_a (\curl\bv)_a\nn\\
 =&(\nabla_a v_b)(\nabla_a v_b)-(\nabla_a v_b)(\nabla_b v_a)\nn\\
 =&(\nabla_a v_b)(\nabla_a v_b)-(\nabla_b v_a)(\nabla_a v_b).
\end{align}
Then
\begin{align}
\sigma_{ab} \sigma_{ab} =&  (\nabla_a v_b)(\nabla_a v_b) -\frac{1}{2}|\nabla \times {\boldsymbol v}|^2 - \frac{1}{3} (\nabla \cdot {\boldsymbol v})^2 \nn \\
  =& \nabla_a[v_b(\nabla_a v_b)] - v_b\nabla_a(\nabla_a v_b) -\frac{1}{2}|\nabla \times {\boldsymbol v}|^2 - \frac{1}{3} (\nabla \cdot {\boldsymbol v})^2.   
\end{align}
Manipulating this to introduce total divergences, we have:
\begin{align}
\eta \sigma_{ab} \sigma_{ab} =&  \nabla_a[\eta v_b(\nabla_a v_b)] - (\nabla_a\eta)v_b \nabla_a v_b - \eta v_a \nabla^2 v_a \nn\\
&- \frac{1}{2}  \eta  |\nabla \times {\boldsymbol v}|^2 - \frac{1}{3} \eta (\nabla \cdot {\boldsymbol v})^2.
\end{align}
Next we use the fact that
\beq
v_b \nabla_a v_b = \frac{1}{2} \nabla_a (v^2) 
\eeq
and re-write the $\nabla^2 v_a$ term using
\beq
\nabla^2 {\boldsymbol v} = \nabla (\nabla \cdot {\boldsymbol v}) - \nabla \times (\nabla \times {\boldsymbol v})
\eeq
to give 
\begin{align}
  \eta \sigma_{ab}\sigma_{ab}
  =&  \nabla_a[\eta v_b(\nabla_a v_b)]  - \eta v_a\{\nabla_a(\div{\boldsymbol v}) - [\nabla \times (\nabla \times {\boldsymbol v})]_a\}\nn\\
& - \frac{1}{2}(\nabla_a\eta) (\nabla_a v^2) - \frac{1}{2}  \eta  |\nabla \times {\boldsymbol v}|^2 - \frac{1}{3} \eta (\nabla \cdot {\boldsymbol v})^2. 
\label{sigsig_mid}
\end{align}
Now we make use of the result
\beq
\nabla\cdot[\eta{\boldsymbol v}\times(\nabla\times{\boldsymbol v})] = [\nabla\times(\eta{\boldsymbol v})]\cdot(\nabla\times{\boldsymbol v}) - \eta{\boldsymbol v}\cdot\nabla\times(\nabla\times{\boldsymbol v}),
\eeq
whose first term we rewrite using
\beq
\nabla\times(\eta{\boldsymbol v}) = \nabla\eta\times{\boldsymbol v} + \eta\nabla\times{\boldsymbol v},
\eeq
finding that
\begin{align}
  \nabla\cdot[\eta{\boldsymbol v} &\times(\nabla\times{\boldsymbol v})] \nn\\
  &= [ \nabla\eta\times{\boldsymbol v}+ \eta\nabla\times{\boldsymbol v}  ]\cdot(\nabla\times{\boldsymbol v})
        - \eta{\boldsymbol v}\cdot\nabla\times(\nabla\times{\boldsymbol v}).
\end{align}
We now use this to eliminate the term $\eta {\boldsymbol v} \cdot  [\nabla \times (\nabla \times {\boldsymbol v})]$ from equation \eqref{sigsig_mid} to give
\begin{align}
\eta \sigma_{ab} \sigma_{ab} =&  \nabla_a[\eta v_b(\nabla_a v_b)] - \nabla\cdot[(\eta{\boldsymbol v})\times(\nabla\times{\boldsymbol v})] \nn\\
& - \eta v_a \nabla_a(\nabla \cdot {\boldsymbol v}) - \frac{1}{2}(\nabla_a\eta) (\nabla_a v^2) \nn \\
& + (\nabla\eta\times{\boldsymbol v})\cdot(\nabla\times{\boldsymbol v}) + \frac{1}{2}  \eta  |\nabla \times {\boldsymbol v}|^2
 - \frac{1}{3} \eta (\nabla \cdot {\boldsymbol v})^2. 
\end{align}
We now rewrite the term $\eta v_a  \nabla_a(\nabla \cdot {\boldsymbol v})$ in the above using
\beq
\eta v_a  \nabla_a(\nabla \cdot {\boldsymbol v}) = \nabla_a(\eta v_a\nabla_b v_b) - \nabla\cdot(\eta{\boldsymbol v})(\nabla\cdot{\boldsymbol v})
\eeq
and note that
\beq
\nabla\cdot(\eta{\boldsymbol v}) = (\nabla\eta \cdot {\boldsymbol v}) (\nabla\cdot {\boldsymbol v}) + \eta(\nabla\cdot {\boldsymbol v})^2
\eeq
to arrive at the expression:
\begin{align}
\dot E_{\rm shear} = -2 \int \Big\{&
\nabla_a \big[ \eta v_b(\nabla_a v_b) -  \eta({\boldsymbol v}\times\nabla\times{\boldsymbol v})_a - \eta v_a \nabla\cdot{\boldsymbol v} \big]\nn\\
&+ (\nabla\eta \cdot {\boldsymbol v}) (\nabla\cdot {\boldsymbol v}) 
- \frac{1}{2}(\nabla_a\eta) (\nabla_a v^2)
   + \frac{2\eta}{3} (\nabla \cdot {\boldsymbol v})^2 \nn\\
&+ (\nabla\eta\times{\boldsymbol v})\cdot(\nabla\times{\boldsymbol v}) 
   + \frac{\eta}{2} |\nabla \times {\boldsymbol v}|^2\Big\}\ \rmd V.
\end{align}
As for the bulk-viscosity case, the first term here may be converted into a surface integral. If, as before, we assume this to give a negligible contribution we arrive at our final result:
\begin{align}
\dot E_{\rm shear}
 = -\int & \bigg[
            \eta |\nabla \times {\boldsymbol v}|^2
            + \frac{4}{3} \eta (\nabla \cdot {\boldsymbol v})^2
            - \nabla\eta\cdot\nabla(v^2) \nn\\
   &+2(\nabla\eta \cdot {\boldsymbol v}) (\nabla\cdot {\boldsymbol v}) 
+2 (\nabla\eta\times{\boldsymbol v})\cdot(\nabla\times{\boldsymbol v})
\bigg]  \ \rmd V.
\end{align}
Dropping the surface integral may not be safe in every situation -- for example, if there are significant surface motions -- but in our problem dissipation is assumed to occur in the stellar core only ($0\leq r\leq 0.9R_*$), so any surface contribution is irrelevant. 
It is aesthetically displeasing that the above dissipation integral is not manifestly negative, but we know it is guaranteed to be so, given our starting point of calculating the work done by the shear-viscosity `force'.  In cases where the gradients in $\eta$ are small we obtain something that \emph{is} manifestly negative:
\beq
\dot{E}_{\rm shear} = -\eta\int\left\{
                               \frac{4}{3}(\div\bv)^2+|\curl\bv|^2
\right\}\ \rmd V.
\eeq
For our numerical work we use the preceding five-term expression for $\dot{E}_{\rm shear}$ rather than the above two-term form, but the three additional terms involving $\nabla\eta$ do in fact prove to be numerically smaller than the others.

\subsection{Bulk viscosity coefficient} \label{sect:viscosity_coefficients}

We now summarise our treatment of bulk viscosity, highlighting a few discrepancies in the literature, and making explicit which results from the literature we have used. Firstly,
the full expression for $\zeta$ can be computed by taking the real part of equation (3.11) of \citet{lindblom_owen_02}, to give:
\begin{equation}
\label{eq:zeta_general_appendix}
\zeta \approx - \frac{n \tau}{1 + (\omega \tau)^2} \left.\frac{\partial P}{\partial x}\right|_n \frac{\rmd x}{\rmd n} ,
\end{equation}
as given by equation (\ref{eq:zeta_general}) in the main text, where the various symbols are defined.

\citet{reisenegger_goldreich_92} give an expression for the proton fraction:
\begin{equation}
\label{eq:x_GR}
x \approx 6 \times 10^{-3} \frac{\rho}{\rho_{\rm nuc}}  \Rightarrow x \approx 6 \times 10^{-3} \frac{n m_{\rm B}}{\rho_{\rm nuc}} 
\end{equation}
from which it follows that
\begin{equation}
\frac{\rmd x}{\rmd n} \approx  6 \times 10^{-3} \frac{m_{\rm B}}{\rho_{\rm nuc}} .
\end{equation}
To compute the derivative of the pressure, make use of equation (26) of \citet{reisenegger_goldreich_92}, which writes the total pressure as the sum of that due to non-relativistic neutrons, and relativistic electrons:
\begin{equation}
P = \frac{2}{5} n_{\rm n} E_{\rm Fn} + \frac{1}{4} n_{\rm e} E_{\rm Fe} ,
\end{equation}
where $E_{\rm Fn}$ and $E_{\rm Fe}$ are the Fermi energies of the neutrons and electrons, given by equations (19) and (20) of \citet{reisenegger_goldreich_92}:
\begin{equation}
E_{\rm Fn} = \frac{\hbar^2}{2m_{\rm n}} (3\pi^2 n_{\rm n})^{2/3} ,
\end{equation}
\begin{equation}
E_{\rm Fe} = \hbar c  (3\pi^2 n_{\rm e})^{1/3} .
\end{equation}
The total pressure is then
\begin{equation}
P = \frac{2}{5} n_{\rm n}  \frac{\hbar^2}{2m_{\rm n}} (3\pi^2 n_{\rm n})^{2/3} + \frac{1}{4} n_{\rm p} \hbar c  (3\pi^2 n_{\rm p})^{1/3} ,
\end{equation}
\skl{where we have invoked charge neutrality to replace $n_e$ with $n_p$.} We want to write the above in terms of $x \approx n_{\rm p} / n$ and $n$.  To convert, we use
\begin{equation}
n_{\rm n} = n (1-x), \hspace{20mm} n_{\rm p} \approx n x ,
\end{equation}
to give 
\begin{equation}
P = \frac{1}{5}   \frac{\hbar^2}{m_{\rm n}} (3\pi^2)^{2/3} n^{5/3} (1-x)^{5/3}  + \frac{1}{4} \hbar c  (3\pi^2)^{1/3} n^{4/3} x^{4/3} .
\end{equation}
Differentiating this, we see that
\begin{equation}
\left.\frac{\partial P}{\partial x}\right|_n \approx  -\frac{1}{3} \frac{\hbar^2}{m_{\rm n}} (3\pi^2)^{2/3} n^{5/3} (1-x)^{2/3}  
+ \frac{1}{3} \hbar c  (3\pi^2)^{1/3} n^{4/3} x^{1/3} .
\end{equation}
Substituting $n = \rho / m_{\rm B}$, using equation (\ref{eq:x_GR}) for $x$, approximating $1-x \approx 1$,  and summing the two terms, we then obtain the result
\begin{equation}
\left.\frac{\partial p}{\partial x}\right|_n \approx  -5.24 \times 10^{33} \, {\rm erg} \, {\rm cm}^{-3} \, 
\left(\frac{\rho}{\rho_{\rm nuc}}\right)^{5/3} .
\end{equation}

Finally, the relaxation timescale $\tau$ is given by  \citet{reisenegger_goldreich_92}, who make use of the calculation of \citet{sawyer_89}.  The relevant reaction is that of modified Urca.  
Strictly, \citet{reisenegger_goldreich_92} discuss a specific type of perturbation: compositional g-modes, and do not specifically mention bulk viscosity.  They do, however, estimate the damping time of g-modes, by writing down the relaxation time associated with modified Urca.  This is the reaction relevant here, so can be used directly. The neutrino efficiency factor $\lambda$ given in their equation (34) agrees that of Sawyer's equation (13), once the latter result is corrected to account for a missing factor of $\rho_{15}^{2/3}$.  The relaxation timescale is proportional to $\lambda^{-1}$, and they give in their equation (35):
\begin{equation}
\label{eq:RG_relaxation_timescale_appendix}
\tau \sim \frac{0.2}{T_9^6} \left(\frac{\rho}{\rho_{\rm nuc}}\right)^{2/3} \, {\rm yr} .
\end{equation}
\citet{reisenegger_goldreich_92}  actually have a power of $-2/3$ on the density factor, but this seems to be a typographical error, as can be seen by tracing through the $\rho$ factors from their equation (34).  

Combining the above results, we can approximate the bulk viscosity and its associated energy-dissipation timescale in the two limiting cases $\omega\tau\ll 1$ and $\omega\tau\gg 1$. First making the approximation $\omega \tau \ll 1$ (not considered by \citet{dallosso09}), we find:
\begin{equation}
\zeta(\omega \tau \ll 1) \approx 2.3 \times 10^{32} \, {\rm g \, cm}^{-1} \, {\rm s}^{-1} \, 
\left(\frac{\rho}{\rho_{\rm nuc}}\right)^{10/3} \frac{1}{T_{10}^6} ,
\end{equation}
insertion of which into equation (\ref{eq:tau_E_both_parameterised}) gives
\begin{equation} 
\tau_{\chi, \, \rm mUrca} (\omega \tau \ll 1)  = 144 {\, \rm seconds \,} \frac{M_{1.4}^5}{R_6^{11}}  \left(\frac{\rho_{\rm nuc}}{\rho}\right)^{10/3} 
\frac{T_{10}^6}{f_{\rm kHz}^4  B_{15}^2} .
\end{equation}
Replacing $\rho$ with the average density of our canonical neutron star we obtain
\begin{equation}
\label{eq:zeta_low_freq_appendix}
\zeta(\omega \tau \ll 1) \approx 4.18 \times 10^{33} \, {\rm g \, cm}^{-1} \, {\rm s}^{-1} \, 
 \frac{1}{T_{10}^6} \left(\frac{M_{1.4}}{R_6^3}\right)^{10/3} ,
\end{equation}
and
\begin{equation} 
\label{eq:tau_E_npe_small_appendix}
\tau_{\chi, \, \rm mUrca} (\omega \tau \ll 1)  = 7.89 {\, \rm seconds \,} \frac{M_{1.4}^{5/3}}{R_6} 
\frac{T_{10}^6}{f_{\rm kHz}^4  B_{15}^2} .
\end{equation}

If instead we make the approximation $\omega \tau \gg1$ (considered by \citet{dallosso09}), then combining the above results we find:
\begin{align}
\label{eq:zeta_Detal}
\zeta (\omega \tau \gg 1) =& 5.5 \times 10^{-59}  \, {\rm g \, cm}^{-1} \, {\rm s}^{-1} \nn\\
&\times \left(\frac{\rho}{1 \, {\rm g \, cm}^{-3}}\right)^2
\left(\frac{T}{1 \, \rm K}\right)^6
\left(\frac{1 \, \rm Hz}{\omega}\right)^2 .
\end{align}
This  agrees with the result of \citet{dallosso09} (see the first equality of their equation (B4), although they write down the units incorrectly).    This equation has the same scalings as the corresponding result of Sawyer (his equation (17)), but is $92$ times bigger.  Given that the two calculations use the same $\lambda$-factor, it is not clear what the cause of the difference is.  In scalings more appropriate for neutron stars we have
\begin{equation}
\zeta (\omega \tau \gg 1) = 4.33 \times 10^{30}  \, {\rm g \, cm}^{-1} \, {\rm s}^{-1} \, 
\left(\frac{\rho}{\rho_{\rm nuc}}\right)^2 
T_{10}^6  
\left(\frac{1 \, \rm Hz}{\omega}\right)^2 .
\end{equation}
Even more usefully, we can eliminate the precession frequency $\omega$ in favour of the magnetic field strength using equation (\ref{eq:omega_parameterised})  to give
\begin{equation}
\label{eq:zeta_high_freq}
\zeta (\omega \tau \gg 1) = 3.01 \times 10^{34}  \, {\rm g \, cm}^{-1} \, {\rm s}^{-1} \!
\left(\frac{\rho}{\rho_{\rm nuc}}\right)^2\!
\frac{T_{10}^6 M_{1.4}^4}{\cos^2\chi B_{15}^4 f_{\rm kHz}^2 R_6^8}.
\end{equation}
Replacing $\rho$ with the average density $M / (4\pi R^3/3)$, we obtain:
\begin{equation}
\zeta (\omega \tau \gg 1) = 1.71 \times 10^{35}  \, {\rm g \, cm}^{-1} \, {\rm s}^{-1} \, 
T_{10}^6
\frac{M_{1.4}^6}{\cos^2\chi B_{15}^4 f_{\rm kHz}^2 R_6^{14}},
\end{equation}
and inserting this into the timescale estimate of equation  (\ref{eq:tau_E_both_parameterised}):
\begin{equation}
\label{eq:tau_E_npe_large_appendix}
\tau_{\chi, \, \rm mUrca} (\omega \tau \gg 1) = 0.192 {\, \rm seconds \,} \frac{R_6^3}{M_{1.4}} \frac{B_{15}^2}{f_{\rm kHz}^2 T_{10}^6} .
\end{equation}
This has exactly the same scalings as equation (13) of \citet{dallosso09}, but is a factor of $17$ longer.  The difference is presumably due to our use of back-of-the-envelope arguments, as opposed to their computation of a dissipation integral (albeit with a knowingly incorrect density perturbation).  Note that \citet{dallosso09} actually give two contradictory versions of their result for the damping time, in their equations (13) and (B20).  The result in their equation (B20) scales as $B^4$, so is clearly incorrect.

\label{lastpage}

\end{document}